\documentclass[aps, prd, twocolumn, twoside, 10pt, flushbottom,
nofootinbib]{revtex4-1}
\pdfoutput=1
\usepackage[dvipsnames]{xcolor}
\usepackage{graphicx}
\usepackage{slashed}
\usepackage{bbm,bm}
\usepackage{mathtools}
\usepackage{dcolumn}
\usepackage{multirow}

\newcommand{\Nf}{{N_\mathrm{f}}}                          
\newcommand{\Nfc}{N_\mathrm{f}^\text{cr}}                 
\newcommand{\iu}{\mathrm i}                               
\newcommand{\bpsi}{\bar\psi}
\newcommand{\bchi}{\bar\chi}
\newcommand{\NL}{N_\text{L}}
\newcommand{\NR}{N_\text{R}}
\DeclareMathOperator{\Tr}{Tr}                             
\DeclareMathOperator{\STr}{STr}
\DeclareMathOperator{\IntX}{\int_0^\infty d\mathit{x}}
\DeclareMathOperator{\TildeDt}{\tilde\partial_\mathit{t}}
\DeclareMathOperator{\Diag}{diag}
\newcolumntype{d}[1]{D{.}{.}{#1}}
\newcommand{\Eqref}[1]{Eq.~\eqref{#1}}

\begin{document}
\title{Critical behavior of the (2+1)-dimensional Thirring model}
\date{August 16, 2012}
\author{Lukas Janssen}
\email{lukas.janssen@uni-jena.de}
\author{Holger Gies}
\email{holger.gies@uni-jena.de}
\affiliation{Theoretisch-Physikalisches Institut,
Friedrich-Schiller-Universit{\"a}t Jena, Max-Wien-Platz 1, 07743
Jena, Germany}

\begin{abstract}
We investigate chiral symmetry breaking in the (2+1)-dimensional Thirring
model as a function of the coupling as well as the Dirac flavor number $\Nf$
with the aid of the functional renormalization group. For small enough flavor
number $\Nf<\Nfc$, the model exhibits a chiral quantum phase transition for
sufficiently large coupling. We compute the critical exponents of this second
order transition as well as the fermionic and bosonic mass spectrum inside the
broken phase within a next-to-leading order derivative expansion. We also
determine the quantum critical behavior of the many-flavor transition which
arises due to a competition between vector and chiral-scalar channel and which
is of second order as well. Due to the problem of competing channels, our
results rely crucially on the RG technique of dynamical bosonization. For the
critical flavor number, we find $\Nfc\simeq 5.1$ with an estimated systematic
error of approximately one flavor.
\end{abstract}

\maketitle

\section{Introduction and Summary}

{\bf Introduction.}
Since the advent of graphene~\cite{novoselov2004electric} there has
been an enormous amount of renewed interest in (2+1)-dimensional
relativistic fermion systems. Various variants of (2+1)d quantum
electrodynamics~\cite{Pisarski:1984dj, Cornwall:1980zw,  Rao:1986jx,
Appelquist:1986fd, Appelquist:1986qw, Appelquist:1988sr, Nash:1989xx,
Maris:1996zg, Kubota:2001kk, Kaveh:2004qa, Hands:2002dv, Hands:2004bh,
Gusynin:2003ww, Fischer:2004nq, Bashir:2009fv, Li:2010zzf, Gusynin:1998kz}
and the (2+1)d Thirring model \cite{Parisi:1975im, Hikami:1976at, Hands:1994kb,
Gomes:1990ed, Hong:1993qk, Itoh:1994cr, Sugiura:1996xk, Kondo:1995jn,
Kim:1996xza, DelDebbio:1997dv, DelDebbio:1999xg, Hands:1999id, Christofi:2007ye,
Chandrasekharan:2011mn, Mesterhazy:2012ei} are actively discussed as effective
descriptions of graphene's exceptional electronic properties
\cite{Semenoff:1984dq, Herbut:2006cs, Herbut:2009qb, drut2009graphene,
Hands:2008id, Armour:2009vj, Gusynin:2007ix, Cortijo:2011aa}.
Moreover, they are likewise intrinsically interesting: in these models the
number of fermion flavors $\Nf$ serves as a control parameter for a quantum
phase transition at a critical value $\Nfc$. 
Several previous works provide a substantial amount of evidence that chiral
symmetry breaking ($\chi$SB) may be prohibited even for arbitrarily large
coupling if 
$\Nf > \Nfc$ \cite{Gomes:1990ed, Itoh:1994cr, Sugiura:1996xk, Kondo:1995jn,
Kim:1996xza, DelDebbio:1997dv, DelDebbio:1999xg, Hands:1999id,
Christofi:2007ye}.
This is a similarity to many-flavor non-abelian gauge theories in four
dimensions which are used for particle physics models for dynamical
electroweak symmetry
breaking~\cite{Miransky:1996pd, Appelquist:1996dq, Appelquist:1998xf}.
However, the search for the quantum critical point has so far
been rather challenging: in the Thirring model different approximate solutions
to the Dyson-Schwinger equations (DSE) yielded values
between $\Nfc \simeq 3.24$ \cite{Gomes:1990ed} and $\Nfc = \infty$
\cite{Hong:1993qk}. By constructing an effective potential for the chiral order parameter, up to leading order of the $1/\Nf$ expansion $\Nfc = 2$ has been 
found \cite{Kondo:1995jn}.
Extensive Monte Carlo simulations point to $\Nfc
\simeq 6.6$ \cite{Christofi:2007ye}. 
The true value of $\Nfc$ 
is of decisive relevance for the applicability of such relativistic
fermion effective theories for condensed matter systems, where the number of
flavors is typically $\Nf = 2$.
If $\Nf=2$ is near $\Nfc$, the properties of the
quantum critical regime could take an important influence on physical effects
corresponding to dynamical mass generation in the effective
(strongly-coupled) theories. 
Understanding such a semimetal-Mott insulator transition is also
desirable from a technological point of view,
e.g., with the ultimate aim of engineering the band gap in
graphene-like systems.

In addition to the significant quantitative discrepancies
observed in the literature, a more detailed comparison of the critical
behavior close to $\Nfc$ reported in those studies reveals our insufficient
understanding of fermionic field theories in the nonperturbative domain:
Kondo~\cite{Kondo:1995jn} reports a \emph{second-order} phase transition with
the usual power-law behavior as a function of the control parameter~$\Nf$. By
contrast, in the DSE studies~\cite{Itoh:1994cr, Sugiura:1996xk} an essential
scaling behavior of the Kosterlitz-Thouless type has been found, that is to
say, a phase transition of \emph{infinite} order. 
As argued in \cite{Christofi:2007ye}, the nature of the transition
in these studies appears to depend on whether the strong-coupling limit
is taken before or after \mbox{$\Nf \nearrow \Nfc$}.
The scaling analysis on the lattice \cite{Chandrasekharan:2011mn,
  Christofi:2007ye} is consistent with a power-law behavior corresponding to a
second-order phase transition which qualitatively confirms, but quantitatively
deviates from the Kondo scenario~\cite{Kondo:1995jn}.

In the present work, we confront the functional renormalization group (RG)
approach with the puzzles given by the (2+1)d chiral fermion systems at
criticality. The question is: What is the nature of the $\chi$SB phase
transition being controlled by either the bare coupling $g$ (for fixed $\Nf <
\Nfc$) or the number of flavors $\Nf$ (for fixed $g>g_\text{cr}$)? What are, if
applicable, the corresponding exponents determining the critical behavior?
Naively, a universal answer may not be apparent as (2+1)d fermion models are
perturbatively nonrenormalizable. However, rather than inherent inconsistencies,
this difficulty may reflect the failure of the perturbative approach
\cite{Redmond:1958pe, Symanzik:1975rz, Gawedzki:1985ed, Felder:1986ub}. In
fact,
there is a large body of evidence for nonperturbative renormalizability 
to all orders in a large-$\Nf$ expansion~\cite{Parisi:1975im,
Hikami:1976at, Hands:1994kb, Gomes:1990ed, Hong:1993qk} and, more recently, also
beyond the $1/\Nf$ expansion from a functional RG approach~\cite{Gies:2010st,
  Janssen:2012phd}. As a bottom line, (2+1)-dimensional fermion models appear
to be a paradigm example for asymptotically safe theories
\cite{Weinberg:1976xy, Weinberg:1980gg, Niedermaier:2006wt, Percacci:2007sz,
  Braun:2010tt}
that are UV complete due to the existence of interacting UV fixed points.

\begin{figure*}
\includegraphics[width=.31\textwidth]{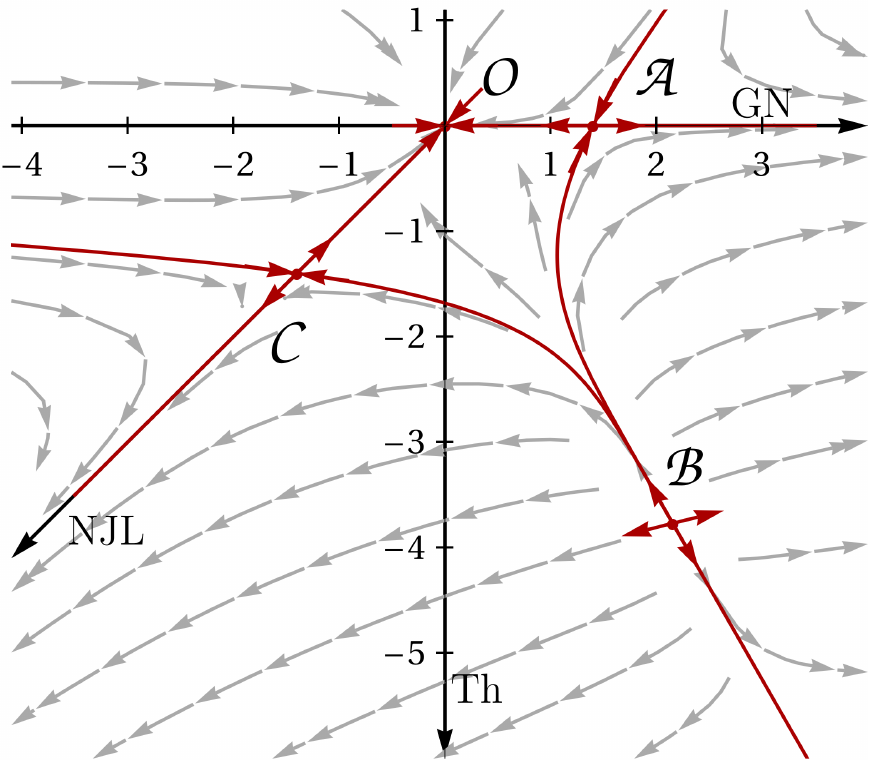}\hfill
\includegraphics[width=.31\textwidth]{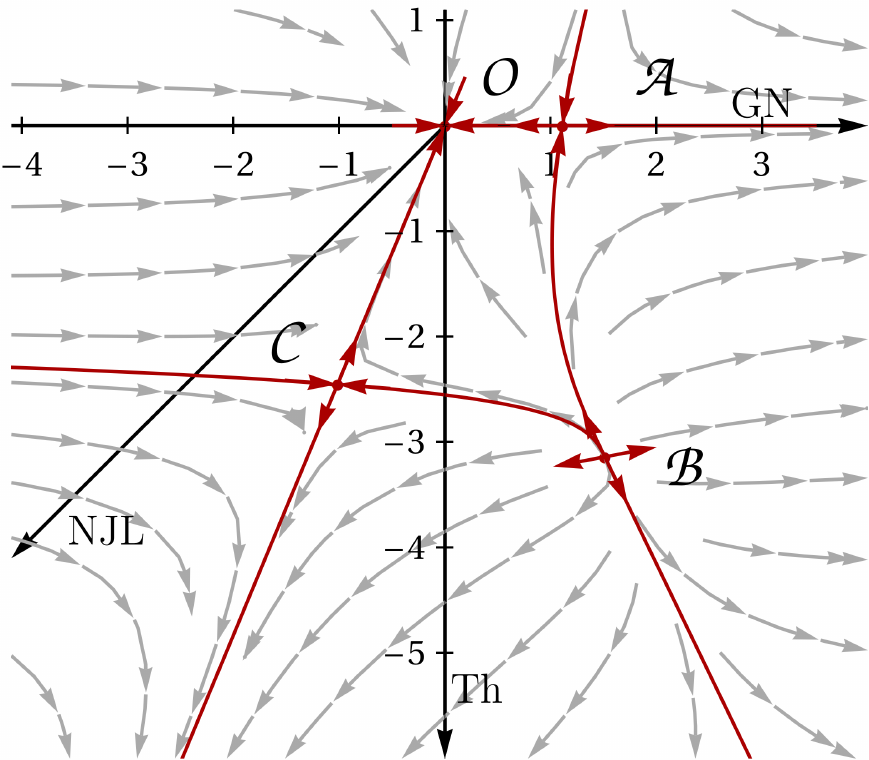}\hfill
\includegraphics[width=.31\textwidth]{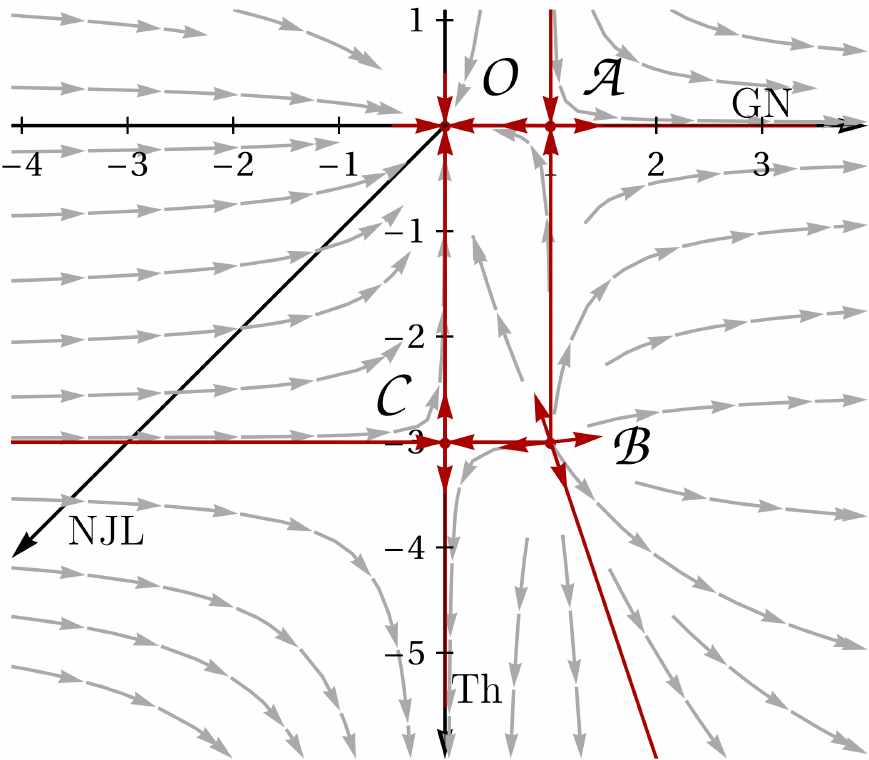}
\caption{Fermionic RG flow in the Gross-Neveu (GN) -- Thirring (Th) coupling 
  plane, compiled from Refs.~\cite{Gies:2010st, Janssen:2012phd}. 
  The angle bisector in the 3rd quadrant defines the 
  theories with pure Nambu-Jona-Lasinio (NJL) interactions.
  Left panel: for $\Nf = 1.75$ the
  Thirring fixed point $\mathcal C$ lies exactly on the NJL axis with $\mathcal
  L_\text{int} \propto (S)^2$. Right panel: in the limit $\Nf \rightarrow
  \infty$, the Thirring fixed point $\mathcal C$ lies on the Thirring
  axis with $\mathcal L_\text{int} \propto (V)^2$. Middle panel: for
  intermediate flavor number (here $\Nf \simeq 5$), we expect a
  competition between the scalar-type NJL channel $(S)^2$ and the
  vector-type Thirring channel $(V)^2$.}
\label{fig:fermionic-flow}
\end{figure*}

In the remainder of this section, we briefly summarize our findings,
starting with a short review of a preceding fermionic RG analysis
\cite{Gies:2010st}. All technical details and a proper embedding and
comparison to the literature is deferred to the following sections.\medskip


{\bf Competing channels in the (2+1)d Thirring model.}
Conventionally, the Thirring model in 2+1 dimensions is defined by the Lagrangian
\begin{align} \label{eq:thirring-lagrangian}
\mathcal L = \mathcal L_\text{kin} + \mathcal L_\text{int} = \bpsi^a \iu \slashed{\partial} \psi^a + \frac{\bar g}{2\Nf} \left(\bpsi^a \gamma_\mu \psi^a\right)^2,
\end{align}
with $\Nf$ flavors of massless four-component Dirac spinors $\psi^a$, i.e., $a
= 1,\dots, \Nf$. The microscopic theory has a chiral symmetry $\mathrm
U(2\Nf)$, cf.\ Sec.~\ref{sec:channels}. Depending on the value of the
four-fermi coupling $\bar g$ and the number of flavors $\Nf$, a fermion mass
can be dynamically generated, which breaks the chiral symmetry
spontaneously. 
Upon integrating out fluctuations,
further interaction terms, being
compatible with the present $\mathrm U(2\Nf)$ symmetry, are generated by the RG
transformations.
On the four-fermi level there are three further interactions
besides the Thirring term $\sim (\bpsi \gamma_\mu \psi)^2$: a flavor-singlet
pseudo-scalar channel (Gross-Neveu interaction), a flavor-nonsinglet scalar
channel (Nambu-Jona-Lasinio interaction), and a flavor-nonsinglet axial
channel. However, not all of these terms are independent in the pointlike
limit: due to the Fierz
identities we can always choose an arbitrary subset of two terms and write the
respective other two as a linear combination of these. A full basis is therefore
given,
e.g., by the Thirring and the Gross-Neveu interaction. The RG flow of the
dimensionless couplings in this basis is depicted in
Fig.~\ref{fig:fermionic-flow}.
The vertical (horizontal) axis defines the theories with pure Thirring
(Gross-Neveu) interaction. We refer to this axis as ``Thirring axis''
(``Gross-Neveu axis''). The angle bisector in the third quadrant
defines the theories with pure Nambu-Jona-Lasinio (NJL) interaction, which we
refer to as ``NJL axis''. In a functional RG analysis (the technique is
sketched in Sec.~\ref{sec:FRG}), we find two interacting
fixed points $\mathcal A$, $\mathcal B$, and $\mathcal C$ besides the
Gau{\ss}ian fixed point $\mathcal{O}$.  For the theories defined by the
microscopic 
Lagrangian \eqref{eq:thirring-lagrangian} being purely on the Thirring
axis, we find two different phases, which are separated by the separatrix
through the fixed points $\mathcal B$ and $\mathcal C$ (red curve in
Fig.~\ref{fig:fermionic-flow}): If we start the RG flow in the UV with a
microscopic coupling on the Thirring axis above the separatrix, the couplings
eventually flow to the noninteracting Gau{\ss}ian fixed point $\mathcal O$; if
we start below this curve, the four-fermi couplings grow large
in the IR. In the
vicinity of the critical coupling, the behavior of the system is governed by
the interacting fixed point $\mathcal C$, as all trajectories are
initially attracted toward this fixed point.
We refer to this fixed point as ``Thirring
fixed point''. We associate all trajectories emanating from $\mathcal C$ with
UV complete fully renormalized versions of the (2+1)d Thirring model. However,
we
emphasize that the Thirring fixed point only in the strict large-$\Nf$ limit
lies directly on the Thirring axis: even if absent on the microscopic scale, a
second coupling besides the Thirring coupling will always be generated by the
fluctuations for finite $\Nf$.

We now give a heuristic argument for the occurrence of the critical flavor
number~\cite{Gies:2010st, Janssen:2012phd}. Let us first consider the
(unphysical) flavor number $\Nf = 1.75$, where the Thirring fixed point
$\mathcal C$ lies exactly on the NJL axis (left panel of
Fig.~\ref{fig:fermionic-flow}). In this specific case, the NJL axis is an IR
attractive hyperplane and the RG flow of the Thirring model for sufficiently
large (negative) coupling is dominated by a divergent NJL coupling. A
divergent scalar-type four-fermi coupling 
at finite RG scale signals bound-state formation in that
channel and can be associated with the
dynamical generation of the corresponding order parameter, in this case the
chiral order parameter $\langle \bpsi \psi \rangle$.  We therefore predict
that spontaneous breaking of chiral symmetry occurs due to a dominance of the
scalar-type NJL channel.  Since the fixed-point positions change only smoothly
with $\Nf$, we expect this conclusion to hold also for small deviations from
$\Nf = 1.75$.  By contrast, for very large flavor number, the IR attractive
hyperplane with the Thirring fixed point $\mathcal C$ is located on the
Thirring axis, i.e., at vanishing NJL coupling (right panel of
Fig.~\ref{fig:fermionic-flow}). In other words, the RG flow is governed by a
strong vector-channel dominance, which generically inhibits $\chi$SB. For
intermediate flavor number (middle panel of Fig.~\ref{fig:fermionic-flow}), we
thus expect a transition between the scalar-type NJL channel, which triggers
$\chi$SB, and the vector-type Thirring channel, inhibiting $\chi$SB.\medskip

\begin{figure}[t]
\includegraphics[scale=1.05]{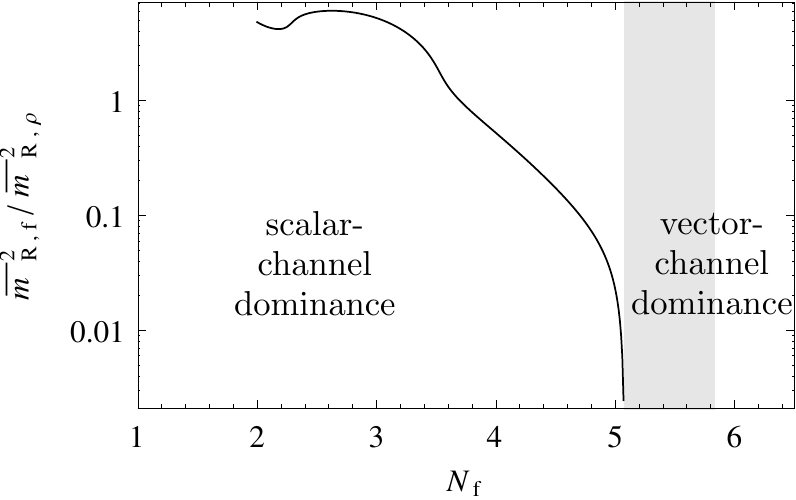}
\caption{Log-plot of the dynamically generated renormalized
  fermion mass $\bar m_{\text{R,f}}^2$ in units of the renormalized mass $\bar
  m_{\text{R},\rho}^2$ of a radial bosonic excitation on top of the
  chiral condensate. Our estimates for the critical flavor number,
  $\Nfc  \simeq 5.1$ for a linear regulator
  is amended by an error bar (gray shaded area) obtained by varying the
  regularization scheme (e.g., $\Nfc \simeq 5.8$ for the sharp cutoff).}
\label{fig:fermion-mass-Nf}
\end{figure}
\begin{figure}
\includegraphics[scale=1.12]{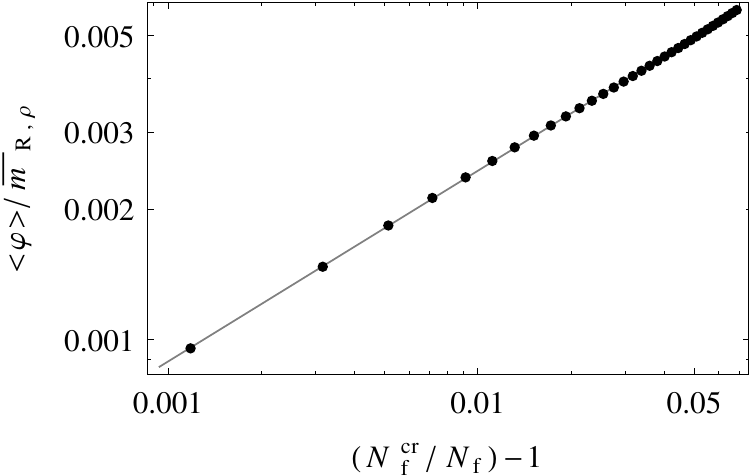}
\caption{Order parameter (black dots) vs.\ $\Nfc/\Nf - 1$ in a double-log
  plot, showing very good compatibility with power-law scaling behavior $\langle
  \varphi \rangle \propto (\Nfc/\Nf - 1)^b$. The slope of the regression line
  (gray) is $b \simeq 0.44$.}
\label{fig:orderparam}
\end{figure}

{\bf Dynamical boson fields.} 
A more quantitative picture of
this quantum phase transition is difficult to obtain in the purely fermionic
language with point-like interactions. A quantitative RG analysis
requires the inclusion of dynamical 
chiral and vector bosonic degrees of freedom, see
Secs.~\ref{sec:DoF}-\ref{sec:thirring-bosonized-RG-fixed},
in order to study the interplay of these competing channels
as a function of $\Nf$. This is the objective of this paper. In particular, we
will show how
the competition between the NJL (dominant for small $\Nf$) and the Thirring
channel (dominant for large $\Nf$) leads to a decrease of the dynamically
generated fermion mass with $\Nf$, culminating in a complete vanishing at
$\Nfc \simeq 5.1$ (linear regulator); cf.\
Fig.~\ref{fig:fermion-mass-Nf}. We can also map out the order of the phase
transition as a function of $\Nf$: in
Fig.~\ref{fig:orderparam} we depict the order parameter $\langle \varphi
\rangle \propto \langle \bpsi \psi \rangle$ vs.\ $\Nfc/\Nf - 1$ in a double-log
plot, showing very good compatibility with
a second order phase transition with
scaling behavior $\langle \varphi \rangle \propto (\Nfc/\Nf - 1)^b$ and the
universal critical exponent $b \simeq 0.44$. To avoid any possibility of
confusion, let us denote the critical exponents for the phase transition as a
function of $\Nf$ (at fixed overcritical coupling) with Latin letters, and the
ones for the phase transition as a function of the coupling (at fixed $\Nf <
\Nfc$) with Greek letters. 
We find that the latter phase transition is also of second
order, the critical behavior of which is determined by the RG flow in the
vicinity of the Thirring fixed point. We give our predictions for the
universal critical exponents $\nu$, $\omega$, and $\eta_\phi^*$ for various
flavor numbers in
Table~\ref{tab:fptvals_dyn-bos} on page~\pageref{tab:fptvals_dyn-bos}.


In addition to our quantitative results, it is one of our most important
observations that a simple partial bosonization {\`a} la Hubbard-Stratonovich is
not sufficient for resolving the competition between the various channels. A
method to deal with this challenging problem is {\em dynamical bosonization},
as detailed in Sec.~\ref{sec:thirring-bosonized-RG-dynamical}. As there is
already a large variety of analytical as well as numerical studies on the (2+1)d
Thirring model in the literature -- with partly contradictory results -- we
finally perform a careful and critical comparison between the literature and
our work in Sec.~\ref{sec:comparison}.


\section{Condensation channels and Fierz basis}
\label{sec:channels}

Let us start by presenting our conventions for
general fermionic models in $d=2+1$ Euclidean
dimensions with local quartic self-interaction, being invariant under
the full symmetry of the noninteracting theory.
We concentrate on representations of the Dirac algebra as they
occur naturally in the language of effective theories describing
electronic interactions on graphene's honeycomb lattice
\cite{Semenoff:1984dq,
  Herbut:2006cs, Herbut:2009qb, drut2009graphene, Hands:2008id, Armour:2009vj,
  Gusynin:2007ix} 
and in cuprates
\cite{Franz:2001zz, Franz:2002qy, Herbut:2002wd, Herbut:2002yq, Herbut:2004ue,
  Mavromatos:2003ss}.

The models shall satisfy Osterwalder-Schrader positivity
\cite{Osterwalder:1973dx}, requiring invariance of the action under
generalized 
complex conjugation defined by $\psi^\dagger \coloneqq \iu \bpsi
\gamma_3$ and a simultaneous reflection of the Euclidean time
coordinate, which we choose to be~$x_3$. For a detailed discussion of
our chiral conventions, see Refs.~\cite{Wetterich:2010ni,
  Wetterich:1990an}.  
Consider the $4\times 4$ \emph{reducible} representation
\begin{equation} \label{eq:gamma-matrix-rep4_A}
\gamma_\mu=
\begin{pmatrix}
0 & -\iu\sigma_\mu\\
\iu \sigma_\mu & 0\\
\end{pmatrix}, \qquad \mu=1,2,3,
\end{equation}
 of the Dirac algebra $\{\gamma_\mu,\gamma_\nu\} = 2\delta_{\mu\nu}$, 
reminiscent of the chiral representation in four dimensions. Here, $\{\sigma_\mu\}_{\mu=1,2,3}$ denote the standard $2\times2$ Pauli
matrices.
In this formulation, the Dirac fermions thus have four components.
There are now \emph{two} other $4\times4$ matrices, which anticommute with all
$\gamma_\mu$ as well as
with each other,
\begin{equation} \label{eq:gamma-matrix-rep4_B}
\gamma_4=
\begin{pmatrix}
0 & \mathbbm{1}_2\\
\mathbbm{1}_2 & 0\\
\end{pmatrix} \quad \text{and} \quad
\gamma_5=\gamma_1\gamma_2\gamma_3\gamma_4=
\begin{pmatrix}
\mathbbm{1}_2 & 0\\
0 & -\mathbbm{1}_2\\
\end{pmatrix}.
\end{equation}
Together with
\begin{equation}
\mathbbm{1}_4, \ \gamma_{\mu\nu}\coloneqq \frac{\iu}{2}[\gamma_\mu,\gamma_\nu] \ (\mu<\nu), \ \iu\gamma_\mu\gamma_4, \ \iu\gamma_\mu\gamma_5,\
\gamma_{45}\coloneqq \iu\gamma_4\gamma_5,
\end{equation}
these $16$ matrices form a complete basis of the $4\times 4$ Dirac algebra,
\begin{equation} \label{eq:basis_dirac}
\left\{\gamma_A\right\}_{A=1,\dots,16} =
\left\{\mathbbm{1}_4,\gamma_\mu,\gamma_4,\gamma_{\mu\nu},\iu\gamma_\mu\gamma_4,
\iu\gamma_\mu\gamma_5,\gamma_{45},\gamma_5\right\}.
\end{equation}
The general fermionic Lagrangian compatible with $\mathrm
U(2\Nf)$ chiral as well as $\mathcal C$, $\mathcal P$, and $\mathcal
T$ discrete symmetry has the form \cite{Kubota:2001kk, Kaveh:2004qa,
Herbut:2009qb, Gies:2010st, Janssen:2012phd}
\begin{align} \label{eq:general-thirring-lagrangian}
\mathcal L & = \bpsi^a \iu\slashed{\partial}\psi^a 
+ \frac{\bar g_1}{2\Nf}
(V)^2
+ \frac{\bar g_2}{2\Nf}
(S)^2
+ \frac{\bar g_3}{2\Nf}
(P)^2
\nonumber \\ &\quad
+ \frac{\bar g_4}{2\Nf}
(A)^2,
\end{align}
with the bare couplings $\bar g_{1,\dots,4}$ carrying an inverse mass
dimension, and with the flavor singlet channels
\begin{align}
(V)^2 & \coloneqq \left(\bpsi^a \gamma_\mu\psi^a\right)^2, &
(P)^2 & \coloneqq \left(\bpsi^a \gamma_{45}\psi^a\right)^2,
\end{align}
and the flavor nonsinglet channels
\begin{align}
(S)^2 & \coloneqq 
\left(\bpsi^a\psi^b\right)^2-\left(\bpsi^a\gamma_4\psi^b\right)^2-
\left(\bpsi^a\gamma_5\psi^b\right)^2
\nonumber \\ &\quad
+\left(\bpsi^a\gamma_{45}\psi^b\right)^2, \\
(A)^2 & \coloneqq
\left(\bpsi^a\gamma_\mu\psi^b\right)^2+
\frac{1}{2}\left(\bpsi^a\gamma_{\mu\nu}\psi^b\right)^2-
\left(\bpsi^a\iu\gamma_\mu\gamma_4\psi^b\right)^2
\nonumber \\ &\quad
-\left(\bpsi^a\iu\gamma_\mu\gamma_5\psi^b\right)^2.
\end{align}
Here, we have abbreviated $(\psi^a\psi^b)^2 \equiv \psi^a\psi^b\psi^b\psi^a$, etc. $\Nf$~denotes the number of four-component (reducible) Dirac spinors, such that $a,b=1,\dots,\Nf$. By means of the Fierz identities,
\begin{align} \label{eq:fierz12}
(V)^2 + (S)^2 + (P)^2 & = 0, &
-4 (V)^2 - 3 (S)^2 + (A)^2 & = 0,
\end{align}
any two four-fermi terms can be rewritten as a linear combination of the remaining two.
Put differently, by adding a linear combination of Eqs.~\eqref{eq:fierz12}
with coefficients $\alpha_i\in \mathbbm{R}$ (in units of some inverse
mass scale) to the
Lagrangian~\eqref{eq:general-thirring-lagrangian}, 
\begin{align} \label{eq:general-thirring-lagrangian_fierz}
\mathcal L & = \bpsi^a \iu\slashed{\partial}\psi^a 
+ \frac{1}{2\Nf}
(\bar g_1 + \alpha_1 - 4\alpha_2) (V)^2
\nonumber \\ &\quad
+ \frac{1}{2\Nf}
(\bar g_2 + \alpha_1 - 3\alpha_2) (S)^2
%
%
+ \frac{1}{2\Nf}
(\bar g_3+\alpha_1) (P)^2
\nonumber \\ &\quad
+ \frac{1}{2\Nf}
(\bar g_4+\alpha_2) (A)^2,
\end{align}
the $\alpha_i$ are redundant parameters:
in a full computation of the functional integral, any physical quantity
has to be independent of $\alpha_i$.
This no longer
necessarily remains true once approximations are employed.
A particular example is given by
 mean-field theory, where this so-called
``Fierz ambiguity'' or ``mean-field ambiguity'' can have
a sizable influence on the results, limiting
its quantitative reliability~\cite{Baier:2000yc}. The ambiguity is
absent in the purely fermionic renormalization group equations
with momentum-independent couplings~\cite{Gies:2010st}. 
A solution of the Fierz ambiguity using the functional RG in a
partially bosonized setting \cite{Jaeckel:2002rm} requires dynamical
bosonization \cite{Gies:2001nw} as will become important below.
An alternative approach in the purely fermionic description employs a new
parametrization of the momentum structure of the four-fermi couplings,
see~\cite{husemann2009efficient}.

The particular choice of couplings $\bar g$ and $\tilde {\bar g}$ as
used in \cite{Gies:2010st} is recovered for $\alpha_1 = -\bar g_2 - 3 \bar
g_4$ and $\alpha_2 = -\bar g_4$ and the definition
\begin{align}
\label{eq:g-tilde-g-A}
\bar g \coloneqq \bar g_1 - \bar g_2 +\bar g_4, \\
\label{eq:g-tilde-g-B}
\tilde {\bar g} \coloneqq - \bar g_2 + \bar g_3 - 3\bar g_4.
\end{align}
The Thirring (vertical) axis in Fig.~\ref{fig:fermionic-flow} corresponds to
$\bar{g}$, whereas the Gross-Neveu (horizontal) axis is associated with $
\tilde {\bar g}$ (more precisely with their dimensionless counterparts).
Upon choosing $\alpha_1 = -\bar g_3$ and $\alpha_2 = -\bar g_4$,
the Lagrangian reads
\begin{align} \label{eq:thirring-NJL_4-comp}
\mathcal L & = \bpsi^a \iu\slashed{\partial}\psi^a 
- \frac{\bar g_V}{2\Nf}
\left(\bpsi^a \gamma_\mu \psi^a\right)^2
+ \frac{\bar g_\phi}{4\Nf}
\Bigl[ \left(\bpsi^a\psi^b\right)^2
\nonumber \\ &\quad
-\left(\bpsi^a\gamma_4\psi^b\right)^2-
\left(\bpsi^a\gamma_5\psi^b\right)^2
+\left(\bpsi^a\gamma_{45}\psi^b\right)^2 \Bigr],
\end{align}
where we have defined the new couplings
\begin{align}
\bar g_V & \coloneqq -\bar g_1 + \bar g_3 - 4 \bar g_4 
= \tilde {\bar g} - \bar g%
, \\
\bar g_\phi & \coloneqq 2(\bar g_2 - \bar g_3 + 3 \bar g_4) 
= -2 \tilde{\bar g}%
.
\end{align}
This form is convenient in order to investigate the competition
between the vector $(V)^2$ and NJL-type $(S)^2$ channel for $\Nf\geq 2$. 
For $\Nf=1$ one might however choose 
yet another basis,
\begin{align} \label{eq:thirring_1flavor}
\mathcal L^{\Nf = 1}
& = \bpsi \iu\slashed{\partial}\psi
+ \frac{2\bar g_1-\bar g_3+3\bar g_4}{4}
\left(\bpsi \gamma_\mu \psi\right)^2
\nonumber \\ &
+ \frac{2\bar g_2-\bar g_3+3\bar g_4}{4}
\left[ \left(\bpsi\psi\right)^2-\left(\bpsi\gamma_4\psi\right)^2-
\left(\bpsi\gamma_5\psi\right)^2\right].
\end{align}
The Dirac spinors can be projected onto their Weyl components, using the
projectors%
\footnote{We note that one could choose more adapted representations
  in which $\gamma_{45} = \mathop{\mathrm{diag}}(\mathbbm 1_2, -
  \mathbbm 1_2)$ and thus the Weyl spinor $\chi^a$ ($\chi^{a+\Nf}$) is
  simply given by the upper (lower) two components of the Dirac spinor
  $\psi^a$. The ``graphene representation'' \cite{Herbut:2006cs,
    Herbut:2009qb} is of this type.}
\begin{align} \label{eq:chiral-projector-thirring}
P_\mathrm{L/R}^{(45)} = \frac{1}{2} (1\pm \gamma_{45}) =
\frac{1}{2}
\begin{pmatrix}
\mathbbm 1_2 & (-\iu) \,\mathbbm 1_2 \\
\iu \, \mathbbm 1_2 & \mathbbm 1_2
\end{pmatrix},
\end{align}
where the last equation holds for the representation
\eqref{eq:gamma-matrix-rep4_A}--\eqref{eq:gamma-matrix-rep4_B}.
This suggests the decomposition~\cite{Kubota:2001kk}
\begin{align} \label{eq:decomposition-two-comp}
\psi^a & = \frac{1}{\sqrt{2}}
\begin{pmatrix}
\chi^a + \chi^{a+\Nf} \\ \iu \left(\chi^a - \chi^{a+\Nf}\right)
\end{pmatrix}, 
\\
\bpsi^a & =
\frac{1}{\sqrt{2}}
\begin{pmatrix}
\bchi^a-\bchi^{a+\Nf}, & (-\iu) \left(\bchi^a + \bchi^{a+\Nf}\right)
\end{pmatrix},
\end{align}
$a = 1,\dots,\Nf$, with the definition of the Dirac adjoint~$\bchi$ chosen such that $\chi^\dagger \coloneqq \iu \bchi \sigma_3$ in agreement with $\psi^\dagger \coloneqq \iu \bpsi \gamma_3$. 
In other words, we can trade the $\Nf$ flavors of four-component spinors $\psi$ for $2\Nf$ flavors of two-component spinors $\chi$.
Therewith, the theory \eqref{eq:thirring-NJL_4-comp} can equivalently be described using an irreducible representation by an action consisting of $2\Nf$ flavors of two-component Weyl spinors $\bchi$, $\chi$,
\begin{align} \label{eq:thirring-NJL}
\mathcal L = \bchi^i \iu\slashed{\partial}\chi^i - \frac{\bar g_V}{2\Nf}
\left(\bchi^i \sigma_\mu \chi^i\right)^2 +
\frac{\bar g_\phi}{2\Nf} \left(\bchi^i\chi^j\right)^2,
\end{align}
where we have introduced the collective indices $i$, $j$, running over $2\Nf$ flavors, $i,j=1,\dots,2\Nf$. It is this representation in which the $\mathrm U(2\Nf)$ symmetry is manifest,
\begin{align}
\mathrm U(2\Nf) &: &\chi^i & \mapsto U^{ij} \chi^j, & \bchi^i & \mapsto \bchi^j \left(U^\dagger\right)^{ji}, & U & \in \mathrm U(2\Nf).
\end{align}
Below, we use this formulation, allowing us to 
conveniently introduce collective low-energy degrees of freedom.


\section{Functional RG approach}
\label{sec:FRG}

The functional RG has become a standard method to investigate
strongly-interacting field theories. In particular for critical
phenomena, it is a versatile tool yielding quantitatively accurate
results in many cases. Conceptually, the functional RG can be
formulated in terms of an RG flow equation for a generating
functional. Among the various formulations, the Wetterich
equation~\cite{Wetterich:1992yh} representing the flow equation for
the effective average action has become the most widely used method
owing to its flexibility, numerical stability and direct applicability
to physics problems. For reviews see~\cite{Berges:2000ew,
  Aoki:2000wm, Polonyi:2001se, Pawlowski:2005xe, Gies:2006wv,
  Delamotte:2007pf, Sonoda:2007av} 
and~\cite{kopietz2010introduction, Metzner:2011cw, Braun:2011pp} for a
particular emphasis on fermionic systems.  Prominent benchmark
examples in three dimensions are 
bosonic O($N$) models \cite{Berges:2000ew,Litim:2010tt,Benitez:2011xx}
or the Gross-Neveu model \cite{Rosa:2000ju,
  Hofling:2002hj,Braun:2010tt}.

The effective average action $\Gamma_k$ is a scale-dependent variant
of the standard generating functional for 1PI correlation
functions. The scale $k$ denotes an IR regulator scale
separating the UV modes with momenta $p^2> k^2$ which have already
been integrated out from the IR modes with momenta $p^2< k^2$ which
still have to be averaged over. $\Gamma_k$ is constructed in such a
way that it can be related to the microscopic bare action
$\Gamma_{k\to\Lambda} \to S_{\Lambda}$ if $k$ approaches the UV cutoff
scale $\Lambda$, while it approaches the standard full effective
action in the IR limit $\Gamma_{k\to0}\to \Gamma$. The effective
average action obeys the Wetterich equation~\cite{Wetterich:1992yh},
\begin{equation} \label{eq:wetterich-equation}
\partial_t \Gamma_k[\Phi] = \frac{1}{2} \STr \left(\frac{\partial_t
R_k}{\Gamma_k^{(2)}[\Phi] + R_k}\right), \qquad \partial_t \equiv
k \frac{\partial}{\partial k},
\end{equation}
where the trace runs over all internal degrees of freedom (flavor,
spinor, momentum) as well as field degrees of freedom. Here, the field
$\Phi$ represents a collective field variable including all bosonic or
fermionic fields under consideration. The denominator contains
 the second functional derivative $\Gamma_k^{(2)}[\Phi]$ with
respect to the field $\Phi$ together with the regulator function $R_k$
which can be thought of as a momentum-dependent mass term. 

Once the initial condition of the flow is fixed in terms of the
microscopic action $S_\Lambda$, the exact solution of
\Eqref{eq:wetterich-equation} provides us with an RG trajectory of
$\Gamma_k$ in the theory space of action functionals, the end-point at
$k=0$ of which is the exact effective action $\Gamma$.

As it is, in practice, difficult to find exact solutions, the
Wetterich equation can also be used to find approximate solutions by
means of systematic and consistent expansion schemes of the effective
action. While perturbation theory constitutes one such expansion
scheme, nonperturbative schemes based on vertex or operator
expansions are equally legitimate and clearly superior at intermediate
or even strong coupling. For critical phenomena and the analysis
of long-range order, derivative expansions in terms of the
order-parameter fields have become a standard tool yielding accurate
results in many nontrivial examples. In the present work, we will
also use a truncation of the effective action in the spirit of the
derivative expansion. However, as the chiral order parameter field is
bosonic, the derivative expansion has to be set up not only for the
microscopic fermionic fields as in \cite{Gies:2010st}, but requires
collective bosonic field variables, as introduced in the next
section. 
As a side remark we note that continuation of the RG flow into regimes
with broken symmetry can also be achieved in purely fermionic descriptions
without Hubbard-Stratonovich transformation by inserting an infinitesimally
small symmetry-breaking component in the initial
action~\cite{salmhofer2004renormalization, 0911.5047v1}.

The above mentioned regulator function $R_k$ is a to some extent
arbitrary function, satisfying a few conditions for implementing a
meaningful regularization \cite{Berges:2000ew, Pawlowski:2005xe,
  Gies:2006wv}. Whereas exact solutions do not at all depend on the
specific choice of $R_k$, approximate solutions and even estimates of
universal (i.e., regularization scheme independent quantities) can
depend on $R_k$. In the present work, we will use this variation of
universal quantities as a function of $R_k$ as an estimate for the
systematic error introduced by our approximations. We consider results
obtained for the {\em linear} regulator as our best estimate, as it
satisfies RG optimization criteria
\cite{Litim:2002cf, Pawlowski:2005xe}. By contrast, the {\em sharp} cutoff is
known to introduce strong regulator artifacts which we therefore use
to maximally span the error bar for our approximations.


\section{Low-energy degrees of freedom}
\label{sec:DoF}
The partition function
of the theory defined by \eqref{eq:thirring-NJL},
\begin{equation} \label{eq:thirring_part-function}
Z = \int \mathcal D \bchi\mathcal D \chi \exp(-S),
\end{equation}
is in fact equivalent to the partition function of the ``mesonic'' theory of a $(2\Nf)\times (2\Nf)$ scalar matrix field $\phi^{ij}$ and a vector field $V_\mu$, coupling via a Yukawa-type interaction to the fermions,
\begin{align} \label{eq:thirring_bosonized}
Z & = \mathcal N \int \mathcal D\chi \mathcal D\bar\chi \mathcal D\phi \mathcal D V
\exp \biggl[-\biggl(
\bar\chi^i\iu\slashed{\partial}\chi^i + \frac{1}{2} \bar m_\phi^2
\phi^{ij}\phi^{ji}
\nonumber \\ &\quad
+ \frac{1}{2} \bar m_V^2 V_\mu^2 - \bar h_V V_\mu \bar\chi^i \sigma_\mu \chi^i + \iu \bar h_\phi
\bar\chi^i \phi^{ij} \chi^j
\biggr)\biggr],
\end{align}
where $i,j=1,\dots,2\Nf$. 
The equivalence can be seen by multiplying Eq.~\eqref{eq:thirring_part-function} with appropriate Gau{\ss}ian (Hubbard-Stratonovich) factors,
\begin{align}
1 & = \mathcal N \int \mathcal D\phi \, \exp \biggl[-
\frac{1}{2}\left(\bar m_\phi \phi^{ij} + \iu \frac{\bar h_\phi}{\bar m_\phi}
\bar\chi^j\chi^i\right) 
\nonumber \\ &\qquad\qquad\qquad\qquad\quad
\label{eq:scalar_one}
\times\left(\bar m_\phi \phi^{ji} + \iu \frac{\bar h_\phi}{\bar m_\phi}
\bar\chi^i\chi^j\right)\biggr], \\
\label{eq:vector_one}
1 & = \mathcal N \int \mathcal D V \, \exp \left[-
\frac{1}{2}\left(\bar m_V V_\mu - \frac{\bar h_V}{\bar m_V}
\bar\chi^i\sigma_\mu\chi^i\right)^2
\right]\!,
\end{align}
with some normalization constants $\mathcal N$, not affecting any expectation values. The scalar matrix field is Hermitian, $\phi^\dagger = \phi$, and the vector field $V$ is real. 
The four-fermi terms in Eq.~\eqref{eq:thirring_part-function}
are then precisely canceled if the constraints
\begin{align}
\frac{\bar h_\phi^2}{2\bar m_\phi^2} & = \frac{\bar g_\phi}{2\Nf}, &
\frac{\bar h_V^2}{2\bar m_V^2} & = \frac{\bar g_V}{2\Nf},
\label{eq:constraints}
\end{align}
are imposed at the microscopic scale. From Eqs.~\eqref{eq:scalar_one}--\eqref{eq:vector_one} we can read off the properties of the boson fields under chiral transformations,
\begin{align} \label{eq:transformation_bosons}
\mathrm U(2\Nf) &: & 
\phi^{ij} & \mapsto U^{ik} \phi^{kl} \left(U^\dagger\right)^{lj}, & 
V_\mu & \mapsto V_\mu,
\end{align}
$U \in \mathrm{U}(2\Nf)$.
The scalar matrix $\phi$ may be decomposed into a traceless part and its trace  \cite{Pisarski:1991kg}
\begin{align}
\Phi^{ij} & \coloneqq \phi^{ij} - \frac{\delta^{ij}}{2\Nf} \Tr \phi, &
\varphi & \coloneqq \Tr \phi.
\end{align}
The $\varphi$ field is parity odd and can be attributed to the parity breaking channel $(P) \sim \bar\psi^a \gamma_{45} \psi^a = \bchi^i \chi^i$. 
By contrast, a vacuum expectation value of the traceless part $\Phi$ corresponds to the dynamical breakdown of chiral symmetry,
\begin{align}
\langle \Phi^{ij} \rangle \neq 0 
\quad \Leftrightarrow \quad
\langle \bpsi^a \psi^a \rangle = \langle \bchi^a \chi^a - \bchi^{a+\Nf} \chi^{a+\Nf}\rangle \neq 0,
\end{align}
$a=1,\dots,\Nf$, with the breaking pattern
\begin{equation} \label{eq:thirring-breaking-pattern-Chap4}
\mathrm{U}(2\Nf) \rightarrow \mathrm{U}\left(\Nf\right) \otimes
\mathrm{U}\left(\Nf\right).
\end{equation}
We can trade the \mbox{$(2\Nf) \times (2\Nf)$} Hermitian traceless matrix $\Phi$ for its independent components~$\Phi_\alpha$,
\begin{align}
\Phi^{ij} & = \sqrt{2} \Phi_{\alpha} (t_\alpha)^{ij},
\end{align}
$i,j=1,\dots,2\Nf, \ \alpha = 1,\dots,(2\Nf)^2-1$,
where the $t_\alpha$ are the generators of $\mathrm{SU}(2\Nf)$ in the fundamental representation, normalized so that $\Tr(t_\alpha t_\beta) = \delta_{\alpha\beta}/2$.

In the one-flavor case $\Nf=1$ this formulation is equivalent to a partial bosonization of the Fierz basis in the four-spinor representation \eqref{eq:thirring_1flavor}: the Hubbard-Stratonovich transformation leads to the equivalent Yukawa-type theory with three scalar modes $(\sigma, \tau, \pi) \sim (\bpsi \psi, \bpsi \gamma_4 \psi, \bpsi \gamma_5 \psi)$ and a vector mode $V_\mu \sim \bpsi \gamma_\mu \psi$ with Lagrangian density
\begin{align}
\mathcal L^{\Nf=1} & = \bar\psi\iu\slashed{\partial}\psi
+ \frac{1}{2} \bar m_\sigma^2 (\sigma^2 + \tau^2 + \pi^2)
\nonumber \\ &\quad
+ \iu \bar h_\sigma \bar\psi\left(
\sigma + \iu \gamma_4 \tau + \iu \gamma_5 \pi
\right)\psi
- \bar h_V V_\mu \bar\psi \gamma_\mu \psi.
\end{align}
From the discussion of the fermionic RG flow \cite{Gies:2010st} 
we expect that the long-range dynamics of this system is
dominated 
by the scalar NJL-type channel. For large enough coupling $\bar
h_\sigma^2/\bar m_\sigma^2$ we thus expect the scalar mode to acquire
a nonvanishing vacuum expectation value (VEV), e.g., in the $\sigma$
direction, and the spectrum in the broken phase consists of two
massless Goldstone modes, e.g., $\tau$ and $\pi$, and a massive radial
mode $\sigma$. The corresponding critical behavior is an interesting
problem by itself: this system can be viewed as an 
effective low-energy theory of spinless electrons on the honeycomb
lattice and thus as a simple model for suspended graphene
\cite{Mesterhazy:2012ei}.
It has also been studied in the context of magnetic
catalysis \cite{Gusynin:1994re}. In the following, we
will focus on the case $\Nf > 1$ where a true competition between
the two channels $(V)$ and $(S)$ is expected.

In contrast to the purely fermionic formulation \cite{Gies:2010st},
the bosonized formulation presented here is well suitable to
quantitatively describe the spontaneous breaking of chiral
symmetry. Loosely speaking, the bosonic fields $\phi^{ij}$ and $V_\mu$
parametrize the possible formation of bound states of the fermionic
fields $\sim \bchi^i \chi^j$ and $\sim \bchi^i \sigma_\mu \chi^i$,
respectively.  The corresponding critical phenomena of such a
strongly-correlated system require nonperturbative approximation
schemes. The functional renormalization group formulated in terms of
the Wetterich equation is such an appropriate tool and has already
shown its quantitative reliability in other (2+1)-dimensional
relativistic fermion systems, see e.g.,~\cite{Rosa:2000ju,
  Hofling:2002hj,Braun:2010tt}. In the effective action we then have
to take into account also higher boson-boson interactions
generated through fluctuations, e.g.,
\begin{align}
\left(\Tr\Phi^2\right)^2 & = \left(\Phi_\alpha\Phi_\alpha\right)^2, \\
\Tr \Phi^4 & = \frac{1}{2\Nf} \left(\Phi_\alpha\Phi_\alpha\right)^2 + 2 d_{\alpha\beta\epsilon}d_{\gamma\delta\epsilon}\Phi_\alpha\Phi_\beta\Phi_\gamma\Phi_\delta,
\end{align}
where the $d_{\alpha\beta\gamma}$'s are the structure constants for
the group $\mathrm{SU}(2\Nf)$. For $\Nf=1$ the
$d_{\alpha\beta\gamma}$'s vanish.  
For computational reasons, we use the representation
\eqref{eq:thirring_bosonized} for the case of general $\Nf$ with
scalar matrix field $\phi^{ij}$ having arbitrary trace.

We parameterize the dynamics of the low-energy degrees of
freedom by an ansatz for
the effective average action $\Gamma_k$. 
This ansatz corresponds to a systematic expansion of the action in powers
of the field gradients to second order. Moreover, we
simplify the discussion by considering interactions only up to fourth
order in the fields. Our ansatz for the effective action then reads
\begin{align} \label{eq:thirring-bosonized-truncation}
\Gamma_k & = \int_x \biggl[ Z_{\chi,k} \bar\chi^i \iu \slashed{\partial}
\chi^i
+ \frac{Z_{\phi,k}}{2}\partial_\mu \phi^{ij}\partial_\mu \phi^{ji} +
U_k (\phi) 
\nonumber \\ &\quad
+ \frac{Z_{V,k}}{4} V_{\mu\nu} V_{\mu\nu} +
\frac{\bar A_{V,k}}{2}
(\partial_\mu V_\mu)^2 + \frac{\bar m_{V,k}^2}{2} V_\mu V_\mu 
\nonumber \\ &\quad
+ \frac{\bar\zeta_k}{6} V_\mu V_\mu \partial_\nu V_\nu
+ \frac{\bar\mu_k}{8}\left(V_\mu V_\mu \right)^2
+ \frac{\bar\nu_k}{4} V_\mu V_\mu \phi^{ij}\phi^{ji}
\nonumber \\ &\quad
 - \bar h_{V,k} V_\mu
\bar\chi^i \gamma_\mu \chi^i + \iu \bar h_{\phi,k} \bar\chi^i \phi^{ij} \chi^j
\nonumber \\ &\quad
 - \frac{\bar g_{V,k}}{2\Nf}
\left(\bchi^i \sigma_\mu \chi^i\right)^2 +
\frac{\bar g_{\phi,k}}{2\Nf} \left(\bchi^i\chi^j\right)^2 \biggr],
\end{align}
where $V_{\mu\nu} \coloneqq \partial_\mu V_\nu - \partial_\nu V_\mu$
and $i,j =1,\dots,2\Nf$. $U_k(\phi)$~describes an effective potential
in the scalar sector. All couplings in the effective action are
understood to be scale-dependent, indicated by the index $k$.
It is straightforward to match this action exactly to the
microscopic models discussed above. For instance, by setting all
bosonic couplings to zero at the UV cutoff scale  $k=\Lambda$, we
return to the fermionic action
\eqref{eq:thirring-NJL}. Alternatively, we can set the fermionic
couplings to zero, $g_{V,\Lambda},g_{\phi,\Lambda}=0$, and choose
the compositeness conditions
\begin{equation}
Z_{\phi,k}, Z_{V,k}, \bar{A}_{V,k}, \bar{\zeta}_k \to 0, \quad
\text{for } k\to\Lambda,\label{eq:compcond}
\end{equation}
which guarantee that the bosonic fields have no kinetic term and are
purely auxiliary at the high scale. Furthermore, satisfying the
constraints \eqref{eq:constraints} and setting the higher-order
bosonic interactions to zero at $k\to\Lambda$ leads us to the
Yukawa-type theory as given in \Eqref{eq:thirring_bosonized}. This
implements the Hubbard-Stratonovich transformation at a fixed scale
$k=\Lambda$ in the functional RG context.
Of course, as soon as we start integrating out modes, the
couplings set to zero at one scale can be generated at lower
scales due to fluctuations. Also the constraints \eqref{eq:constraints}
will generally not be satisfied at lower scales.
Whereas the purely fermionic flow has been extensively discussed
in \cite{Gies:2010st}, we will now concentrate on the partially
bosonized variants.
We start with computing the RG flow in
Sec.~\ref{sec:thirring-bosonized-RG-fixed} for the fixed-field
variables introduced above. 
As the Hubbard-Stratonivich transformation demonstrates, the
effective action \eqref{eq:thirring-bosonized-truncation} contains
some redundancy, as field reparametrizations can trade various
operators for one another. In Sec.~\ref{sec:thirring-bosonized-RG-dynamical},
we exploit this
redundancy to introduce scale-dependent fields that adjust the
fermion-boson couplings dynamically in order to effectively perform
Hubbard-Stratonovich transformations on each scale (dynamical bosonization).
In the following, we will mostly ignore the
momentum-dependent terms in the vector channel, i.e., we will consider
the pointlike limit
\begin{align}
Z_{V,k} & \rightarrow 0, & \bar A_{V,k} & \rightarrow 0, & \bar \zeta_k &
\rightarrow 0, &\text{for all $k$.}
\end{align}
The beta functions will however be computed for general $Z_{V,k},\bar A_{V,k} \geq 0$ (but $\bar\zeta_k = 0$).


\section{Scalar mass spectrum}
\label{sec:spectrum}

\begin{table*}
\begin{center}
\caption{Spectrum of the scalar mass matrix $\delta^2 U_k / \delta
  \phi \delta \phi$ for 
  $\partial^2 U_k/\partial \tau^2 = \partial^2 U_k/\partial \rho \partial
  \tau=0$ and
  $\Nf>1$. Primes denote partial derivatives
  with respect to $\rho$, $U_{k}^{(n)} \equiv \partial^n U_k/\partial
  \rho^n$, and $U_{k,\tau} \equiv \partial U_k / \partial \tau$. For
  $\widehat m \equiv \widehat m(\rho,\tau)$ and $\epsilon \equiv
  \epsilon(\rho,\tau)$ see Eqs.~\eqref{eq:hatm}--\eqref{eq:hatepsilon}.}
\label{tab:scalar-mass-spectrum}
\begin{tabular*}{\textwidth}{l@{\extracolsep{\fill}}l}
\hline \hline
Eigenvalue & Degeneracy \\ \hline
$U_k'+\frac{\widehat m^2}{2} U_{k,\tau}
\left[3\epsilon^2 -1 + \frac{1-\epsilon^2}{\Nf}\right] $
 & $1$\\
$U_k'+\frac{\widehat m^2}{4\Nf} 
\biggl\{
4\Nf(\Nf-1)U_k''+(4-\Nf) U_{k,\tau} + \left[4\Nf U_k'' +
(3\Nf-4) U_{k,\tau}\right]\epsilon^2
$ & $1+1$ \\ \qquad $
\pm\Bigl[\left(4\Nf(\Nf-1)U_k''+(\Nf+2) U_{k,\tau}\right)^2
$ & \\ \quad \qquad $
+2\bigl(16\Nf^2(\Nf-1)U_k''{}^2-4\Nf(\Nf+2)(3\Nf-2)U_k'' U_{k,\tau}
$ & \\ \quad \qquad $
-(3\Nf^2-4\Nf+4)U_{k,\tau}^2\bigr)\epsilon^2
+\left(4\Nf U_k''+(3\Nf-2) U_{k,\tau}\right)^2\epsilon^4
\Bigr]^{1/2}\biggr\}$ \\
$U_k'+\frac{\widehat m^2}{2}  U_{k,\tau}
\frac{1-\Nf}{\Nf}(1-\epsilon^2) $
 & $2$\\
$U_k'+\frac{\widehat m^2}{2}  U_{k,\tau}
\left[\epsilon^2 \pm \epsilon +\frac{1-\epsilon^2}{\Nf}\right] $
 & $4(\Nf-1)+ 4(\Nf-1)$\\
$U_k'+\frac{\widehat m^2}{2}  U_{k,\tau}
\left[2+\frac{1-\epsilon^2}{\Nf}\right] $
 & $2\Nf^2-4\Nf+1$\\
$U_k'+\frac{\widehat m^2}{2}  U_{k,\tau}
\frac{1-\epsilon^2}{\Nf} $
 & $2\left(\Nf-1\right)^2$ \\ \hline\hline
\end{tabular*}
\end{center}
\end{table*}

Due to $\mathrm U(2\Nf)$ symmetry,
the effective potential $U_k(\phi)$ necessarily has to be a
pure function of $\mathrm U(2\Nf)$-invariant quantities. 
The Hermitian scalar matrix field can be diagonalized by a
$\mathrm U(2\Nf)$ rotation [cf.\ Eq.~\eqref{eq:transformation_bosons}],
\begin{align}
\phi & \mapsto U \phi U^\dagger =
\begin{pmatrix}
\widehat m_1 & 0 & \ldots & 0 \\
0 & \widehat m_2 & \ldots & 0 \\
\vdots & & \ddots & \vdots \\
0 & 0 & \ldots & \widehat m_{2\Nf} 
\end{pmatrix},
\end{align}
with real eigenvalues $\widehat m_i$.  In a quartic approximation
$U_k(\phi)$ 
can then be parametrized in terms of the two invariants
$\rho$ and $\tau$,
\begin{align}
\rho & = \frac{1}{2}\Tr \phi^2 = \frac{1}{2} \sum_i \widehat m_i^2, \\
\tau &  = \frac{1}{2}\Tr\left(\frac{1}{2}\phi^2 - \frac{\rho}{2\Nf}\right)^2
 = \frac{1}{8} \sum_i \widehat m_i^4 - \frac{1}{2}\left(\frac{\sum_i \widehat m_i}{4\Nf}\right)^2.
\end{align}
At higher field orders the potential can depend on additional
(suitably defined 
\cite{Jungnickel:1995fp}) higher order invariants
constructed from $\tilde \tau_n \sim \Tr(\phi^2/2 - \rho/2\Nf)^n$ for
$n\geq 3$. We expand the scalar potential about its $k$-dependent
minimum $(\rho_{0,k},\tau_{0,k})$. In the symmetric (SYM) regime the
potential is minimal at the origin $(\rho_{0,k},\tau_{0,k})=(0,0)$,
whereas in the chiral symmetry broken ($\chi$SB) regime we allow for
a nonvanishing VEV which we assume to be acquired along the
$\rho$ direction, i.e., $\rho_{0,k}>0$ and $\tau_{0,k}=0$:
\begin{align} \label{eq:eff-pot-expansion}
U_k(\rho,\tau) = 
\begin{cases}
\bar m_{\phi,k}^2 \rho + \frac{\bar\lambda_{1,k}}{2} \rho^2 +
\bar\lambda_{2,k} \tau, &
\text{SYM regime,} \\
\frac{\bar\lambda_{1,k}}{2} \left(\rho - \rho_{0,k}\right)^2 +
\bar\lambda_{2,k} \tau, &
\text{$\chi$SB regime.}
\end{cases}
\end{align}
If the potential solely depends on $\rho$ and $\tau$, it is sufficient
to evaluate its flow equation in a two-dimensional subspace of all
possible scalar configurations. We consider the traceless class, valid
for $\Nf > 1$,
\begin{align} \label{eq:field-configuration}
(\phi^{ij})
\eqqcolon \widehat{m}
\Diag(\epsilon,\underbrace{+1,\ldots,+1}_{\Nf-1 \text{ times}
},-\epsilon,\underbrace{
-1 , \ldots,-1}_{\Nf-1 \text{ times}}),
\end{align}
where $\epsilon \geq 1, \ \widehat m \in \mathbbm R$.
For any $\rho$ and $\tau$ with $\tau / \rho^2 < (\Nf-1)/4\Nf $, 
the parameters $\widehat m$ and $\epsilon$ are given by
\begin{align} \label{eq:hatm}
\widehat m^2 & =
\frac{\rho}{\Nf}\left(1-\sqrt{\frac{4\Nf}{\Nf-1}{\frac{\tau}{\rho^2}}}\right),
\\ \label{eq:hatepsilon}
\epsilon^2 & =
1+\Nf\left(\frac{1}{1-\sqrt{\frac{4\Nf}{\Nf-1}{\frac{\tau}{\rho^2}}}} -1
\right).
\end{align}
If the flow eventually chooses a vacuum configuration with $\epsilon = 1$ and $\widehat m > 0$, the chiral symmetry is spontaneously broken while parity symmetry remains preserved,
\begin{multline}
\left(\phi_0^{ij}\right) = \frac{\rho_0}{\Nf}
\begin{pmatrix}
\mathbbm 1 & 0 \\
0 & -\mathbbm 1
\end{pmatrix} \neq 0
\\ \quad
\ \Leftrightarrow \ 
\langle \bchi^a \chi^a - \bchi^{a+\Nf}\chi^{a+\Nf} \rangle = 
\langle \bpsi^a \psi^a \rangle \neq 0.
\end{multline}
For simplicity, we assume in the following that $\partial^2 U_k/\partial
\tau^2= \partial^2 U_k/\partial \rho \partial \tau=0$, which holds in the case
of the quartic approximation \eqref{eq:eff-pot-expansion}.  The spectrum of
the scalar mass matrix $\delta^2 U_k/\delta \phi \delta \phi$ for $\Nf>1$ is
given in Table~\ref{tab:scalar-mass-spectrum}.

For a vacuum configuration with $\epsilon = 1$ we have $\tau_{0,k}=0$.
In the SYM regime $\rho_{0,k}=0$, all modes are degenerate and have
mass $\bar m_\phi^2 = \partial U_k / \partial
\rho|_{(\rho,\tau)=(0,0)}$. In the $\chi$SB regime with $\rho_{0,k}>0$
and $\partial U_k/ \partial \rho |_{(\rho,\tau)=(\rho_{0,k},0)}=0$ we
find $2\Nf^2$ massless scalar modes corresponding exactly to the
number of broken generators in the symmetry breaking pattern
\begin{equation}
\mathrm{U}(2\Nf) \rightarrow \mathrm{U}\left(\Nf\right) \otimes
\mathrm{U}\left(\Nf\right),
\end{equation}
in accordance with Goldstone's theorem. Additionally, we obtain one massive
radial mode with $\bar m_\rho^2 = 2 \rho_{0,k} \partial^2 U_k / \partial \rho^2$
and $2\Nf^2-1$ massive modes in $\tau$ direction with $\bar m_\tau^2 =
(\rho_{0,k}/\Nf) \partial U_k / \partial \tau$.
Since $\delta \tau / \delta \phi |_{\phi_0} = 0$ we expect the $2\Nf^2 -
1$ degeneracy of the masses $\bar m_\tau^2$
to be a general result, holding
also beyond our quartic approximation~\eqref{eq:eff-pot-expansion}, as long as
no higher-order invariants $\sim \tilde \tau_n$ become important.


\section{Partially bosonized RG flow}
\label{sec:thirring-bosonized-RG-fixed}

Next, we determine the flow of the model in a partially
bosonized description. This corresponds to start at the high scale with the
Hubbard-Stratonovich transformed action of
\eqref{eq:thirring_bosonized} and the compositeness condition
\eqref{eq:compcond}. As is standard in this type of truncation,
higher-order fermionic interactions are set to zero also on all
lower scales.
 
We fix the standard RG invariance of field rescalings by defining the
renormalized fields as
\begin{align}
\tilde{\phi}^{ij} & \coloneqq Z_{\phi,k}^{1/2} \phi^{ij}, &
\tilde{\chi}^i & \coloneqq  Z_{\chi,k}^{1/2} \chi^i, \\
 \tilde{V}_{\mu} & \coloneqq  Z_{V,k}^{1/2} V_\mu, &
\tilde{\bar\chi}^i & \coloneqq  Z_{\chi,k}^{1/2} \bar\chi^i.
\end{align}
The dimensionless effective potential for space-time dimension $d$ then reads
\begin{align}
u(\tilde\rho,\tilde\tau) & \coloneqq k^{-d} U_k(Z_{\phi,k}^{-1} k^{d-2}\tilde\rho, Z_{\phi,k}^{-2} k^{2(d-2)}\tilde\tau),
\end{align}
where $\tilde\rho \coloneqq Z_{\phi,k} k^{2-d} \rho$ and $\tilde\tau \coloneqq Z_{\phi,k}^{2} k^{2(2-d)} \tau$, and the dimensionless renormalized couplings are
\begin{align}
A_{V,k} & \coloneqq Z_{V}^{-1} \bar A_{V,k}, &
\nu & \coloneqq Z_{\phi,k}^{-1} Z_{V,k}^{-1} k^{d-4} \bar\nu_k, \\
m_V^2 & \coloneqq Z_{V,k}^{-1} k^{-2} \bar m_{V,k}^2, &
h_\phi^2 & \coloneqq Z_{\phi,k}^{-1} Z_{\chi,k}^{-2} k^{d-4} \bar h_{\phi,k}^2,
\\
\mu & \coloneqq Z_{V,k}^{-2} k^{d-4} \bar\mu_k, &
h_V^2 & \coloneqq Z_{V,k}^{-1} Z_{\chi,k}^{-2} k^{d-4} \bar h_{V,k}^2.
\end{align}
By evaluating the Wetterich equation \eqref{eq:wetterich-equation}
for a constant scalar background field 
\eqref{eq:field-configuration}, we 
obtain the flow of the dimensionless scalar potential
\begin{align} \label{eq:flow-potential}
\partial_t u & = -d u + (d-2+\eta_\phi) \tilde\rho  u' + (2d-4+2\eta_\phi)
\tilde\tau  u_{,\tilde\tau}
\nonumber \\ &\quad 
+ 2 v_d 
\sum_{i=1}^{2\Nf} \ell_0^{\mathrm{(B)}d}(m_i^2; \eta_\phi)
%
%
+2 v_d d\,\ell_0^{\mathrm{(B)}d}(m_V^2 + \nu \tilde\rho;\eta_V)
\nonumber \\ &\quad 
- 4 v_d d_\gamma 
\Bigl[
(\Nf-1)\,\ell_0^{\mathrm{(F)}d}(\widetilde m^2 h_\phi^2;\eta_\chi)
\nonumber \\ &\qquad \qquad\quad
+ \ell_0^{\mathrm{(F)}d}(\widetilde m^2 \epsilon^2 h_\phi^2;\eta_\chi)
\Bigr]
\end{align}
with $u\equiv u(\tilde\rho,\tilde\tau)$, $u'\equiv \partial u/\partial
\tilde \rho$, $u_{,\tilde\tau}\equiv \partial u/\partial \tilde \tau$
and $\widetilde m^2 {\coloneqq} Z_{\phi,k} k^{2-d} \widehat{m}^2$. The
dimensionless scalar masses $m_i^2\equiv m_i^2(\tilde
\rho,\tilde\tau)$ 
can straightforwardly be deduced from
Table~\ref{tab:scalar-mass-spectrum}. We have further defined the
anomalous dimensions
\begin{align}
\eta_{\phi/\chi/V} = -\partial_t \ln Z_{\phi/\chi/V,k}.
\end{align}
The flow involves the threshold functions
$\ell_0^\mathrm{(B/F)}(\dots)$, which encode the details of the
regularization scheme; their definitions and explicit forms for the
linear and the sharp cutoff are listed in
the Appendix.  We have abbreviated
$v_d^{-1}=2^{d+1} \pi^{d/2} \Gamma(d/2)$, i.e., $ v_3^{-1} =
  8\pi^2$.  As discussed above, we work here with the two-component
Weyl spinors such that the dimension of the gamma matrices is
$d_\gamma=2$.  By suitable differentiation of
Eq.~\eqref{eq:flow-potential} we obtain the flow of the scalar
couplings occurring in Eq.~\eqref{eq:eff-pot-expansion}.  In the SYM
regime,
\begin{align}
\partial_t m_\phi^2 & = \partial_t u' \Bigr|_
{(\tilde\rho,\tilde\tau)=(0,0)}, 
\\
\partial_t \lambda_1 & = \partial_t u''\Bigr|_{(\tilde\rho,\tilde\tau)=(0,0)}, &
\partial_t \lambda_2 & = \partial_t u_{,\tilde\tau}
\Bigr|_{(\tilde\rho,\tilde\tau)=(0,0)},
\end{align}
whereas in the $\chi$SB regime,
\begin{align}
\partial_t \kappa & = - \frac{1}{\lambda_1}\partial_t u'
\Bigr|_{(\tilde\rho,\tilde\tau)=(\kappa,0)},
\\
\partial_t \lambda_1 & = \partial_t u''
\Bigr|_{(\tilde\rho,\tilde\tau)=(\kappa,0)}, &
\partial_t \lambda_2 & = \partial_t u_{,\tilde\tau}
\Bigr|_{(\tilde\rho,\tilde\tau)=(\kappa,0)},
\end{align}
with the dimensionless VEV $\kappa \coloneqq Z_{\phi,k} k^{2-d}
\rho_{0,k}$ and the abbreviations $u^{(n)} \equiv \partial^n u / \partial
{\tilde\rho}^n$ and $u_{,\tilde\tau} \equiv \partial u / \partial
\tilde\tau$.

Similarly, the flow equations for all other couplings present in the
effective action are straightforwardly obtained by suitable
projections of the Wetterich equation \eqref{eq:wetterich-equation}.
This amounts to a summation of all possible 1-loop diagrams where the
vertices are given by full (though truncated) 
vertex functions and the inner lines correspond to the full propagators. 
With the useful Mathematica package DoFun~\cite{Huber:2011qr},
the evaluation of the RG flow equation can be automated easily.
The beta function for the vector mass reads
\begin{widetext}
\begin{align}
\partial_t m_V^2 & = 
(- 2+\eta_V) m_V^2
- 2v_d (d+2)\,\ell_1^{\mathrm{(B)}d}(m_V^2+\nu \kappa; \eta_V) \mu
\nonumber \\& \quad 
- 2 v_d \left[2\Nf^2\,\ell_1^{\mathrm{(B)}d}(u';\eta_\phi)
+ \left(2\Nf^2-1\right) \ell_{1}^{\mathrm{(B)}d}(u'+\tfrac{\kappa}{\Nf}
u_{,\tilde\tau};\eta_\phi) +
\ell_{1}^{\mathrm{(B)}d}(u'+ 2\kappa u'';\eta_\phi) \right] \nu
\nonumber \\& \quad 
+ \frac{8 v_d (d-2)d_\gamma \Nf}{d}
\,\ell_1^{\mathrm{(F)}d}(\tfrac{\kappa}{\Nf} h_\phi^2;\eta_\chi)\,h_V^2
+ \frac{16 v_d d_\gamma}{d} \kappa
\,\ell_2^{\mathrm{(F)}d}(\tfrac{\kappa}{\Nf} h_\phi^2;\eta_\chi)\,h_\phi^2
h_V^2,
\end{align}
with the derivatives of the potential $u^{(n)}$ and $u_{,\tilde\tau}$ being evaluated at the minimum $(\tilde\rho,\tilde\tau)=(\kappa,0)$. In the symmetric regime we have of course $\kappa=0$.
For the vector-vector interaction $\mu$ and the vector-scalar interaction $\nu$ we get
\begin{align}
\partial_t \mu & = (d-4+2 \eta_V) \mu
%
%
+ \frac{2v_d(d^2+10d+12)}{d+2} \ell_2^{\mathrm{(B)}d}(m_V^2+\nu\kappa; \eta_V) \mu^2
\nonumber \\ &\quad 
+ 2 v_d \left[2\Nf^2\,\ell_{2}^{\mathrm{(B)}d}(u';\eta_\phi)
+ \left(2\Nf^2-1\right) \ell_{2}^{\mathrm{(B)}d}(u'+\tfrac{\kappa}{\Nf}
u_{,\tilde\tau};\eta_\phi) +
\ell_{2}^{\mathrm{(B)}d}(u'+ 2\kappa u'';\eta_\phi) \right] \nu^2
\nonumber \\ &\quad 
+ \frac{16 v_d (d-2)(4-d) d_\gamma \Nf}{d(d+2)}
\,\ell_2^{\mathrm{(F)}d}(\tfrac{\kappa}{\Nf} h_\phi^2; \eta_\chi)\, h_V^4
%
%
+ \frac{128 v_d (4-d) d_\gamma}{d(d+2)} \kappa
\,\ell_3^{\mathrm{(F)}d}(\tfrac{\kappa}{\Nf} h_\phi^2; \eta_\chi)\, h_\phi^2
h_V^4
\nonumber \\ &\quad 
- \frac{384 v_d d_\gamma}{d(d+2) \Nf} \kappa^2
\,\ell_4^{\mathrm{(F)}d}(\tfrac{\kappa}{\Nf} h_\phi^2; \eta_\chi)\, h_\phi^4
h_V^4,
\displaybreak[0] \\
%
%
\partial_t \nu & = (d-4+\eta_\phi+\eta_V)\nu
+{2(d+2)v_d}
\,\ell_2^{\mathrm{(B)}d}(m_V^2+\nu\kappa; \eta_V) \mu \nu
\nonumber \\ & \quad
+ v_d \biggl\{  
\left(\frac{2\Nf^2+1}{\Nf^2} u'' + \frac{2\Nf^2-1}{2\Nf^3} u_{,\tilde\tau} \right)
\biggl[
2\Nf^2\,\ell_{2}^{\mathrm{(B)}d}(u';\eta_\phi)
+ \left(2\Nf^2-1\right)
\ell_{2}^{\mathrm{(B)}d}(u'+\tfrac{\kappa}{\Nf}u_{,\tilde\tau};\eta_\phi)
\nonumber \\ &\qquad\qquad
+ \ell_{2}^{\mathrm{(B)}d}(u'+2\kappa u'';\eta_\phi)
\biggr]
+ \frac{1}{\Nf} u_{,\tilde\tau}
\,\ell_2^{\mathrm{(B)}d}(u'+\tfrac{\kappa}{\Nf}u_{,\tilde\tau};\eta_\phi)
\biggr\} \nu
\nonumber \\ &\quad
+ \frac{2 v_d}{\Nf^2} \biggl[ 
2\Nf^2\,
\ell_{1,1}^{\mathrm{(BB)}d}(u',m_V^2+\nu\kappa;\eta_\phi,\eta_V)
%
%
+ \left(2\Nf^2-1\right)
\ell_{1,1}^{\mathrm{(BB)}d}(u'+\tfrac{\kappa}{\Nf} u_{,\tilde\tau}, m_V^2+\nu\kappa;\eta_\phi, \eta_V)
\nonumber \\ &\qquad\qquad
+\ell_{1,1}^{\mathrm{(BB)}d}(u'+2\kappa
u'',m_V^2+\nu\kappa;\eta_\phi,\eta_V) \biggr] \nu^2
+ \frac{8v_d (4-d) d_\gamma}{d}\,\ell_2^{\mathrm{(F)}d}(\tfrac{\kappa}{\Nf} 
h_\phi^2;\eta_\chi)
\,h_V^2 h_\phi^2
\nonumber \\ &\quad
-\frac{96 v_d d_\gamma}{d \Nf} \kappa \,\ell_3^{\mathrm{(F)}d}(\tfrac{\kappa}{\Nf}
h_\phi^2;\eta_\chi)\,h_V^2 h_\phi^4
%
%
-\frac{32 v_d (4-d)d_\gamma}{d \Nf^2}\kappa^2
\,\ell_4^{\mathrm{(F)}d}(\tfrac{\kappa}{\Nf} h_\phi^2;\eta_\chi)\,h_V^2
h_\phi^6.
\end{align}
In the symmetric regime the flow of the Yukawa coupling $h_\phi$ is
unambiguous. In the broken regime however the diverse scalar modes in
general can develop different couplings. We will focus on the
Goldstone-mode coupling to the fermions, which is expected to give the
dominant contribution for aspects of criticality.  However, this
Goldstone-mode projection may introduce artifacts deeply in the broken
regime, see below.  The flow equation reads
\begin{align}
\partial_t h_\phi^2 & = (d-4+\eta_\phi+2\eta_\chi) h_\phi^2 
\nonumber \\ &\quad
+ \frac{4 v_d}{\Nf} \left[
\ell_{1,1}^{\mathrm{(FB)}d}(\tfrac{\kappa}{\Nf}
h_\phi^2,u'+
\tfrac{\kappa}{\Nf} u_{,\tilde\tau};\eta_\chi,\eta_\phi)
- \ell_{1,1}^{\mathrm{(FB)}d}(\tfrac{\kappa}{\Nf}
h_\phi^2,u'+2\kappa u'';\eta_\chi,\eta_\phi)
\right] h_\phi^4
\nonumber \\ &\quad
+ \frac{8 v_d}{\Nf^2} \kappa 
\Bigl[
2\Nf^2 u'' \,\ell_{1,2}^{\mathrm{(FB)}d}(\tfrac{\kappa}{\Nf}
h_\phi^2,u';\eta_\chi,\eta_\phi)
%
%
+ \left((2\Nf^2-1) u_{,\tilde\tau}
+ 2(\Nf-1) u''\right)
\ell_{1,1,1}^{\mathrm{(FBB)}d}(\tfrac{\kappa}{\Nf}
h_\phi^2,u',u'+\tfrac{\kappa}{\Nf} u_{,\tilde\tau};\eta_\chi,\eta_\phi)
\nonumber \\ &\qquad\qquad
+ \left(2 u'' + u_{,\tilde\tau}\right)
\ell_{1,1,1}^{\mathrm{(FBB)}d}(\tfrac{\kappa}{\Nf}
h_\phi^2,u',u'+2\kappa u'';\eta_\chi,\eta_\phi)
\Bigr]
 \, h_\phi^4
%
%
- 8 v_d d\,\ell_{1,1}^{\mathrm{(FB)}d}(\tfrac{\kappa}{\Nf}
h_\phi^2,m_V^2+\nu\kappa;\eta_\chi,\eta_V)\, h_\phi^2 h_V^2.
\end{align}
For the flow of the fermion-vector coupling we find
\begin{align}
\partial_t h_V^2 & = (d-4 + \eta_V + 2\eta_\chi) h_V^2 
\nonumber \\ & \quad 
- \frac{4 v_d(d-2)}{d \Nf}
\biggl[
2\Nf^2\,\ell_{1,1}^{\mathrm{(FB)}d}(\tfrac{\kappa}{\Nf}
h_\phi^2,u';\eta_\chi,\eta_\phi)
+ \left(2\Nf^2-1\right)
\ell_{1,1}^{\mathrm{(FB)}d}(\tfrac{\kappa}{\Nf} h_\phi^2,u'+
\tfrac{\kappa}{\Nf} u_{,\tilde\tau};\eta_\chi,\eta_\phi)
\nonumber \\ & \qquad\qquad
+ \ell_{1,1}^{\mathrm{(FB)}d}(\tfrac{\kappa}{\Nf}
h_\phi^2,u'+2\kappa u'';\eta_\chi,\eta_\phi)
\biggr]
h_\phi^2 h_V^2
\nonumber \\ & \quad 
- \frac{8 v_d}{d \Nf^2} \kappa 
\biggl[
2\Nf^2\,\ell_{2,1}^{\mathrm{(FB)}d}(\tfrac{\kappa}{\Nf}
h_\phi^2,u';\eta_\chi,\eta_\phi)
+ \left(2\Nf^2-1\right)
\ell_{2,1}^{\mathrm{(FB)}d}(\tfrac{\kappa}{\Nf} h_\phi^2,u'+
\tfrac{\kappa}{\Nf} u_{,\tilde\tau};\eta_\chi,\eta_\phi)
\nonumber \\ & \qquad\qquad
+ \ell_{2,1}^{\mathrm{(FB)}d}(\tfrac{\kappa}{\Nf}
h_\phi^2,u'+2\kappa u'';\eta_\chi,\eta_\phi)
\biggr]
h_\phi^4 h_V^2
\nonumber \\ & \quad 
- \frac{8 v_d (d-2)^2}{d} \,\ell_{1,1}^{\mathrm{(FB)}d}(\tfrac{\kappa}{\Nf}
h_\phi^2,m_V^2+\nu\kappa;\eta_\chi,\eta_V) 
\, h_V^4 
%
%
- \frac{16 v_d(d-2)}{d \Nf} \kappa
\,\ell_{2,1}^{\mathrm{(FB)}d}(\tfrac{2}{n}\kappa
h_\phi^2,m_V^2+\nu\kappa;\eta_\chi,\eta_V)\, h_\phi^2 h_V^4,
\end{align}
and the anomalous dimensions read
\begin{align}
\eta_\phi & = 
\frac{8 v_d d_\gamma}{d}
\,m_4^{\mathrm{(F)}d}(\tfrac{\kappa}{\Nf} h_\phi^2;\eta_\chi)
\,h_\phi^2
+ \frac{8 v_d d_\gamma}{d \Nf} \kappa
\,m_2^{\mathrm{(F)}d}(\tfrac{\kappa}{\Nf} h_\phi^2;\eta_\chi)
\,h_\phi^4
\nonumber \\ &\quad
+ \frac{16 v_d}{d \Nf^2} \kappa 
\left[\Nf^2 u''{}^2+(\Nf-1) u_{,\tilde\tau}{}^2\right] 
m_{2,2}^{\mathrm{(B)}d}(u', u'+2\kappa u'';\eta_\phi),
\\ 
\label{eq:flow-eqs_etapsi}
\eta_\chi & = \frac{4 v_d}{d \Nf} \biggl[
2\Nf^2\, m_{1,2}^{\mathrm{(FB)}d}(\tfrac{\kappa}{\Nf}
h_\phi^2,u';\eta_\chi,\eta_\phi)
+ \left(2\Nf^2-1\right) m_{1,2}^{\mathrm{(FB)}d}(\tfrac{\kappa}{\Nf}
h_\phi^2,u'+\tfrac{\kappa}{\Nf}u_{,\tilde\tau};
\eta_\chi,\eta_\phi)
\nonumber \\ & \qquad \qquad
+ m_{1,2}^{\mathrm{(FB)}d}(\tfrac{\kappa}{\Nf}
h_\phi^2,u'+2\kappa u'';\eta_\chi,\eta_\phi)\biggr]
h_\phi^2
- 8v_d m_{1,2}^{\mathrm{(FB)}d}(\tfrac{\kappa}{\Nf}
h_\phi^2,m_V^2+\nu\kappa;\eta_\chi,\eta_V)\,
h_V^2,
\\ 
%
\eta_V & = \frac{16 v_d (d-2) d_\gamma \Nf}{d}
\,m_4^{\mathrm{(F)}d}(\tfrac{\kappa}{\Nf} h_\phi^2;\eta_\chi) \, h_V^2 +
16v_d d_\gamma \kappa
\,m_2^{\mathrm{(F)}d}(\tfrac{\kappa}{\Nf} h_\phi^2;\eta_\chi)
\,h_\phi^2 h_V^2.
\end{align}
\end{widetext}

Within our truncation we find that the flows of the two possible
vector-field kinetic terms are in fact equivalent, i.e., $\partial_t
\bar A_k = \partial_t Z_{V,k}$. The vector propagator thus is
diagonal, $G_{\mu\nu,k}^{(V)} = \delta_{\mu\nu} / (Z_{V,k} p^2 + \bar
m_{V,k}^2)$.  The definitions of the threshold functions
$\ell_{\dots}^{\mathrm{(B/F)}d}(\dots)$ and
$m_{\dots}^{\mathrm{(B/F)}d}(\dots)$ are listed in
the Appendix, together with their explicit forms
for linear and sharp cutoff. The flow equations have been
independently verified for the symmetric regime in local potential
approximation with the DoFun package~\cite{Huber:2011qr}.

As is the case in the fermionic formulation, we expect the flow to be
dominated by the vector channel for sufficiently large flavor number
$\Nf$. Whether for large coupling the $V$ field then can develop a
finite vacuum expectation value is an interesting question on its own
right: e.g., (2+1)-dimensional models exhibiting spontaneous
breaking of Lorentz symmetry have been investigated in
\cite{Hosotani:1993dg, Hosotani:1994sc, Wesolowski:1995xz,
  Higashijima:2001sq, Khalilov:2004wf}. By contrast, one could suspect
that the vector mass eventually flows to zero, triggering a close
resemblance of the strongly coupled Thirring model for large flavor
number to a $\mathrm U(1)$ gauge theory. This is in fact the
prediction of the large-$\Nf$ studies
\cite{Hands:1994kb, Hands:1997xv}. 
In the present work, we are mainly
interested in a possible scalar condensation corresponding to chiral
symmetry breaking and thus leave this issue for future studies.

As long as the vector mass $m_V^2$ does not become too small it is
then sufficient to consider the pointlike approximation in the vector
channel, $Z_{V,k} \rightarrow 0$, corresponding to $m_V^2 \rightarrow
\infty$. In this limit, the beta functions no
longer depend on $h_V^2$ and $m_V^2$ separately but only on the ratio
$g_V = \Nf h_V^2/m_V^2$, reflecting the RG invariance of fields
rescalings. In the remaining flow equations the vector anomalous
dimension $\eta_V$ completely drops out. The flow of $Z_{\chi,k}$
given by Eq.~\eqref{eq:flow-eqs_etapsi} is driven by a competition
between a positive scalar loop term $\propto h_\phi^2$ and a negative
vector loop term $\propto h_V^2$. In the pointlike approximation of
the vector sector, $m_V^2\rightarrow \infty$ and hence the vector
loop term vanishes. For reasons of consistency, we will therefore
suppress also the scalar loop term in Eq.~\eqref{eq:flow-eqs_etapsi},
i.e., we treat the fermionic sector in the leading-order derivative
approximation $\eta_\chi \equiv 0$. This is compatible with the
observation that the flow of the fermionic wave function
renormalization in 3d fermion systems at criticality is usually very
small \cite{Gracey:1993kc, Vasiliev:1992wr, Rosa:2000ju,
  Hofling:2002hj, Gies:2009da, Braun:2010tt}. As we shall see in
Sec.~\ref{sec:thirring-bosonized-RG-dynamical}, this assumption is
exactly fulfilled for large number of fermion flavors $\Nf
\rightarrow \infty$, where the fixed-point equations can be solved
analytically; in fact, this is the known result of the $1/\Nf$
expansion \cite{Hikami:1976at}.

In comparison, the scalar anomalous dimension is nonvanishing in this
limit, and we thus expect the flow of $Z_{\phi,k}$ to be crucial also
for finite $\Nf$. We note that similar observations have also been
made in other Yukawa-type systems in three dimensions, where a
nonvanishing $\eta_\phi \sim \mathcal O(1)$ is essential in order to
find the correct critical behavior~\cite{Rosa:2000ju, Hofling:2002hj,
  Gies:2009da, Braun:2010tt}. Also the recent results for the
$\Nf=1$ model \cite{Mesterhazy:2012ei} fit into this scheme.

\begin{figure*}
\includegraphics[scale=.98]{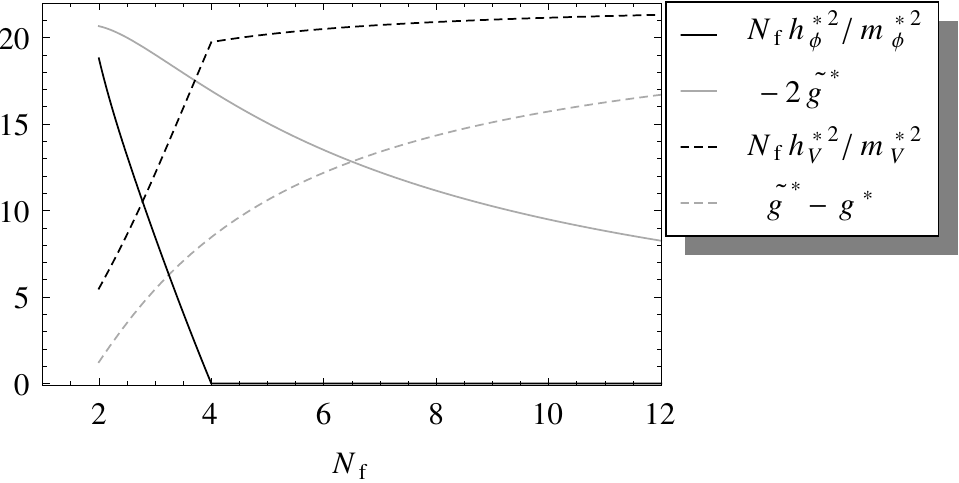}\hfill 
\includegraphics[scale=.9]{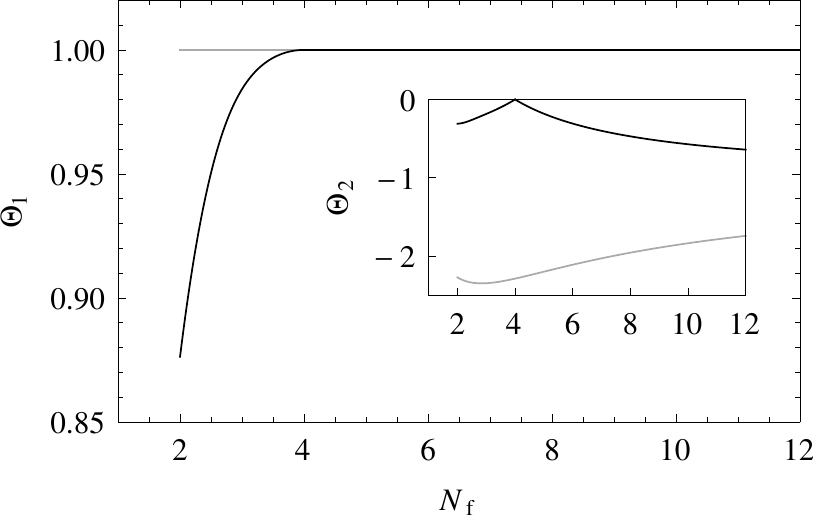}
\caption{Left panel: nonuniversal UV fixed-point values
  of the Thirring fixed point for the
  bosonic (black) and fermionic RG flow (gray, compiled
  from~\cite{Gies:2010st}) for the 
  linear regulator. In the bosonic
  language already for $\Nf \geq 4$ the fixed point is located in the
  pure fermion-vector sector, while in the fermionic language this is
  the case only for large $\Nf$. Fermionic and bosonic descriptions
  coincide only for $\Nf\rightarrow \infty$ but approach each
  other again for small $\Nf$. Right panel: largest
  critical exponent $\Theta_{1}$ and subleading exponent $\Theta_2$ (inset).}
\label{fig:gVgphi_fixedFields}
\end{figure*}

For the remainder of this section, we 
omit the vector-vector interaction $\mu$ and
the vector-scalar interaction $\nu$ for simplicity. 
We concentrate on the UV structure only 
in order to compare it with our
previous results in the fermionic language. 
A full analysis of the IR properties 
of the dynamically bosonized
RG flow follows in Sec.~\ref{sec:thirring-bosonized-RG-dynamical}. For the
search of a fixed point, we hence end up with a system of six coupled
nonlinear equations for the five couplings in the SYM regime
$(m_\phi^2, \lambda_1, \lambda_2, h_\phi^2, \Nf h_V^2/m_V^2)$ and the
anomalous dimension $\eta_\phi$.  
The stability matrix $\partial \beta_i / \partial g_j |_{g^*}$ at a fixed
point $g^* = (m_\phi^{*2}, \lambda_1^*, \lambda_2^*, h_\phi^{*2}, \Nf
h_V^{*2}/m_V^{*2})$ has five eigenvalues, which we refer to as $-\Theta_1,
\dots, - \Theta_5$, in ascending order according to their real part.
In the large-$\Nf$ limit the
fixed-point structure can be mapped out analytically and we discover
precisely the known structure from the fermionic
flow~\cite{Gies:2010st}: the Thirring universality class is governed
by a UV fixed point having one IR relevant direction with critical
exponent $\Theta_1=1$ and $\Theta_{i\geq 2} < 0$.
It is located in the pure fermion-vector sector,
i.e., $\Nf h_\phi^{*2} / m_\phi^{*2} = 0$ and $\Nf h_V^{*2}/m_V^{*2} >
0$.  Moreover, the fixed-point position exactly coincides with the
Thirring fixed point in the fermionic RG~\cite{Gies:2010st},
\begin{align}
g_\phi^* & =\Nf \frac{h_\phi^{*2}}{m_\phi^{*2}} = 0,\qquad  &
\text{for } \Nf &\rightarrow \infty,\\
g_V^* & =\Nf \frac{h_V^{*2}}{m_V^{*2}} =
\frac{3\pi^2}{2\,\ell_{1}^{\mathrm{(F)}}(0)}, &
\text{for } \Nf &\rightarrow \infty,
\end{align}
for the dimensionless renormalized fermionic couplings $g_\phi =
  Z_\chi^{-2} k^{d-2} \bar g_\phi$ and $g_V = Z_\chi^{-2}
  k^{d-2} \bar g_V$. For finite $\Nf$ we discover deviations from the
fermionic fixed-point structure. In Fig.~\ref{fig:gVgphi_fixedFields},
results from the bosonized description are shown as black lines, and
the corresponding fermionic results from \cite{Gies:2010st} are
plotted as gray lines. These deviations are not
unexpected, since the fermionic description differs from the partially
bosonized description in two respects: first, the fluctuation-induced
kinetic terms of the bosons describe momentum-dependencies of the
fermionic four-point functions. These momentum-dependencies are
neglected in a fermionic derivative expansion used in
\cite{Gies:2010st}. In addition to this advantage of the partially
bosonized description, there is also a disadvantage arising from the
fact that the fluctuations are no longer Fierz-complete in the
partially bosonized description even in the point-like limit, as will
be demonstrated explicitly in the next section.

The question which description is
quantitatively more accurate cannot be answered a priori. There are,
however, a couple of indications that the fermionic description is
more reliable at least in the UV: First, as already observed in
\cite{Gies:2010st}, the universal aspects of the UV fixed-point
structure in the fermionic language do not depend on the choice of
regulator, as it should be. If momentum-dependencies were
important, we would expect a strong scheme dependence of the
universal observables on approximate RG calculations. This UV
stability of fermionic flows has been observed in a variety of
contexts \cite{Gies:2005as, Braun:2006jd}. Second, the fixed-point
properties in the fermionic and partially bosonized description
differ both quantitatively and qualitatively, as is visible in
Fig.~\ref{fig:gVgphi_fixedFields} for the Thirring fixed point. Also, in the
bosonized  language, we find two further
interacting fixed points with two or more relevant
directions (while there is only one fixed point with two relevant
directions in the fermionic description). In fact, we observe
a qualitative breakdown of the partially bosonized description for
intermediate
$\Nf$: for
$\Nf<4$ the additional fixed points are located in the pure fermion-scalar
sector $h_V^{2} / m_V^2 = 0$ and the pure fermion-vector sector 
$h_\phi^{2} / m_\phi^2 = 0$, respectively. 
Since $\partial_t h_\phi^2 \propto \mathcal O(h_\phi^2)$ 
and $\partial_t h_V^2 \propto \mathcal O(h_V^2)$ both
sectors are in fact invariant under RG transformations. 

The Thirring
fixed point, which for $\Nf<4$ 
has components in both sectors,
$h_\phi^2/m_\phi^2 > 0$ and $h_V^2/m_V^2>0$, hits the fixed point in
the pure fermion-vector sector once $\Nf \rightarrow 4$. The fixed-point
position as a function of $\Nf$ is therefore nonanalytic at
$\Nf=4$. For $\Nf >4$ the Thirring fixed-point stays in the
fermion-vector subspace $h_\phi^2 / m_\phi^2 = 0$, coinciding with the
fermionic-RG fixed point only for $\Nf \rightarrow \infty$.  In
Fig.~\ref{fig:gVgphi_fixedFields} we have plotted the critical
exponents for the RG relevant ($\Theta_1$) and the RG irrelevant
direction ($\Theta_2$), showing good agreement in the former case
while in the latter case, in particular for small $\Nf$, large
deviations occur.

To summarize: for large $\Nf \gg 1$, we find a qualitative
behavior which is consistent between the fermionic and the partially
bosonized description. The scalar sector decouples for $\Nf \gg 1$
and only the fermion-vector interactions $\Nf h_V^2 / m_V^2 \sim -
g$ matter. For smaller $\Nf$, the quantitative deviations between
both descriptions increase and particularly the fixed-point
structure becomes rather different, culminating at a nonanalyticity
at $\Nf=4$. We interpret this behavior as a breakdown of the
partially bosonized description arising from the Fierz
incompleteness as shown below. Incidentally, the agreement between
both descriptions improves again for smaller $\Nf$, where the flow
in both cases is dominated by a strong fermion-scalar coupling $\Nf
h_\phi^2 / m_\phi^2 \sim -2\tilde g$. This single-channel dominance
appears to alleviate the problem of Fierz incompleteness. This
observation also lends support to the recent single-channel analysis of the
$\Nf=1$ system \cite{Mesterhazy:2012ei}.

We conclude that the ``Fierz ambiguity'' is a severe problem for
partially bosonized formulations of the Thirring model precisely in
the region where different channels compete with each other. For the
location of the quantum phase transition in terms of a critical
fermion number, a solution of this problem is mandatory. For
applications of the Thirring model at a fixed given fermion number,
say $\Nf=2$ for graphene, a solution of this problem may still exert
a sizable quantitative influence. A solution in terms of a
dynamically bosonized RG flow will be presented in the next section.


\section{Dynamically bosonized RG flow} 
\label{sec:thirring-bosonized-RG-dynamical}

\begin{figure*}
\includegraphics[width=0.30\textwidth]{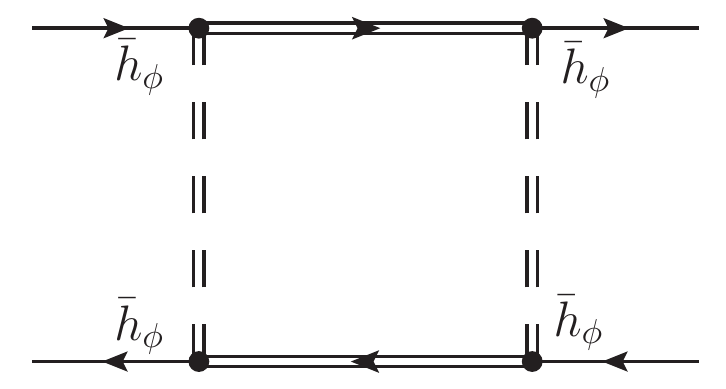}\hfill
\includegraphics[width=0.30\textwidth]{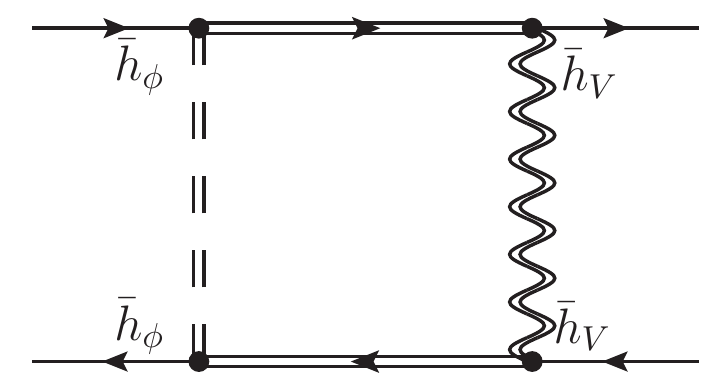}\hfill
\includegraphics[width=0.30\textwidth]{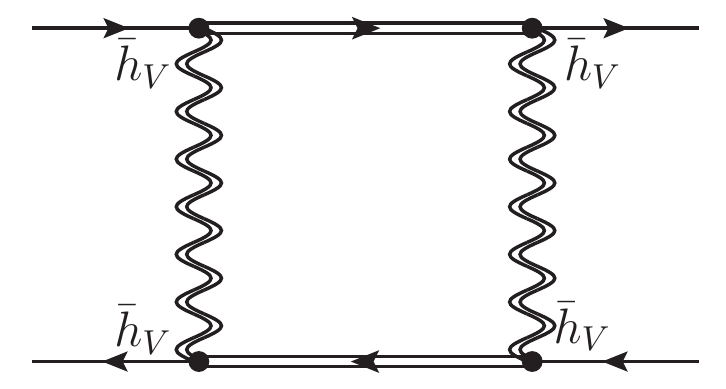}\\[0.75\baselineskip]
\includegraphics[width=0.30\textwidth]{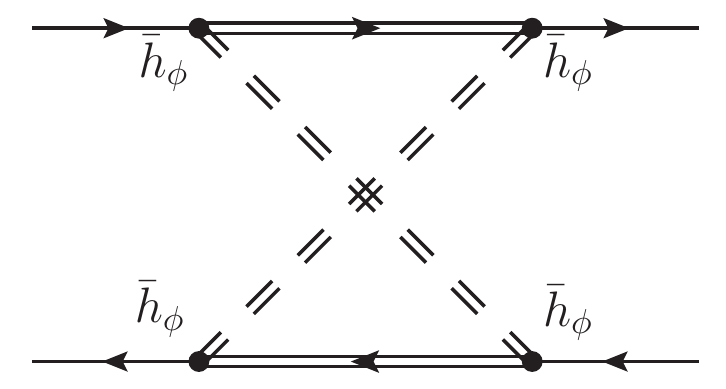}\hfill
\includegraphics[width=0.30\textwidth]{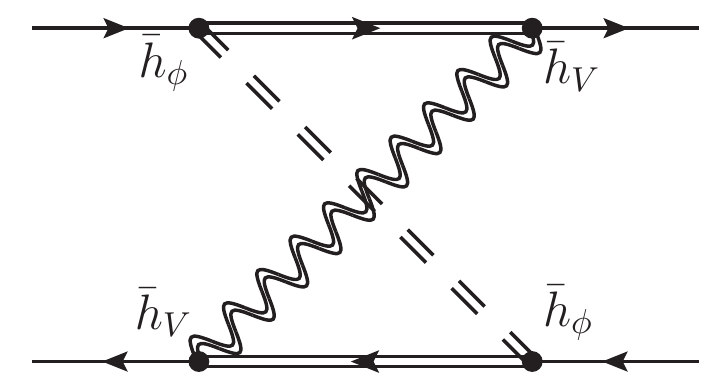}\hfill
\includegraphics[width=0.30\textwidth]{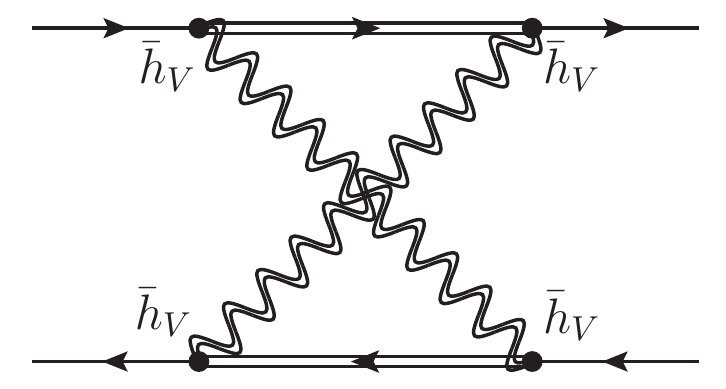}
\caption{Box diagrams contributing to the flow of $g_\phi$ and
  $g_V$. Solid lines are fermions, dashed lines are scalar fields, and
  wiggly lines vector fields. Doubled inner lines denote full
  propagators $G_k = (\Gamma_{k,0}^{(2)}+R_k)^{-1}$.}
\label{fig:4-fermi-diagrams}
\end{figure*}

Although the bosonic partition function \eqref{eq:thirring_bosonized}
is fully equivalent to the original fermion theory
\eqref{eq:thirring-NJL}, the corresponding leading-order truncations
of the effective action are not. The reason is that the four-fermi
couplings, though absent in the bare action due to the
Hubbard-Stratonovich trick, are again generated by
the box diagrams displayed in Fig.~\ref{fig:4-fermi-diagrams}.
However, an inclusion of $\bar g_{\phi,k}$ and $\bar g_{V,k}$ in the
truncation \eqref{eq:thirring-bosonized-truncation} does not seem very
appealing, as it introduces a redundancy in the effective action.
Instead, we employ dynamical bosonization
\cite{Gies:2001nw, Pawlowski:2005xe, Gies:2006wv, Floerchinger:2009uf}
as it has already been successfully used to resolve mean-field
ambiguities \cite{Jaeckel:2002rm} or implement particle-hole
fluctuations in ultracold fermi gases \cite{Floerchinger:2009pg}.
Here we use the simplified approach%
\footnote{This approach follows from an
  approximation to an exact equation \cite{Pawlowski:2005xe} where the
  neglected terms are parametrically suppressed for the present
  application \cite{Gies:2006wv}.}
proposed in~\cite{Gies:2001nw}. The idea is to perform a Hubbard-Stratonovich
transformation at \emph{each} RG step, such that all newly generated
four-fermi interactions are again reexpressed in terms of the
bosonic interactions.
In this way, the four-fermi couplings $\bar g_{\phi,k}$ and $\bar g_{V,k}$
vanish at all scales.
The bosonic fields then necessarily become scale dependent. 
We use the following field redefinitions
\begin{align}
\phi^{ij}_{k-dk} & = \phi^{ij}_k - \iu (\bchi^j \chi^i) \delta\omega_{\phi,k}, &
\phi_{\Lambda} & \equiv \phi, \\
V_{\mu,k-dk} & = V_{\mu,k} + (\bchi^i \sigma_\mu \chi^i) \delta\omega_{V,k},&
V_{\Lambda} & \equiv V,
\end{align}
with to be determined functions $\omega_{\phi/V,k}$. Note that we keep the
fermion fields fixed. For scale-dependent bosonic fields the flow equation for
the effective average action is modified,
\begin{widetext}
\begin{align}
\nonumber
\partial_k \Gamma_k[\phi_k,V_k] & = 
\partial_k \Gamma_k [\phi_k,V_k] \bigr|_{\phi_k,V_k}
%
%
+ \int \frac{\delta \Gamma_k [\phi_k,V_k]}{\delta \phi^{ij}_k} \partial_k \phi^{ij}_k + \int  \frac{\delta \Gamma_k [\phi_k,V_k]}{\delta V_{\mu,k}} \partial_k V_{\mu,k}
\displaybreak[0] \\
& = \frac{1}{2} \STr \frac{\partial_k R_k}{\Gamma_k^{(2)}[\phi_k,V_k]+ R_k}
+ \iu \int \frac{\delta \Gamma_k [\phi_k,V_k]}{\delta \phi_k^{ij}} (\bchi^j \chi^i) \partial_k\omega_{\phi,k} 
%
%
- \int  \frac{\delta \Gamma_k [\phi_k,V_k]}{\delta V_{\mu,k}} (\bchi^i \sigma_\mu \chi^i) \partial_k\omega_{V,k},
\end{align}
\end{widetext}
where the first term is evaluated for fixed fields and hence leads to
the standard flow of $\Gamma_k$ with $\phi_\Lambda$ and $V_\Lambda$
replaced by $\phi_k$ and $V_k$, respectively \cite{Gies:2001nw}. We
have suppressed the additional dependence of $\Gamma_k$ on the fermion
fields $\bchi$ and $\chi$ for brevity.
Projecting onto the boson couplings, we find the beta functions
\begin{align}
\partial_t u & = \partial_t u \bigr|_{\phi_k,V_k}, &
\partial_t \nu^2 & = \partial_t \nu^2 \bigr|_{\phi_k,V_k}, \\
\partial_t m_V^2 & = \partial_t m_V^2 \bigr|_{\phi_k,V_k}, &
\partial_t h_\phi^2 & = \partial_t h_\phi^2 \bigr|_{\phi_k,V_k}
+ u' \partial_t \omega_{\phi,k}, \\
\partial_t \mu^2 & = \partial_t \mu^2 \bigr|_{\phi_k,V_k}, &
\partial_t h_V^2 & = \partial_t h_V^2 \bigr|_{\phi_k,V_k}
+ m_V^2 \partial_t \omega_{V,k},
\end{align}
i.e., the scale-dependent bosonization changes only the flow of $h_\phi^2$ and $h_V^2$ and leaves the other beta functions in the bosonic sector invariant. For the four-fermi couplings we obtain
\begin{align}
\partial_t g_\phi & = \partial_t g_\phi \bigr|_{\phi_k,V_k} - h_\phi \partial_t \omega_{\phi,k}, \\
\partial_t g_V & = \partial_t g_V \bigr|_{\phi_k,V_k} - h_V \partial_t \omega_{V,k}.
\end{align}
Choosing 
\begin{align}
\partial_t \omega_{\phi,k} & \equiv \frac{\beta_{g_\phi}}{h_\phi}, &
\partial_t \omega_{V,k} & \equiv \frac{\beta_{g_V}}{h_V},
\end{align}
where $\beta_{g_\phi} \coloneqq \partial_t g_\phi \bigr|_{\phi_k,V_k}, \ 
\beta_{g_V} \coloneqq \partial_t g_V \bigr|_{\phi_k,V_k}$, establishes that
$g_{V,k}$ and $g_{\phi,k}$ vanish at all scales, if absent at the UV scale
$k=\Lambda$. The beta functions $\beta_{g_\phi}$ and $\beta_{g_V}$ are
straightforwardly obtained by suitable projections of the box diagrams in
Fig.~\ref{fig:4-fermi-diagrams},
\begin{widetext}
\begin{align}
\beta_{g_{\phi/V}} & = 
4 v_d a_{\phi/V}^{(1)}
\biggl[
2\Nf^2\,\ell_{1,2}^{\mathrm{(FB)}d}(\tfrac{\kappa}{\Nf}
h_\phi^2,u';\eta_\chi,\eta_\phi)
+ \left(2\Nf^2-1\right)
\ell_{1,2}^{\mathrm{(FB)}d}(\tfrac{\kappa}{\Nf} h_\phi^2,u'+
\tfrac{\kappa}{\Nf} u_{,\tilde\tau};\eta_\chi,\eta_\phi)
\nonumber \\ & \qquad\qquad
+ \ell_{1,2}^{\mathrm{(FB)}d}(\tfrac{\kappa}{\Nf}
h_\phi^2,u'+2\kappa u'';\eta_\chi,\eta_\phi)
\biggr]
h_\phi^4
+ 4 v_d a_{\phi/V}^{(2)} \, \ell_{1,2}^{\mathrm{(FB)}d}(\tfrac{\kappa}{\Nf}
h_\phi^2,m_V^2+\nu\kappa;\eta_\chi,\eta_V)h_V^4
\nonumber \\ & \quad
+ 4 v_d a_{\phi/V}^{(3)} 
\biggl[
2\Nf^2\,\ell_{1,1,1}^{\mathrm{(FBB)}d}(\tfrac{\kappa}{\Nf}
h_\phi^2,u',m_V^2;\eta_\chi,\eta_\phi)
%
%
+ \left(2\Nf^2-1\right)
\ell_{1,1,1}^{\mathrm{(FBB)}d}(\tfrac{\kappa}{\Nf} h_\phi^2,u'+
\tfrac{\kappa}{\Nf} u_{,\tilde\tau},m_V^2;\eta_\chi,\eta_\phi)
\nonumber \\ & \qquad\qquad
+ \ell_{1,1,1}^{\mathrm{(FBB)}d}(\tfrac{\kappa}{\Nf}
h_\phi^2,u'+2\kappa u'',m_V^2;\eta_\chi,\eta_\phi)
\biggr]
h_\phi^2 h_V^2
\end{align}
\end{widetext}
with to be determined (possibly $\Nf$-dependent) constants
$a_{\phi/V}^{(i)}$. In the $\chi$SB regime the fermions couple also to
the expectation value $\kappa$ and there are more terms $\sim \kappa
\,\ell_{2,2}^{\mathrm{(FB)}d}(\kappa h_\phi^2/\Nf,\dots)\,h_{\phi/V}^4
h_\phi^2$. They are suppressed for small $\kappa \ll 1$ and bounded
from above for large $\kappa \gg 1$ since
$\kappa\,\ell_{2,2}^{\mathrm{(FB)}d}(\dots) \sim \kappa/(1+\kappa
h_\phi/\Nf) \cdot \ell_{1,2}^{\mathrm{(FB)}d}(\dots)$. 
We expect that 
such terms do not take a significant influence on the flow.
For simplicity, we will omit them in the following.  Instead of
evaluating the diagrams in Fig.~\ref{fig:4-fermi-diagrams} explicitly,
we can determine the $a_{\phi/V}^{(i)}$ by taking advantage of the
fact \cite{Gies:2001nw, Jaeckel:2002rm} that the dynamically bosonized
flow in the pointlike limit $Z_{V,k} \rightarrow 0$ and $Z_{\phi,k}
\rightarrow 0$ exactly coincides with the fermionic flow computed
in~\cite{Gies:2010st},
\begin{align}
\partial_t \left(\Nf \frac{h_\phi^2}{m_\phi^2}\right) &\equiv - 2\partial_t \tilde g \bigr|_{\phi_k,V_k} , \\
\partial_t \left(\Nf \frac{h_V^2}{m_V^2}\right) &\equiv \partial_t (\tilde g -g) \bigr|_{\phi_k,V_k}.
\end{align}
This fixes $a_{\phi/V}^{(i)}$ uniquely and
establishes an exact mapping of the fermionic fixed-point
structure~\cite{Gies:2010st} onto the bosonized language in the
pointlike limit.
Beyond the pointlike approximation, the bosonized RG permits to
reliably run toward and into the $\chi$SB regime, allowing us to
predict the desired IR values of, for instance, fermion mass or order
parameter as a function of $\Nf$.


\section{UV structure and fixed points} 
\label{sec:fp-structure-phase-transition}
\subsection{Large-$\Nf$ limit}

In the limit of infinite flavor number $\Nf\rightarrow\infty$, the
flow equations simplify considerably. For the dynamically bosonized
flow in the symmetric regime we have
\begin{align}
\partial_t \left(\frac{m_\phi^2}{\Nf^2}\right) & = 
(-2+\eta_\phi) \left(\frac{m_\phi^2}{\Nf^2}\right), 
\label{eq:flow-mphi-largeN} \\
\partial_t m_V^2 & =
(-2+\eta_V) m_V^2 + \frac{2}{3\pi^2}\,\ell_1^{\mathrm{(F)}}(0;\eta_\chi)\,\left(
\Nf h_V^2\right) , \displaybreak[0] \\
\partial_t \left(\Nf \lambda_1\right) & = 
(-1+2\eta_\phi) \left(\Nf \lambda_1\right) -
\frac{1}{\pi^2}\,\ell_2^{\mathrm{(F)}}(0;\eta_\chi)\,h_\phi^4, \displaybreak[0]
\\
\partial_t \lambda_2 & = 
(-1+2\eta_\phi) \lambda_2 - 
\frac{2}{\pi^2}\,\ell_2^{\mathrm{(F)}}(0;\eta_\chi)\,h_\phi^4, \displaybreak[0]
\\
\partial_t \left(\Nf \mu \right) & = 
(-1+2\eta_V)\left( \Nf \mu \right)
\nonumber \\ &\quad 
+ \frac{4}{15\pi^2}\,\ell_2^{\mathrm{(F)}}(0;\eta_\chi)\,\left(\Nf
h_V^2\right)^2, \displaybreak[0] \\
\partial_t \left(\Nf \nu \right)& =
(-1+\eta_\phi+\eta_V) \left(\Nf \nu \right)
\nonumber \\ &\quad 
+ \frac{2}{3\pi^2}\,\ell_2^{\mathrm{(F)}}(0;\eta_\chi)\, \left(\Nf h_V^2\right)
h_\phi^2, \displaybreak[0] \\
\partial_t h_\phi^2 & = (-1+\eta_\phi+2\eta_\chi) h_\phi^2
\nonumber \\ &
-\frac{4}{\pi^2}\left(\frac{m_\phi^2}{\Nf^2} \right)
\ell_{1,2}^{\mathrm{(FB)}}(0,m_V^2;\eta_\chi,\eta_V) \left(\Nf h_V^2 \right)^2,
\label{eq:flow-hphi2-largeN} \displaybreak[0] \\
\partial_t \left(\Nf h_V^2 \right) & = 
(-1+\eta_V+2\eta_\chi) \left(\Nf h_V^2 \right), \label{eq:flow-hV2-largeN}
\displaybreak[0] \\
\eta_\phi & = \frac{2}{3\pi^2}\,m_4^{\mathrm{(F)}}(0;\eta_\chi)\, h_\phi^2,
\displaybreak[0] \\
\eta_V & = \frac{4}{3\pi^2}\,m_4^{\mathrm{(F)}}(0;\eta_\chi)\, \left( \Nf h_V^2
\right), \\
\eta_\chi & = 0, \label{eq:etapsi-largeN}
\end{align}%
where we have multiplied the flow equations with suitable factors of
$\Nf$ in order to simplify the large-$\Nf$ counting of orders. Hence,
in this limit the fermion wave function renormalization does not flow,
$\eta_\chi = - \partial_t \ln Z_{\chi,k} \equiv 0$, whereas the flow
of the bosonic wave function renormalizations is nonzero. 

\begin{table*}[t]
\begin{center}
\caption{Left columns: nonuniversal fixed-point couplings for
  the linear regulator and various flavor numbers $\Nf$. Right columns:
  universal correlation length exponent $\nu$, subleading
  exponent $\omega$, and anomalous dimension $\eta_\phi^*$. Rough error
  estimates arise from the comparison to sharp cutoff results. 
  For $\Nf \gtrsim 6$
  we do not expect chiral symmetry breaking due to vector channel
  domination, such that the critical exponents of the Thirring
  fixed point do not have the same physical meaning for the
  long-range physics. This is indicated by the gray font of the corresponding
  critical exponents; see Sec.~\ref{sec:thirring-IR-Nf-QPT}. For
  $\Nf=2$, a meaningful estimate of the systematic error for $\nu$ is
  not available due to sharp-cutoff artifacts, see
  text. }
\label{tab:fptvals_dyn-bos}
\begin{tabular*}{\textwidth}{@{\extracolsep{\fill}}
d{3.0}|d{2.4}d{3.3}d{3.3}d{3.3}d{3.3}|d{2.8}d{2.7}d{2.7}} %
\hline \hline
\multicolumn{1}{c|}{$\Nf$} & \multicolumn{1}{c}{$m_\phi^2/\Nf^2$} &
\multicolumn{1}{c}{$\Nf \lambda_1$} & \multicolumn{1}{c}{$\lambda_2$} &
\multicolumn{1}{c}{$h_\phi^{*2}$} & \multicolumn{1}{c|}{$\Nf h_V^{*2} /
m_V^{*2}$} & \multicolumn{1}{c}{$\nu$} & \multicolumn{1}{c}{$\omega$} &
\multicolumn{1}{c}{$\eta_\phi^*$} \\ \hline
 2 & 0.132 & 5.85 & 24.68 & 15.96 & 1.99 & 2.4(?) & 1.0(2) & 1.4(7)
\\
 3 & 0.265 & 16.94 & 39.84 & 19.36 & 6.11 & 1.22(2) & 1.5(2) & 1.6(3) \\
 4 & 0.271 & 20.95 & 44.45 & 20.61 & 8.83 & 1.084(5) & 1.7(1) & 1.7(2) \\
 5 & 0.266 & 22.70 & 46.55 & 21.41 & 10.83 & 1.043(8) & 1.74(7) & 1.8(1) \\
 6 & 0.259 & 23.56 & 47.70 & 21.95 & 12.35 &
\textcolor{gray}{1}.\textcolor{gray}{026(6)} & 
\textcolor{gray}{1}.\textcolor{gray}{80(4)} & 
\textcolor{gray}{1}.\textcolor{gray}{85(8)} \\
 8 & 0.247 & 24.34 & 48.86 & 22.59 & 14.44 &
\textcolor{gray}{1}.\textcolor{gray}{012(3)} &
\textcolor{gray}{1}.\textcolor{gray}{84(2)} &
\textcolor{gray}{1}.\textcolor{gray}{91(5)} \\
 10 & 0.238 & 24.68 & 49.42 & 22.93 & 15.79 &
\textcolor{gray}{1}.\textcolor{gray}{007(2)} &
\textcolor{gray}{1}.\textcolor{gray}{79(4)} &
\textcolor{gray}{1}.\textcolor{gray}{94(3)} \\
 12 & 0.231 & 24.85 & 49.73 & 23.14 & 16.74 &
\textcolor{gray}{1}.\textcolor{gray}{004(9)} &
\textcolor{gray}{1}.\textcolor{gray}{72(4)} &
\textcolor{gray}{1}.\textcolor{gray}{95(2)} \\
 25 & 0.209 & 25.16 & 50.33 & 23.54 & 19.39 &
\textcolor{gray}{1}.\textcolor{gray}{0006(1)} &
\textcolor{gray}{1}.\textcolor{gray}{423(1)} &
\textcolor{gray}{1}.\textcolor{gray}{988(2)} \\ 
100 & 0.187 & 25.26 & 50.52 & 23.68 & 21.45 &
\textcolor{gray}{1}.\textcolor{gray}{0000(1)} &
\textcolor{gray}{1}.\textcolor{gray}{126(1)} &
\textcolor{gray}{1}.\textcolor{gray}{999(1)} \\
\hline
\multicolumn{1}{c|}{\multirow{2}{*}{$\infty$}} &  
\multicolumn{1}{c}{$\frac{8}{45}$}
& \multicolumn{1}{c}{$\frac{64\pi^2}{25}$} &
\multicolumn{1}{c}{$\frac{128\pi^2}{25}$} &
\multicolumn{1}{c}{$\frac{12\pi^2}{5}$} &
\multicolumn{1}{c|}{$\frac{9\pi^2}{4}$} &
\multicolumn{1}{c}{\multirow{2}{*}{\textcolor{gray}{1}}} &
\multicolumn{1}{c}{\multirow{2}{*}{\textcolor{gray}{1}}} &
\multicolumn{1}{c}{\multirow{2}{*}{\textcolor{gray}{2}}}
\\
&  \simeq 0.178 & \simeq 25.27 & \simeq 50.53 & \simeq 23.69 & \simeq 22.21
\\
\hline\hline
\end{tabular*}
\end{center}
\end{table*}

Let us search for fixed points: assuming a fixed-point value
$m_\phi^{*2} \neq 0$, 
we are forced to conclude
from Eq.~\eqref{eq:flow-mphi-largeN}
that the scalar anomalous dimension at the fixed point obeys
$\eta_\phi^* = 2$. Similarly, Eq.~\eqref{eq:flow-hV2-largeN}
requires
for an interacting fixed 
point with $h_V^{*2} \neq 0$ for the vector anomalous dimension
$\eta_V^* = 1$. In fact, this is the 
large-$\Nf$ result found in
\cite{Hikami:1976at}.  With these values of the anomalous dimensions,
the full set of fixed-point values can be computed analytically, see
Table~\ref{tab:fptvals_dyn-bos}.

At this point, it may be worthwhile to make a few comments.  For
taking the limit $\Nf\rightarrow \infty$, we have implicitly assumed a
specific $\Nf$-scaling of the couplings, defining the
corresponding 't Hooft couplings. 
Since all rescaled couplings 
acquire finite (in particular nonzero) fixed-point values,
the corresponding rescalings are self-consistent and the
large-$\Nf$ limit is well-defined. For the original four-fermi
couplings $g_\phi = \Nf h_\phi^2/m_\phi^2$ and $g_V = \Nf h_V^2/m_V^2$
this implies the large-$\Nf$ scaling $g_\phi = \mathcal O(1/\Nf)$ and
$g_V = \mathcal O(1)$, already known from the fermionic
flow~\cite{Gies:2010st}. In fact, the fixed-point positions of
$g_\phi$ and $g_V$ for the dynamically bosonized flow in the
large-$\Nf$ limit and fermionic flow exactly coincide, provided the
same regulator is employed. We note, however, that the equivalence in
general does not hold for the flow beyond the fixed-point regime,
since kinetic terms of the bosonic fields are generated during the
flow according to their finite anomalous dimensions. For instance, the
large anomalous dimension $\eta_\phi^* = 2$ indicates a rapid
flow of the scalar kinetic term.

Consider the second contribution to the beta function of the
scalar-fermion coupling $h_\phi^2$ in
Eq.~\eqref{eq:flow-hphi2-largeN}, being proportional to
$m_\phi^2$. This is exactly the contribution from the diagrams in
Fig.~\ref{fig:4-fermi-diagrams}, i.e., the large-$\Nf$ reminiscence of
the scale-dependent Hubbard-Stratonovich transformation. Without this
contribution, the large-$\Nf$ fixed-point equations cannot be solved
with finite couplings. In other words, dynamical bosonization is
crucial to find the correct large-$\Nf$ behavior in the scalar sector
in the present model. We have already seen an indication for
this in Sec.~\ref{sec:thirring-bosonized-RG-fixed}, where the scalar
sector completely decoupled for $\Nf \geq 4$ using only standard
partial bosonization.

For the given set of integer anomalous dimensions, the large-$\Nf$
fixed-point values for the vector-vector coupling $\mu$ and the
vector-scalar coupling $\nu$ are in fact negative. We attribute this
to the fact that we neglected the momentum-dependent contribution
$\propto \bar\zeta_k$ in Eq.~\eqref{eq:thirring-bosonized-truncation}
as well as higher order terms in the effective potential. In any case,
in the large-$\Nf$ limit $\mu$ and $\nu$ do not feed back into the
flow of the remaining couplings. For consistency, we again evaluate in
what follows the flow equations for a pointlike current-current
interaction, $Z_{V,k} \rightarrow 0$, and neglect the vector-vector
selfinteractions $\mu$ and vector-scalar interactions $\nu$, analogous
to Sec.~\ref{sec:thirring-bosonized-RG-fixed}. This is expected to be
a reasonable approximation as long as the vector mass $m_V^2$ does not
become too small.

\subsection{Fixed-point structure at general $\Nf$}

\begin{figure*}
\includegraphics[scale=1]{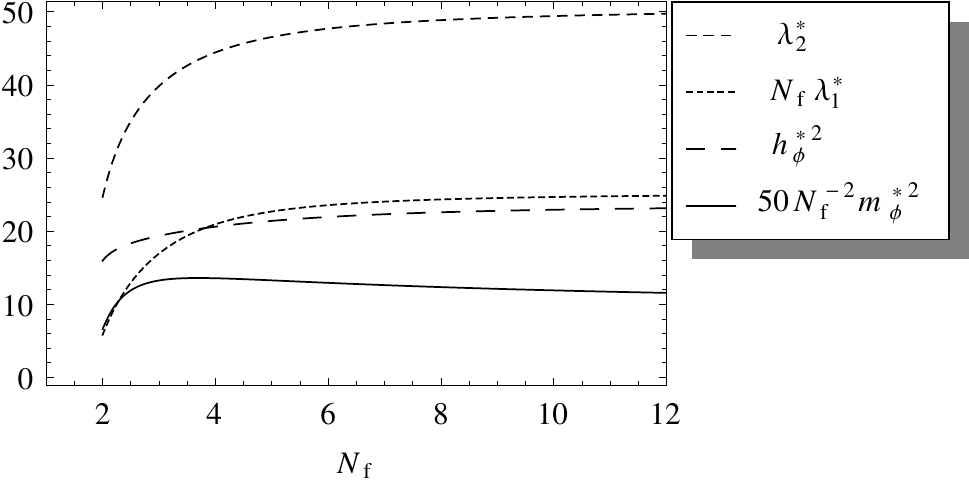}\hfill
\includegraphics[scale=.98]{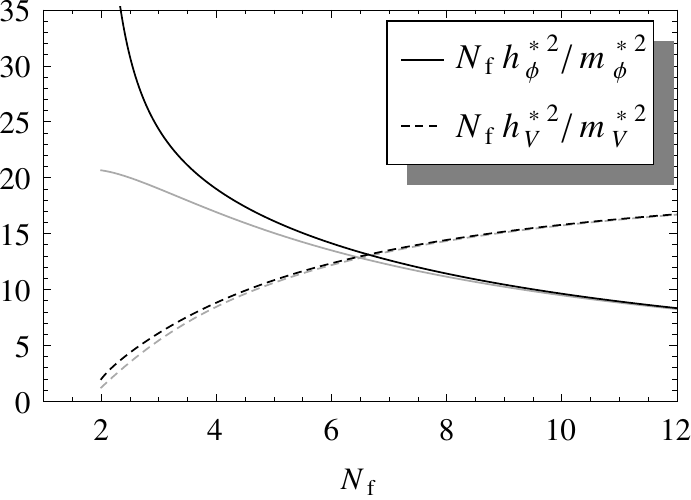}
\caption{Left panel: nonuniversal UV fixed-point values of dynamically
  bosonized RG flow for the linear regulator. The couplings have
  been multiplied with suitable $\Nf$ factors; for better visibility,
  the scalar mass has been multiplied with an additional factor
  $50$. Right panel: Comparison of dynamically bosonized RG (black) with
  fermionic RG (gray). The improvement due to {\em dynamical}
    bosonization becomes obvious by comparing this plot with the left
  panel of Fig.~\ref{fig:gVgphi_fixedFields}.}
\label{fig:fptvals_dyn-bos}
\end{figure*}
\begin{figure*}
\includegraphics[scale=1.05]{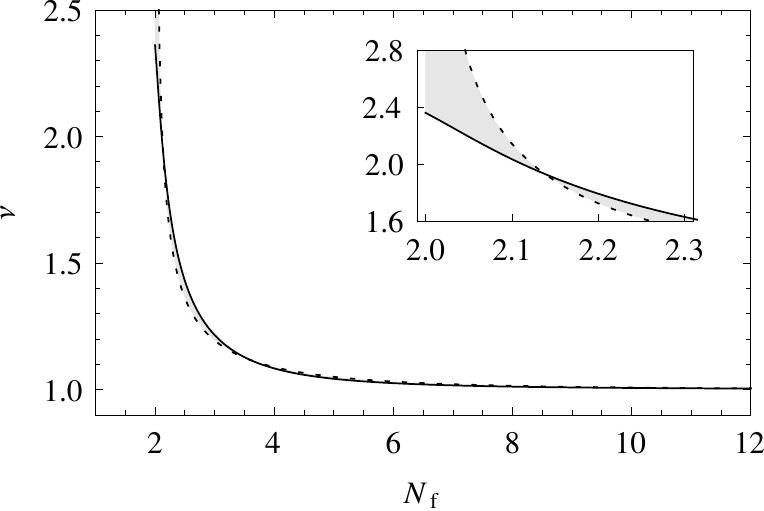} \hfill
\includegraphics[scale=1.05]{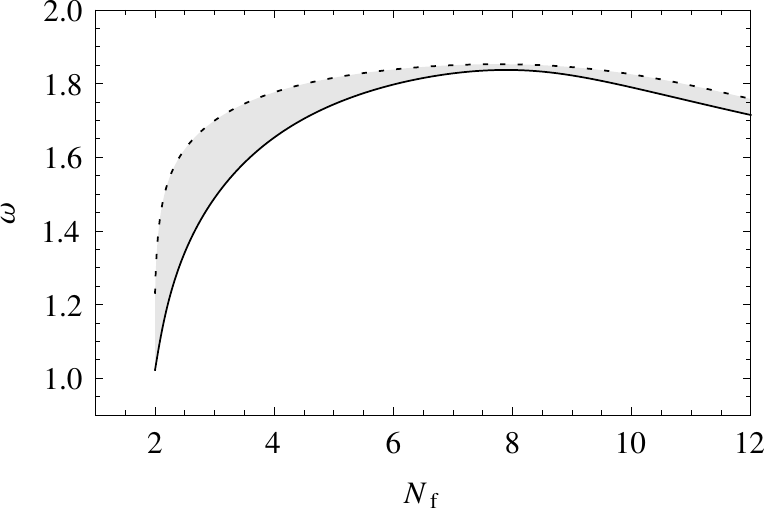}
\caption{Left panel: correlation length exponent $\nu$ extracted from
  the linearized flow in the fixed-point regime with linear
  (solid) and sharp regulator (dashed). The difference (gray
  shaded band) serves as rough error estimate. The critical exponent is to
  a large extent independent of the regulator, while the uncertainty
  increases for $\Nf \searrow 2$, as shown in the inset. For
  increasing $\Nf$ it approaches rapidly the fermionic result
  $1/\Theta_1 = 1$. Right panel: corrections-to-scaling exponent $\omega$.}
\label{fig:critical-exponents1}
\end{figure*}

\begin{figure}[t]
\includegraphics[scale=1]{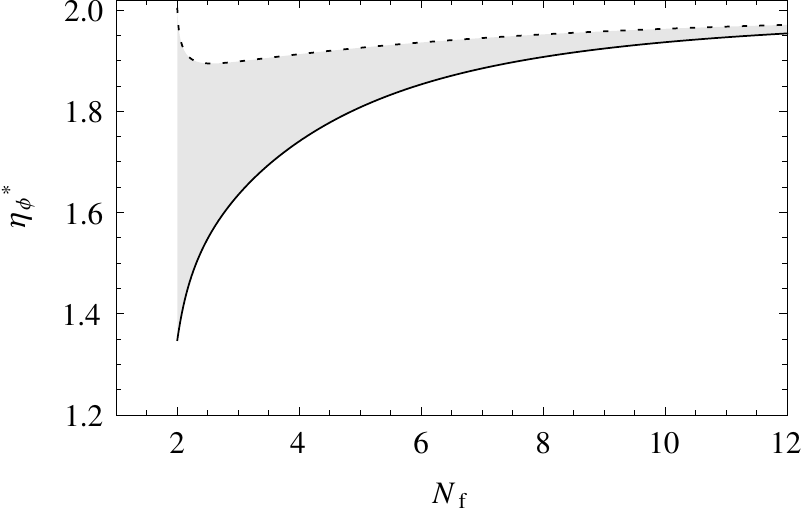}
\caption{Anomalous dimension $\eta_\phi$ at the fixed point, with
  linear regulator (solid) and sharp regulator (dashed).}
\label{fig:critical-exponents2}
\end{figure}

Beyond the large-$\Nf$ limit, we evaluate the fixed-point equations
numerically, both for linear and sharp cutoff. Again, we
recover the known UV structure: there is one interacting Thirring
fixed point for all $1<\Nf\leq\infty$, having only one IR relevant
direction. [Recall that the case $\Nf=1$ has been explicitly excluded
  in the derivation of our bosonic flow equations,
  cf.\ Eq.~\eqref{eq:field-configuration}.] For small $\Nf$ 
the Thirring fixed point is
located close to the pure scalar channel subspace, whereas for
$\Nf\to\infty$ the scalar-fermion coupling becomes negligible. As we
have also observed in the partially bosonized formulation in
Sec.~\ref{sec:thirring-bosonized-RG-fixed}, the
fixed-point position in this limit 
again exactly coincides with the
fermionic fixed point. For small $\Nf$, we find deviations from the
fermionic UV structure. As the dynamically bosonized flow and
the fermionic one have an identical fixed-point structure in the
point-like limit (by construction), these deviations
are fully related to the momentum dependence in the
bosonized scalar channel. 
The latter is particularly important once the scalar
coupling becomes long range, i.e., for small $\bar m_\phi^2$. In
other words, the bosons dynamically become fluctuating relevant
degrees of freedom. We depict the fixed-point positions for the
linear regulator in Fig.~\ref{fig:fptvals_dyn-bos}, together with
the corresponding values for the fermionic flow compiled
from~\cite{Gies:2010st} for comparison.  Explicit values are given in
Table~\ref{tab:fptvals_dyn-bos}.
If the fixed point corresponds to a second-order phase
transition as we expect it for small $\Nf$, the critical
behavior is uniquely determined by the fixed-point regime, where the
flow can be linearized. 
Let $g=(m_\phi^2, \lambda_1, \lambda_2, h_\phi^2, \Nf h_V^2/m_V^2)$
denote the vector of our (generalized) couplings.
The scaling of 
the correlation length in the vicinity of the critical point is
\begin{align}
\xi = \bar m_{\phi,\text{R}}^{-1} \propto |\delta g|^{-\nu}\left(1+b_\pm |\delta g|^{\omega\nu}+\dots\right),
\end{align}
with $\delta g\coloneqq g_\Lambda-g_\text{cr}$ measuring the distance from criticality (``reduced temperature'') and the renormalized scalar mass $\bar m_{\phi,\text{R}}^2 = \lim_{k\rightarrow 0}  m_\phi^2 k^2$.
By again denoting the smallest eigenvalue (being negative) of the
stability matrix $\partial \beta_i/\partial g_j |_{g^*}$ with
$-\Theta_1<0$ and the second smallest eigenvalue (being positive) with
$-\Theta_2>0$ we have for the correlation length exponent and the first
subleading exponent~\cite{ZinnJustin:1996cy}
\begin{align} \label{eq:nu-omega-theta12-again}
\nu = 1/\Theta_1 > 0, \qquad \text{and} \qquad \omega =-\Theta_2 > 0.
\end{align}
At the critical point $\delta g=0$, where the correlation length diverges, the asymptotic behavior of the scalar two-point function is determined by the anomalous dimension $\eta_\phi^*=\eta_\phi(g^*)$ as
\begin{align}
\langle \varphi(x) \varphi(0) \rangle \propto \frac{1}{|x|^{d-2+\eta_\phi^*}}.
\end{align}
The critical exponents extracted from the flow in the fixed-point
regime are shown in Figs.~\ref{fig:critical-exponents1}
and~\ref{fig:critical-exponents2}. We expect the values, as
listed in Table~\ref{tab:fptvals_dyn-bos}, obtained with
the linear regulator to represent our most accurate results
and use the difference to the sharp-cutoff results as a rough estimate on the
truncation-induced error. 
We find that the correlation-length exponent
is to a large extent regulator-independent, $\Delta \nu/\nu \lesssim
1\dots 2\%$ for $\Nf\gtrsim 2.1$. Near $\Nf=2$, the sharp-cutoff
results deviate from the linear regulator by up to a factor of
two. We interpret this behavior as a large artifact of the sharp
cutoff and consider the result using the linear regulator as our
best estimate. For this particular case of $\Nf=2$, it is however difficult to
estimate the systematic error. As expected, slightly larger
but controlled deviations
between linear and sharp cutoff occur for the anomalous dimension
and the corrections-to-scaling exponent, $\Delta
\eta_\phi/\eta_\phi,\Delta\omega/\omega \lesssim 10\dots
15\%$. Explicit values are again given in
Table~\ref{tab:fptvals_dyn-bos}. This hierarchy of accuracy is
well-known from RG studies of scalar models based on the derivative
expansion \cite{Berges:2000ew, Litim:2010tt, Blaizot:2005xy}.

\section{IR behavior and $\Nf$-controlled quantum phase transition} 
\label{sec:thirring-IR-Nf-QPT}

\subsection{Critical fermion number}
If we start the flow for small $\Nf$ with initial UV couplings close
to the fixed point, we find that the scalar mass eventually vanishes
at some scale $k^*$, indicating the spontaneous breakdown of chiral
symmetry. In the following, we will refer to $k^*$ as ``$\chi$SB
scale''. Continuing the flow for $k<k^*$ in the $\chi$SB regime the
fermions become massive with renormalized mass $\bar m_{\text{R,f}}^2
= \Nf^{-1} \kappa h^2 k^2$ and the scalar sector consists of one
radial mode with renormalized mass $\bar m_{\text{R},\rho}^2 = 2
\kappa \lambda_1 k^2$, $2\Nf^2-1$~massive modes with $\bar
m_{\text{R},\tau}^2 = \Nf^{-1} \kappa \lambda_2 k^2$, and $2\Nf^2$
massless Goldstone modes. In the deep IR $k\to 0$, we expect all
massive modes as well as eventually the Goldstone modes to decouple,
leading to IR predictions for the mass spectrum.

\begin{figure}
\includegraphics[scale=.97]{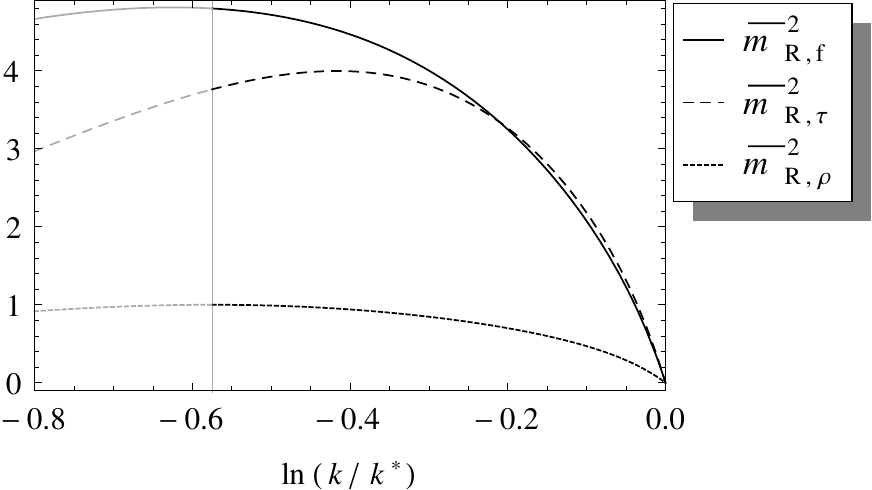}
\caption{RG evolution of renormalized masses in $\chi$SB regime for $k<k^*$ and
$\Nf=2$. We stop the flow when the radial mass $\bar m_{\text{R},\rho}^2$
approaches its maximum (vertical line).}
\label{fig:massflow}
\end{figure}

However, our truncation is not able to resolve
this decoupling, as all couplings among the different modes are
approximated by the same expansion coefficient of the effective
potential.
In fact, we naively find that the dimensionless parameters run into an
attractive IR fixed point, such that the dimensionful masses run into
a maximum and eventually decrease again for small $k \ll k^*$, see
Fig.~\ref{fig:massflow}. This effect is a well-known artifact
of our treatment of the effective potential, as it
artificially couples the Goldstone modes to the flow of all other
scalar operators.
This problem can technically be solved
by an adapted choice of
field coordinates
\cite{Lamprecht:diploma}.
Here, we follow a more immediate strategy, and simply stop the
flow before it
enters the artificial IR fixed-point regime.
For our quantitative estimates,
we stop the flow at the maximum of the radial
mass $\bar m_{\text{R},\rho}^2$.
We have checked that the following results to a large
extent do not depend on this choice. We note that the critical
behavior in terms of the exponents $\nu$ and $\eta_\phi^*$ remains, of
course, unaffected by this, since it is solely determined by the UV
structure of the theory.

\begin{figure}
\includegraphics[scale=1.04]{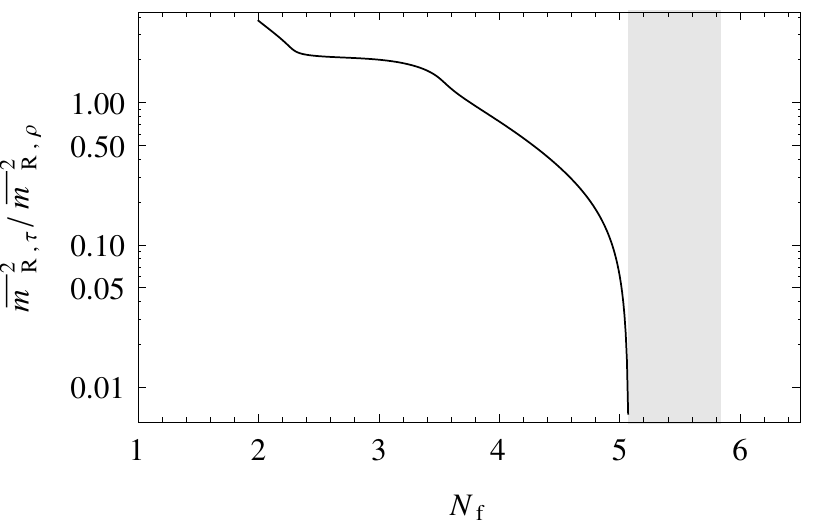}
\caption{Scalar mass $\bar m_{\text{R},\tau}^2$ in units of the radial mass
$\bar
  m_{\text{R},\rho}^2$ on logarithmic scale (linear cutoff). Gray shaded area: 
  estimates for critical flavor number $\Nfc \simeq 5.1$ (linear cutoff) and 
  $5.8$ (sharp cutoff).}
\label{fig:scalar-mass-Nf}
\end{figure}

Once the physical scale has been set, for instance, by measuring the
value of the radial mass $\bar m_{\text{R},\rho}^2$, we can compare
the dynamically generated masses among different $\Nf$. 
We have already shown 
the renormalized fermion mass $\bar m_{\text{R,f}}^2$ in units of the radial
mass $\bar m_{\text{R},\rho}^2$ in Fig.~\ref{fig:fermion-mass-Nf}. 
In Fig.~\ref{fig:scalar-mass-Nf} we now depict the renormalized scalar mass
$\bar m_{\text{R},\tau}^2$ in the same units.
For increasing flavor number we observe that
both $\bar m_{\text{R,f}}^2$ and $\bar m_{\text{R},\tau}^2$ decrease
(apart from some interesting nonmonotonic behavior of the fermion mass
for $\Nf \sim 3$) and eventually vanish for some critical flavor
number $\Nf \nearrow \Nfc$. This is a clear and expected
signature for a quantum phase transition governed by flavor number
$\Nf$.

Directly at $\Nf=\Nfc$ we find that the vector-fermion coupling $\Nf
h_V^2/m_V^2$ diverges at the same time as the scalar mass
$m_{\phi,k}^2$ approaches zero, i.e., at the $\chi$SB scale $k^*$. We
expect that this divergence is an artifact of our truncation which
will be stabilized by higher terms in the vector sector [for instance,
  the terms $\propto \bar\mu_k$, $\bar\nu_k$, $\bar\zeta_k$ in
  Eq.~\eqref{eq:thirring-bosonized-truncation}]. Here, we interpret
the divergence as an indication for ``vector-boson-dominance'' in
the IR, inhibiting $\chi$SB. If the divergence was real, the model
could exhibit dynamical Lorentz symmetry breaking
\cite{Hosotani:1993dg, Hosotani:1994sc, Wesolowski:1995xz,
  Higashijima:2001sq, Khalilov:2004wf}.  For larger $\Nf > \Nfc$ we
observe the divergence already at higher scales $k>k^*$, inhibiting
the flow to enter the $\chi$SB regime. 

The value of $\Nfc$ can thus be obtained in two ways: 
First, we observe for increasing $\Nf$ a sharp decrease of the
logarithm of the fermion mass once $\Nf \nearrow \Nfc$.
Second, we look for the largest $\Nf$ below
which the vector-fermion coupling $\Nf h_V^2 / m_V^2$ remains finite for all
scales in the interval $k \in [k^*,\Lambda]$. Above this $\Nf$, the
vector-fermion coupling diverges before the scalar mass $m_{\phi,k}$
approaches zero.
For the linear cutoff we find $\Nfc \simeq 5.07$ with the
former method and $\Nfc \simeq 5.1$ with the latter, agreeing
with each other on the level of numerical precision. For the sharp
cutoff, our truncation unfortunately does not allow to enter the
$\chi$SB regime: we find $\lambda_{1,k^*} < 0$ at the $\chi$SB scale,
indicating the requirement of higher terms in the polynomial expansion
of the effective potential \eqref{eq:eff-pot-expansion}. 
This is in line with optimization considerations \cite{Litim:2002cf,
Pawlowski:2005xe}:
for a nonoptimized regulator (as is the sharp cutoff) 
a higher expansion order is needed to achieve similar predictive
power. For the scalar and fermion sectors, the linear regulator
by contrast satisfies standard optimization criteria
\cite{Litim:2002cf, Pawlowski:2005xe}.
Nonetheless, also in the sharp-cutoff scheme, we can determine
the flavor number above which the vector-fermion coupling diverges
before $m_{\phi,k} \rightarrow 0$.
From this criterion, we find $\Nfc \simeq
5.8$. Identifying the cutoff dependence with a rough error estimate
yields our result for the critical flavor number
\begin{align}
\Nfc \simeq 5.1(7).
\end{align}
This prediction represents one of the central results of this work.

\subsection{Thirring universality class}

\begin{figure}
\includegraphics[scale=.95]{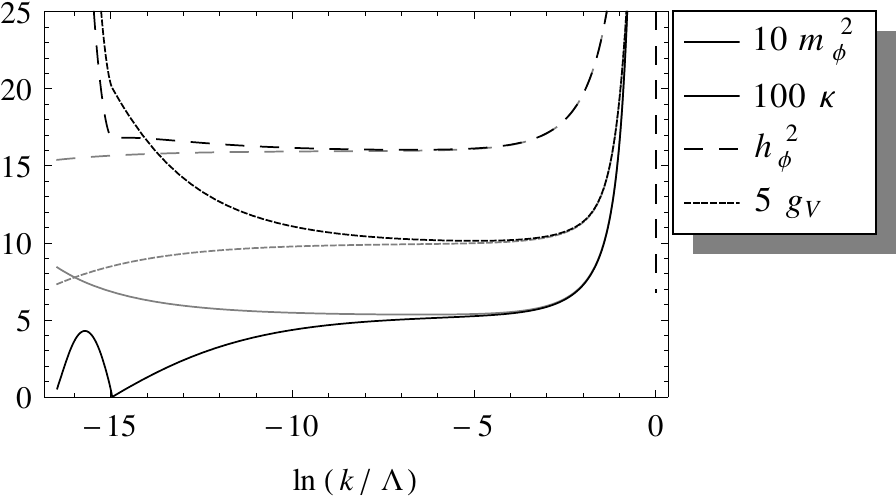}
\caption{RG evolution of dimensionless couplings for the ``pure''
  Thirring model with $\Nf=2$ and initial couplings
  $m_{\phi,\Lambda}^2 = 10^6$, $\lambda_{i,\Lambda}=0$,
  $h^2_{\phi,\Lambda}=0$, and two different $g_{V,\Lambda} = \Nf
  h_{V,\Lambda}^2/m_{V,\Lambda}^2$ just above (black) and below (gray)
  $g_\text{cr}$. 
  Near $\ln (k/\Lambda) 
  \simeq -5$ the flow approaches
  the UV fixed-point regime. The black solid line corresponds to
  $m_\phi^2$  above and  $\kappa$ below the $\chi$SB scale $\ln
  (k^*/\Lambda) \simeq -15$, respectively.}
\label{fig:rgflowNf2}
\end{figure}

\begin{figure}
\includegraphics[scale=1]{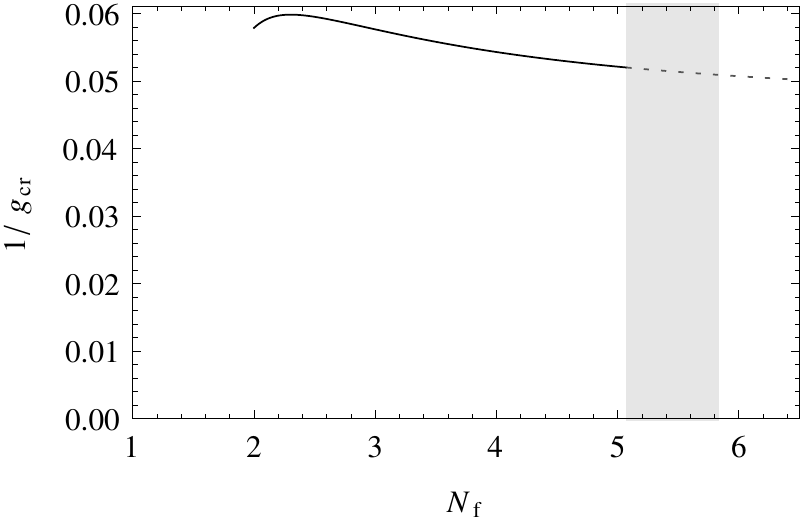}
\caption{Critical coupling $1/g_\text{cr}$ for the ``pure'' Thirring
  model with initially decoupled scalar sector as a function of
  $\Nf$. For $\Nf \gtrsim 5.1$ (gray area) we
  find no signature of chiral symmetry breaking (dashed line).}
\label{fig:gcr}
\end{figure}

In the fermionic RG \cite{Gies:2010st}, we have shown that the
``pure'' Thirring model, defined by a microscopic action including
only current-current interaction, is in the universality class defined
by the interacting Thirring fixed point with one IR relevant
coupling. Of course, this statement ultimately only holds for our
specific regularization scheme, since fixed-point positions itself are
nonuniversal. In the bosonized formulation developed here we can 
further explore the universality class of the ``pure'' Thirring model by
starting the flow with initial couplings
\begin{align}
Z_{\phi}^{-1} \propto m_{\phi,\Lambda}^2 & \gg 1, &
\lambda_{i,\Lambda} & \ll 1 \quad (i=1,2), &
h_{\phi,\Lambda}^2 & \ll 1,
\end{align}
i.e., with a decoupled scalar sector. In our explicit computations we
use $Z_{\phi,\Lambda} = 10^{-6}$, $m_{\phi,\Lambda} = 10^6$,
$\lambda_{i,\Lambda} = 0$, $h_{\phi,\Lambda}^2 = 0$. However,
universality guarantees that our results are independent of the exact
values for the initial couplings as long as the system is
governed by the corresponding fixed point.
With the functional RG flow, we can explicitly 
verify universality.
Fig.~\ref{fig:rgflowNf2} shows the RG evolution
of the dimensionless couplings for $\Nf=2$ for two different initial 
values of the
vector-fermion coupling $g_V = \Nf h_V^2/m_V^2$ just above and below
the critical value $g_\text{cr}$.
Though we start with a decoupled scalar sector, the scalar couplings
are rapidly generated by the RG flow, as already known from the purely
fermionic RG.
For initial couplings close to criticality, the flow runs from
the pure Thirring axis into the fixed-point regime, which is
exactly given by the Thirring fixed-point couplings in
Table~\ref{tab:fptvals_dyn-bos}, cf.\ Fig.~\ref{fig:rgflowNf2}. Thus,
the critical behavior of a conventionally defined ``pure'' Thirring
model without scalar channel in the bare action is indeed given by the
universality class of our Thirring fixed point. For completeness, we
show in Fig.~\ref{fig:gcr} the critical coupling $g_\text{cr}$ for
which the flow approaches the Thirring fixed point as a function of
$\Nf$, to be compared with previous results in
Ref.~\cite{Gies:2010st}.%
\footnote{We use the opportunity to
  point out that the values for $1/g_{\text{cr}}$ plotted in
  \cite{Gies:2010st} include an erroneous factor of 2. This has
  already been corrected in \cite{Janssen:2012phd}.}
As in the fermionic calculation, we do not find a sharp decrease of
$1/g_\text{cr}$ as $\Nf \nearrow \Nfc$. This observation is in
agreement with lattice simulations \cite{Kim:1996xza, DelDebbio:1997dv,
DelDebbio:1999xg, Hands:1999id, Christofi:2007ye} but disagrees with
earlier analytical estimates \cite{Itoh:1994cr, Sugiura:1996xk}.
A rapid variation of $1/g_\text{cr}$ would have been expected if
the many-flavor quantum phase transition was induced by a change in
the UV fixed-point structure. This, however, is not the case in the
Thirring model where the transition is
rather due to a competition between the vector and scalar
channel. Since the Thirring fixed point is present for all $\Nf$, we
can still determine a ``would-be'' critical coupling $g_\text{cr}$
even for $\Nf> \Nfc$ (dashed line in Fig.~\ref{fig:gcr}).

\begin{figure*}
\includegraphics[scale=1.1]{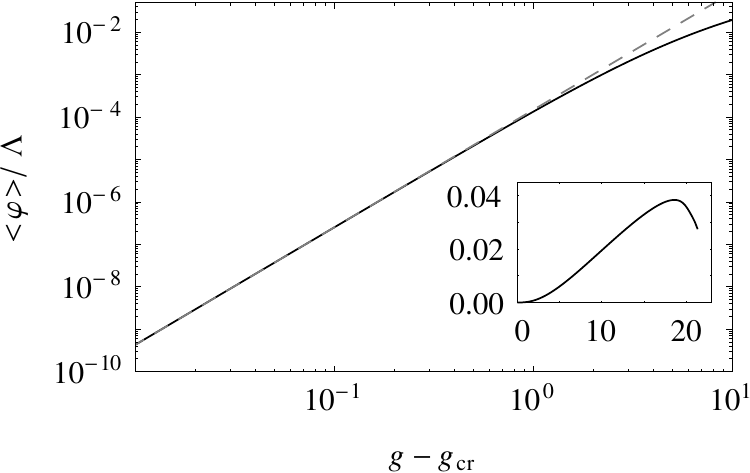} \hfill
\includegraphics[scale=1.1]{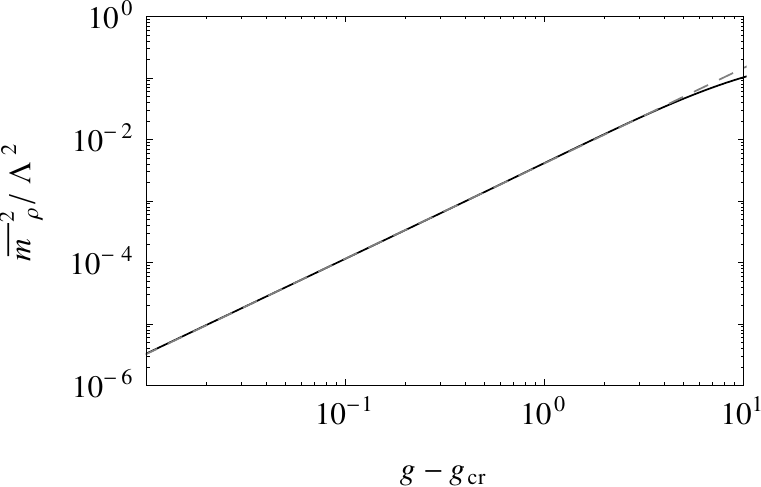}
\caption{Critical behavior of ``pure'' Thirring model with $\Nf=2$:
  order parameter $\langle \varphi \rangle \propto |\delta g|^{\beta}$
  (left panel) and inverse susceptibility $\chi^{-1} = \bar m^2_{\rho}
  \propto |\delta g|^{\gamma}$ (right panel) in double-log plot, showing the
  expected power law. The slope of the regression line (gray dashed)
  is $\beta = 2.769$ and $\gamma = 1.553$, respectively. Left inset:
  for very large coupling on the pure Thirring axis the order
  parameter decreases again, see Sec.~\ref{sec:comparison}.}
\label{fig:crit-exp-beta-gamma}
\end{figure*}

In order to establish the connection between the negative eigenvalues of the
stability matrix $\Theta_i$ and the critical exponents in
Eq.~\eqref{eq:nu-omega-theta12-again} we have assumed that the propagator has a
scaling form, from which one infers the hyperscaling
relations~\cite{ZinnJustin:1996cy}
\begin{align}
\beta & = \frac{\nu}{2} (d-2+\eta_\phi^*), &
\gamma & = \nu (2-\eta_\phi^*).
\end{align}
The hyperscaling assumption can in fact explicitly be 
checked by computing directly the critical behavior of the
order parameter $\langle \varphi \rangle \propto |\delta g|^\beta$,
inverse susceptibility (unrenormalized mass) $\chi^{-1} = \bar
m_{\rho}^2 \propto |\delta g|^{\gamma}$, and inverse correlation
length (renormalized mass) $\xi^{-1} = \bar m_{\text{R},\rho} \propto
|\delta g|^{\nu}$. As a function of the distance to criticality, we
indeed find that the expected linear behavior on a double-log plot is
excellently fulfilled; see Fig.~\ref{fig:crit-exp-beta-gamma} for the
case of the ``pure'' Thirring model (initially decoupled scalar
sector) and $\Nf=2$.
The slopes of the regression lines for initial couplings on the pure Thirring axis are (for $\Nf=2$)
\begin{align} \label{eq:beta-gamma-nu-thirringAxis}
\beta & = 2.769, &
\gamma & = 1.553, &
\nu & = 2.364,
\end{align}
whereas if we start the flow 
directly near the Thirring fixed point, we obtain
\begin{align} \label{eq:beta-gamma-nu-thirringFP}
\beta & = 2.771, &
\gamma & = 1.539, &
\nu & = 2.361.
\end{align}
The values should be compared with the critical exponents obtained from the
linearized flow in the Thirring fixed-point regime (see
Table~\ref{tab:fptvals_dyn-bos}), together with the hyperscaling assumption:
\begin{align} \label{eq:beta-gamma-nu-thirringFP-linflow}
\frac{\nu}{2} (d-2+\eta_\phi^*) & = 2.771, &
\nu (2-\eta_\phi^*) & = 1.539, &
\nu & = 2.360,
\end{align}
in excellent agreement with Eqs.~\eqref{eq:beta-gamma-nu-thirringFP}.
However, hyperscaling is to a large extent also fulfilled for the
conventionally defined ``pure'' Thirring model,
cf.\ Eqs.~\eqref{eq:beta-gamma-nu-thirringAxis} with
\eqref{eq:beta-gamma-nu-thirringFP-linflow}. We attribute the small
(but significant) hyperscaling violations
to the presence of other
fixed points: in particular the critical behavior could be influenced
by fixed point $\mathcal B$, which (at least in the pointlike limit, 
see Fig.~\ref{fig:fermionic-flow})
is located in the broader vicinity of the
pure Thirring axis and potentially describes a different universality
class. We will further elaborate on this aspect in the following
section, see also Fig.~\ref{fig:deltaFP-B}.


\section{Comparison with previous studies} \label{sec:comparison}
The diversity of results for the 3d Thirring model
obtained so far in the literature calls for a careful comparison of
our findings with those derived with other techniques. With
hindsight, this variety of partly contradictory results arises from
the somewhat unexpected complexity of the model. We have identified
several sources for this complexity: (i) an involved UV fixed point
structure (three instead of just one UV fixed point with the
Thirring fixed point being off the pure Thirring axis), (ii) a
complex bound state spectrum giving rise to effective vector and
scalar degrees of freedom, (iii) a symmetry breaking
mechanism with competing vector and scalar channels. All issues need
to be taken care of in order to arrive at conclusive answers.
 
Let us start with a comparison with
the large-$\Nf$ expansion. With standard large-$\Nf$ techniques,
renormalizability of the ``pure'' 3d Thirring model with only
current-current interaction $\propto g_V (\bar\psi \gamma_\mu \psi)^2$
(i.e., without scalar channel) has been shown to hold if and only if a
regularization scheme is employed in which the vector field propagator
remains purely transverse on the quantum level, defining an
interacting UV fixed point \cite{Parisi:1975im, Hikami:1976at,
  Gomes:1990ed, Yang:1990ki, Hands:1994kb}. In
Ref.~\cite{Hikami:1976at} it is found to leading order in $1/\Nf$ (in
terms of our notation)
\begin{align} \label{eq:hikami-largeN}
\partial_t g_V & = g_V\left(1 -\frac{g_V}{g_V^*} \right), &
\eta_V & = \frac{g_V}{g_V^*}, &
\eta_\chi & = \mathcal O(1/\Nf),
\end{align}
which is exactly the large-$\Nf$ behavior of our flow equations
\eqref{eq:flow-mphi-largeN}--\eqref{eq:etapsi-largeN}: if the fixed
point corresponds to a second-order phase transition, the
corresponding correlation-length exponent is $\nu = 1/\Theta = 1$
where $\Theta = - \partial \beta /\partial g_V |_{g_V^*}$ and the
vector field anomalous dimension is $\eta_V(g_V^*) = 1$.

The seeming resemblance to gauge theories has been taken even more
seriously: Yang \cite{Yang:1990ki} claims that the partially bosonized
Thirring model is equivalent to (a gauge-fixed version of) a $\mathrm
U(1)$ gauge theory, in which the mass of the vector boson is generated
by the Higgs mechanism. Insisting on the gauge symmetry, it is
therewith argued that the coupling $g_V$ \emph{cannot} be
renormalized. Consequently, the beta function would be vanishing for
any value of $g_V$~\cite{Hands:1994kb, DelDebbio:1997dv}.

From our functional RG viewpoint, nonperturbative renormalizability
does neither rely on a specific regularization scheme nor a
resemblance or equivalence to a gauge theory. All that is needed is a
non-Gau\ss ian UV fixed point (or a line of fixed points) with
suitable properties to render the system asymptotically safe
\cite{Weinberg:1976xy,Weinberg:1980gg}. We observe 
such a fixed
point both in the purely fermionic description which does not have a
gauge symmetry as well as in the bosonized language where the vector
mass term breaks gauge symmetry manifestly. Still, it is interesting
to see that our large-$\Nf$ flow
equations~\eqref{eq:flow-mphi-largeN}--\eqref{eq:etapsi-largeN},
resemble that of a gauge-fixed theory: the vector mass does not feed
back into the vector sector and the second kinetic term ($\sim
\bar{A}_{V,k}$) looks like a gauge-fixing term that is even locked to
the kinetic term in our truncation.
  
Whether or not the RG flow is eventually attracted by a
gauge-invariant theory in the long-range limit still remains an open
question. In our approach, this could be investigated by measuring
the flow with respect to the hyperplane of actions in theory space
that satisfy the (regulator-modified) Ward-Takahashi identity
\cite{Ellwanger:1994iz, Pawlowski:2005xe, Gies:2006wv}. We believe, however,
that this question is unrelated to that of chiral symmetry breaking,
as the latter occurs in the chiral scalar sector which does not
participate in the local symmetry.

The formulation of the Thirring model as a gauge theory is also
at the basis of the analytical studies \cite{Gomes:1990ed, Hong:1993qk,
  Itoh:1994cr, Sugiura:1996xk}, which attempt to solve the
Dyson-Schwinger equations (DSE) by setting the full vector propagator
to its large-$\Nf$ form, and the full vertex to the bare vertex (known
as ladder approximation). All such studies 
observe the existence of a
nontrivial solution corresponding to chiral symmetry breaking for
small values of $\Nf < \Nfc$. However, their predictions for the value
of $\Nfc$ and the critical behavior of fermion mass and chiral order
parameter differ significantly: Ref.~\cite{Gomes:1990ed} reports $\Nfc
\simeq 3.24$. Hong and Park \cite{Hong:1993qk} claim that symmetry
breaking should persist for any value of $\Nf$, i.e., $\Nfc =
\infty$. In Ref.~\cite{Itoh:1994cr} an essentially singular behavior
close to $\Nfc \simeq 4.32$ is found,
\begin{align} \label{eq:essential-scaling}
\bar m_\text{R,f} \propto \Lambda \exp\left(-\frac{2\pi}{\sqrt{\Nfc/\Nf-1}}\right), \qquad \text{for } \Nf < \Nfc,
\end{align}
that is to say, a phase transition of \emph{infinite} order.  Such an
essential scaling law 
is known from a number of gauge theories,
such as quenched QED$_4$ \cite{Fomin:1984tv,
  Miransky:1984ef,Miransky:1996pd} and QED$_3$ with $\Nf$ fermion
flavors \cite{Appelquist:1986qw, Appelquist:1988sr, Nash:1989xx,
  Gusynin:1998kz, Fischer:2004nq}, as well as many-flavor QCD$_4$
\cite{Miransky:1996pd, Appelquist:1996dq, Harada:2003dc,
  Iwasaki:2003de,Dietrich:2006cm}. In this context, the scaling law is often referred to
as Miransky scaling; the scenario has also been termed
\mbox{(pseudo-)}conformal phase transition
\cite{Miransky:1996pd}. Essential scaling has also been found in
two-dimensional statistical systems, such as the XY model, where the
scenario is called Kosterlitz-Thouless phase transition
\cite{Berezinsky:1970fr,berezinsky1972destruction, Kosterlitz:1973xp,
  Kosterlitz1974critical}. As has been pointed out recently
\cite{Kaplan:2009kr,Braun:2011pp}, the general mechanism being responsible for
essential scaling is the merger of two RG fixed points \cite{Gies:2005as, Braun:2009ns, Braun:2010qs} and their
subsequent disappearance into the complex plane, or a running of a
fixed point off to zero or infinite coupling. By contrast, we find that
the UV Thirring fixed point persists for any $\Nf$, hence there
is no basis for and consequently no observation of essential scaling in
our work.

By constructing an effective potential for the chiral order parameter
$\langle \varphi \rangle \propto \langle \bar\psi\psi\rangle$,
to leading order in $1/\Nf$, Kondo~\cite{Kondo:1995jn} reports a
\emph{second-order}
phase transition,
\begin{align} \label{eq:orderparam-scaling-Kondo}
\langle \varphi \rangle \propto \left(\frac{\Nfc}{\Nf} -1 \right)^{b}, \qquad \text{for } \Nf < \Nfc,
\end{align}
where for $d=3$ he finds the critical exponent $b=1$ and
$\Nfc=2$. We have been motivated by these considerations to attempt a
fit to our results for order parameter and fermion mass. In
Fig.~\ref{fig:orderparam}, we depict $\langle \varphi \rangle / \bar
m_{\mathrm{R},\rho}$ versus the combination $(\Nfc/\Nf) -1$ in a
double-log plot. In fact, the behavior is very well compatible with a
power-law scaling corresponding to a second-order phase transition at
$\Nfc$. However, the linear fit (gray line) gives $b \simeq 0.44$ in
qualitative agreement with the Kondo scenario
\cite{Kondo:1995jn} but a quantitatively differing estimate for
the exponent.  We emphasize that for a quantum phase transition at
$\Nfc$ occurring due to competing condensation channels, the
corresponding critical exponents do not 
have to coincide with
those determining the phase transition with the coupling as control
parameter. In particular, there is no reason that 
$\left.\beta\right|_{\Nfc}$ and $b$ should coincide. In any case, a fit to an
essential scaling behavior analogous to \eqref{eq:essential-scaling} both for
order parameter and fermion mass is much less successful and is not
supported by our results.

Chiral symmetry breaking in the ``pure'' 3d Thirring model has also
been investigated on the lattice \cite{Chandrasekharan:2011mn,
  Christofi:2007ye, DelDebbio:1999xg, Hands:1999id, DelDebbio:1997dv,
  Kim:1996xza}. One extensive study \cite{Christofi:2007ye} finds
$\Nfc = 6.6(1)$, the bare value of which we consider as still
consistent with our result, see below. Moreover, the study also
analyzes the critical behavior close to the phase
transition, both for fixed $\Nf<\Nfc$ as a function of the bare
coupling~$g$ and for fixed $g > g_\text{cr}$ as a function of $\Nf$.
In the latter case, the data is fitted to an equation of state of the
form
\begin{align} \label{eq:EoS-lattice}
m = A \left[ \left( \Nf - \Nfc \right) + C L^{-\frac{1}{{n}}}
\right] \langle \bpsi \psi \rangle^p + B \langle \bpsi \psi
\rangle^{{p+\frac{1}{b}}},
\end{align}
with $m$ being the bare fermion mass (explicit symmetry-breaking
parameter), $L$ the linear extent of the system, and $A$, $B$, $C$
some constants. Close to criticality in the chiral ($m\rightarrow 0$)
and thermodynamic ($L \rightarrow \infty$) limit,
Eq.~\eqref{eq:EoS-lattice} in fact reduces to the
Kondo-formula~\eqref{eq:orderparam-scaling-Kondo}.
It is interesting to notice that the fit
reported in \cite{Christofi:2007ye} yields $b \simeq 0.37$,
which could be considered as roughly compatible with our result $b \simeq
0.44$.

However, striking discrepancies occur, when one compares the critical behavior for fixed $\Nf < \Nfc$ with the bare coupling as control parameter.
In particular, in the sequence of studies \cite{DelDebbio:1997dv, DelDebbio:1999xg, Hands:1999id, Christofi:2007ye} it is found that the exponent $\delta$ lies between $\delta \approx 2.8$ ($\Nf=2$) and $\delta \approx 5.8$ ($\Nf=6$), i.e., most notably, $\delta$~\emph{increases} with $\Nf$. 
The result for $\Nf=2$ has been confirmed in a 
recent work \cite{Chandrasekharan:2011mn} based on a new
algorithmic approach, where the same lattice action has been
employed directly in the massless-fermion limit with manifest $\mathrm
U(1) \otimes \mathrm U(1)$ chiral symmetry. The resulting anomalous
dimension reads $\eta_\phi^* = 0.65(1)$, which (under the hyperscaling
assumption) is equivalent to $\delta = (5-\eta_\phi^*)/(1+\eta_\phi^*) =
2.64(2)$.  
By contrast, we find (cf.\ Table~\ref{tab:fptvals_dyn-bos}) $\delta
\approx 1.5$ ($\Nf=2$) and $\delta \approx 1.1$ ($\Nf=6$), that is to
say, $\delta$~\emph{decreases} with $\Nf$ and the values lie well
below the above-mentioned lattice predictions%
\footnote{
  As a side remark we note that the DSE studies \cite{Itoh:1994cr,
    Sugiura:1996xk} point to $\delta = 2$ ($\Nf<\Nfc$) and $\delta =
  1$~($\Nf=\Nfc$). (This fact has been debated in 
  Ref.~\cite{Christofi:2007ye}.) 
  The numerical similarity to our results is probably accidental,
  since the nature of the transition reported in
  these studies substantially differs from ours; see
  Eq.~\eqref{eq:essential-scaling} above. 
}.

\begin{figure}
\centering \includegraphics[scale=1.05]{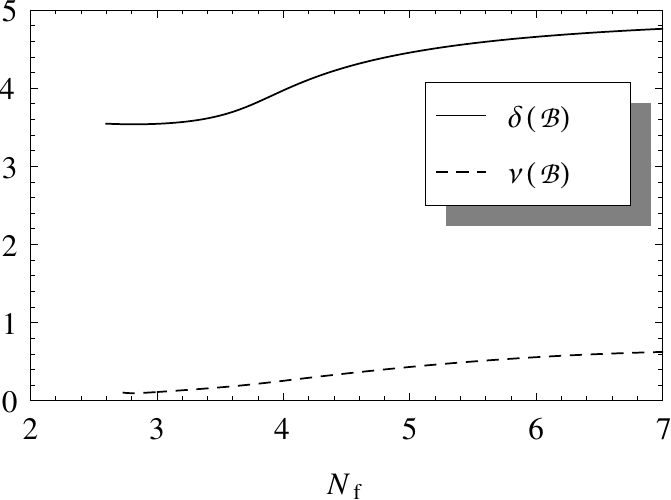}
\caption{Critical exponents $\delta$
and $\nu$ for a theory defined at the fixed point $\mathcal B$ 
with more than one RG relevant
direction; to be compared with Fig.~5 of Ref.~\cite{Christofi:2007ye}.}
\label{fig:deltaFP-B}
\end{figure}

The lattice simulations rely on a microscopic definition of the
Thirring model which is fixed by a bare action including only the
Thirring-like current-current interaction. It is therefore \emph{a
  priori} not 
clear, whether the critical behavior in the
lattice models is in fact given by the Thirring fixed point, cf.\ the
discussion in Ref.~\cite{Gies:2010st}. Our analysis in the preceding
section \ref{sec:fp-structure-phase-transition} supports the
implicit assumption that bare actions on the Thirring axis
belong to the Thirring universality class for all relevant $\Nf$
(cf.\ Fig.~\ref{fig:rgflowNf2}). Nonetheless, since fixed-point
positions are scheme-dependent, our results may not be directly
transferable to the lattice theory. In particular, it is known from
the flow in the pointlike limit,
that a microscopic formulation,
being fixed on the Thirring axis, could, for instance, be influenced
by the fixed point $\mathcal B$,
which has more than one RG relevant direction (see
Fig.~\ref{fig:fermionic-flow}).
In order to get a glimpse of how such a situation could change the
corresponding critical behavior we compute in the dynamically
bosonized formulation the critical exponents associated with fixed
point $\mathcal B$. We emphasize however that a theory defined at
$\mathcal B$ in any case cannot be fixed by just one bare parameter.
Naively using the hyperscaling relation%
\footnote{
  Of course, hyperscaling may not hold at a fixed point with two relevant directions.
}
$\delta(\mathcal B) = [5-\eta_{\phi}^*({\mathcal B})]/[1+\eta^*_{\phi}({\mathcal
B})]$ we can relate the critical exponent $\delta(\mathcal B)$ to the anomalous
dimension $\eta^*_{\phi}(\mathcal B)$ at $\mathcal B$, see
Fig.~\ref{fig:deltaFP-B}.
The values for $\delta(\mathcal B)$ lie well above those for the
Thirring fixed point. Most notably, $\delta(\mathcal B)$ in fact now
\emph{increases} with $\Nf$, in qualitative 
consistency with the behavior shown in Fig.~5 of \cite{Christofi:2007ye}.
For comparison, we also depict the correlation length exponent $\nu =
1/\Theta_1$, associated with the largest negative eigenvalue of the
stability matrix, i.e., the strongest RG relevant direction. 
In order to test this scenario lattice simulations with actions containing both
invariant operators $\sim g_V$ as well as $g_\phi$ are needed.

However, we would like to propose yet another way 
to resolve this
seeming contradiction between our results and lattice
studies: 
All quoted lattice studies implement \emph{staggered fermions}, in which
case the chiral limit is more easily accessible.
Exactly the same lattice action employed in the Thirring studies
has also been used in simulations of the chiral $\mathrm
U(1) \otimes \mathrm U(1)$ Gross-Neveu ($\chi$GN) model in three dimensions
\cite{Barbour:1998yc},
yielding $\nu = 0.88(8)$ and $\eta^*_\phi = 0.46(11)$, probably
consistent with the $\Nf = 2$ lattice Thirring results $\nu = 0.85(1)$,
$\eta^*_\phi = 0.65(1)$ \cite{Chandrasekharan:2011mn} and $\nu =
0.71(4)$, $\eta^*_\phi = 0.60(2)$ \cite{DelDebbio:1997dv}.
In fact, the symmetry breaking pattern of the lattice
Thirring model in the staggered fermion formulation
is \cite{DelDebbio:1999xg} 
\begin{align} \label{eq:breaking-pattern-lattice}
\mathrm U(N_\text{stagg}) \otimes \mathrm U(N_\text{stagg}) \rightarrow \mathrm U(N_\text{stagg}),
\end{align}
where $N_\text{stagg}$ is the number of staggered fermion flavors.
It is related to the number of continuum four-component fermions 
by $\Nf =2N_\text{stagg}$ \cite{Burden:1986by}. It is an open question whether
the breaking pattern of the continuum Thirring model, 
\begin{align} \label{eq:breaking-pattern-continuum}
\mathrm U(2\Nf) \rightarrow \mathrm U(\Nf) \otimes \mathrm U(\Nf),
\end{align}
is restored in the continuum limit of the lattice Thirring model; see
\cite{DelDebbio:1997dv, DelDebbio:1999xg} for a discussion. This
scenario could be checked by a careful analysis of the low-energy
spectrum in the broken phase: the number of Goldstone modes for the
breaking pattern \eqref{eq:breaking-pattern-lattice} would be
$N^2_\text{stagg} = \Nf^2/4$, whereas for
\eqref{eq:breaking-pattern-continuum} one expects $2\Nf^2$ massless
modes. 
Alternatively, one could examine the spectrum of the Dirac operator on the
lattice: a necessary condition for the restoration of the full $\mathrm
U(2\Nf)$ symmetry in the continuum limit is that the lattice Dirac operator
has to exhibit the same four-fold degeneracy of the continuum Dirac operator.
In fact, for a similar strongly-coupled (2+1)-dimensional model (in the
context of graphene), such a comparison reveals significant breaking of 
the $\mathrm U(2\Nf)$ symmetry, even in the vicinity of the second-order phase
transition point~\cite{Giedt:2009td}.
Unfortunately, we do not know of any such study for the
Thirring model so far. 
If the
continuum Thirring pattern \eqref{eq:breaking-pattern-continuum} is
not restored in the simulations, then the lattice Thirring result
could possibly describe the $\chi$GN universality class with breaking
pattern \eqref{eq:breaking-pattern-lattice}. In other words, the
results for critical exponents would differ, simply because the
continuum and the lattice models would not be in the same
universality class.

We have been motivated
by these consideration to compare the lattice results also to results
for a fermionic $\mathrm U(\NL) \otimes \mathrm U(\NR)$
model~\cite{Gies:2009da, Janssen:2012phd}. For $\NL = \NR = 1$, 
a functional RG study has provided the result $\delta_{\chi\text{GN}} =
2.58(9)$. Interestingly
enough, we observe that these $\chi$GN findings are in fact well
consistent with the general trend reported for the lattice Thirring
model, e.g., $\delta = 2.75(9)$ for $N_\text{stagg}=1$
\cite{DelDebbio:1997dv}. [We 
emphasize however that $N_\text{stagg}=1$
corresponds to two four-component continuum flavors, whereas the
$\mathrm U(1) \otimes \mathrm U(1)$ model in Refs.~\cite{Gies:2009da,
  Janssen:2012phd}
is defined with solely one four-component fermion.]  In
Refs.~\cite{Barbour:1998yc, DelDebbio:1999xg} the question has been
raised whether for $\Nf=2$ the distinction between $\chi$GN and
Thirring model might be unimportant and both models might actually
coincide.  In such a scenario the fixed point pertaining to the
$\chi$SB phase transition would incidentally lie in a $\mathrm
U(4)$-symmetric IR attractive subspace of the theory space. While
certainly possible, we know of no argument supporting such a
conjecture; it is also not favored by our results for the $\chi$GN and
Thirring model.

In order to clarify these 
questions, it would be very interesting
to simulate the Thirring model in a formulation in which either the
restoration of the $\mathrm U(2\Nf)$ symmetry can be explicitly
verified in the continuum limit or in which it is manifest. In the
two-component fermion formulation [\Eqref{eq:thirring-NJL}] the latter
option might be pursued using standard methods (e.g., Wilson fermions
\cite{Wilson:1975new}), since there is no notion
of chirality in this formulation and the $\mathrm U(2\Nf)$ is just a flavor
symmetry. Alternatively, in the four-component
language [\Eqref{eq:thirring-NJL_4-comp}], one could
employ more advanced chiral-fermion techniques, such as domain-wall
\cite{Kaplan:1992bt, Shamir:1993zy, Furman:1994ky} or overlap
\cite{Neuberger:1997fp} fermions, both of which are explicitly chirally
symmetric and deserve to be studied on their own rights due to their
relevance to lattice QCD$_4$.

We would finally like to comment on the unorthodox behavior of the
chiral condensate far from the critical point, as reported in the
lattice study \cite{Christofi:2007ye}. There, it has been found that
$\langle \bpsi \psi \rangle$ decreases again for large
(current-current) coupling $g_V$. The corresponding peak position at
$g_V = g_{V,\text{max}}$ is independent of both lattice volume and
bare fermion mass, indicating that its origin is at the UV scale. The
effect has then been attributed to the fact that the vector propagator
in the lattice regularization violates the transversity condition, the
latter being crucial to the renormalizability of the $1/\Nf$
expansion. The impact of the extra divergence can however be absorbed
by a coupling renormalization $g_{V}' = g_V (1- g_V /
g_{V,\text{lim}})^{-1}$, such that the strong coupling limit is
recovered at $g_V = g_{V,\text{lim}}$. $g_{V,\text{lim}}$ is thereupon
identified with the peak position $g_{V,\text{max}}$.

By contrast, our RG results suggest that the 3d Thirring model can in
fact be renormalized nonperturbatively without insisting on a
transverse vector propagator, if one allows for a microscopic
definition in the two-dimensional coupling plane spanned, for
instance, by the couplings $g_V$ and $g_\phi$.  If this remains robust
beyond our approximation, it could provide a natural explanation for
the nonmonotonic behavior of the condensate as a function of~$g_V$:
Since the numerical value of any IR observable (in units of a fixed UV
cutoff $\Lambda$) is first and foremost given by the ``duration'' of
RG time $t = \ln (k/\Lambda)$ before the flow freezes out or enters
the IR regime, it could well be that for a theory defined on the pure
Thirring axis with bare $g_\phi = 0$ an IR quantity decreases again
for large $g_V$ far from criticality. Loosely speaking, just because
we start the flow with a large current-current interaction does not
necessarily mean that
we are closer to the $\chi$SB regime. In order to check these
considerations we have computed the order parameter for different bare
couplings $g_V$ on the Thirring axis and an initially decoupled scalar
sector, showing indeed a maximum far from criticality $g_V \gg
g_\text{cr}$ and a subsequent decrease; see inset of
Fig.~\ref{fig:crit-exp-beta-gamma}.


\section{Conclusions}
We have studied chiral symmetry breaking in the 3d Thirring
model with $\Nf$ fermion flavors. Using the functional RG equation for
the effective average action, we have investigated the RG flow of the
system parametrized in terms of the fundamental fermionic as well as
composite bosonic degrees of freedom. We have analyzed both UV
structure and IR behavior of the theory in the usual formulation with
fixed fields as well as for dynamically bosonized fields, that is to
say, by applying a scale-dependent Hubbard-Stratonovich
transformation. Both formulations show the existence of a Thirring
fixed point with one RG relevant direction. For small $\Nf$ it is
located close to the scalar-channel subspace, while it approaches the
vector-channel subspace (the pure Thirring axis) for
increasing~$\Nf$. All this is well compatible with previous findings
or indications in the purely fermionic
description~\cite{Gies:2010st}. 

For entering the symmetry-broken regime, however, the
description using composite bosonic fields is highly advantageous.
Whereas partial bosonization has become a standard tool for
functional RG analyses of strongly-correlated fermion systems, a
quantitatively reliable study of the 3d Thirring model appears to
require {\em dynamical bosonization} \cite{Gies:2001nw,
  Pawlowski:2005xe, Gies:2006wv, Floerchinger:2009uf}. The reason is
that the physics of competing channels is technically affected by
the Fierz ambiguity representing a challenge for many techniques
based on Hubbard-Stratonovich transformations. The 3d Thirring
model therefore serves as a paradigm example for the
competing-channel problem and its resolution through dynamical
bosonization.
 
For the present model, we have confirmed previous results from
DSE studies and Monte-Carlo simulations indicating that chiral
symmetry breaking, corresponding to a condensation in the scalar
channel, occurs for large enough coupling if the number of fermion
flavors $\Nf$ is smaller than a critical value $\Nfc$, while it is
shown to be absent for $\Nf>\Nfc$, independently of the coupling.
We obtain the estimate $\Nfc \simeq 5.1(7)$, where we
have used regulator dependencies as an indicator for the systematic
error of our approximation.

Since all previous studies we know of rely on a microscopic definition
of the 3d Thirring model with pure current-current (Thirring-like)
interaction, the mechanism behind the quantum phase transition at
$\Nfc$ has so far been rather unclear. The present work has shown for
the first time that the critical flavor number may occur due to a
competition between different condensation channels: the NJL-type
scalar channel on the one hand, dominating for small $\Nf$ and
triggering $\chi$SB, and the Thirring-type vector channel on the other
hand, dominating for large~$\Nf$ and 
inhibiting $\chi$SB.
In particular, our results exclude a quantum phase transition mechanism
based on fixed-point annihilation that would imply essential scaling
of order parameters as a function of $\Nfc-\Nf$.

The RG approach is particularly useful for predicting
the critical phenomena associated with a given universality class. We have
computed the critical behavior for fixed $\Nf < \Nfc$ as a function of the bare
coupling in terms of the exponents $\nu$, $\eta^*_\phi$, $\beta$, and $\gamma$
both for a microscopic definition of the model at the Thirring fixed point as
well as on the pure Thirring axis (``pure'' Thirring model).
For $\Nf=2$, we have 
verified that the hyperscaling relations are fulfilled for UV
complete models defined at the Thirring fixed point. Models with
initial conditions on the pure Thirring axis necessarily inherit a certain
degree of nonuniversality which becomes visible in small violations
of hyperscaling, presumably induced by the vicinity of another fixed
point (fixed point $\mathcal B$ in our notation).
Strictly speaking, the model starting on the
pure Thirring axis is not itself a UV complete quantum field theory. We can,
however, think of it as belonging to an RG trajectory (line of constant physics)
that emanates from fixed point $\mathcal B$ and thus has a definite UV
completion. For the theory defined at the Thirring fixed point, we have also
computed the corrections-to-scaling exponent $\omega$. 

Close to the quantum critical point $\Nf \nearrow \Nfc$, we have
discussed the scaling behavior as a function of $\Nf$ (for fixed
couplings above their critical values). For the quantum phase
transition as a function of $\Nfc-\Nf$, we have computed the
exponent $b\simeq0.44$ (``magnetization exponent''), implying that
there is clear evidence for the transition to be of second order. We
have also computed the dynamically generated fermion mass $\bar
m_\text{R,f}^2$ (fermion gap) and the $\tau$-mode mass $\bar
m_{\mathrm{R},\tau}^2$ in units of the radial mass $\bar
m_{\mathrm{R},\rho}^2$ (inverse healing length) as a function of
$\Nf<\Nfc$. In principle, these are inherent predictions of our
analysis and could be verified by correlator measurements in lattice
simulations. Since our IR analysis is affected by an artificial
non-decoupling of Goldstone modes, our quantitative estimate should,
however, not be taken too literally. We consider our results on
critical exponents as our most accurate quantitative predictions.

At the present stage, the long-range behavior in the
vector-dominated phase for large $\Nf>\Nfc$ is difficult to
resolve. Our approximation with pointlike vector channel does not
allow a reliable prediction of the IR properties of the system for
$\Nf \gtrsim \Nfc$, since for small vector mass $m_V^2 \lesssim
\mathcal O(1)$ momentum-dependent terms $\propto Z_{V,k}, \bar
A_{V,k}, \bar \zeta_k$ in the effective action
\eqref{eq:thirring-bosonized-truncation}
become important as we can read off from the vector anomalous
dimension growing large. In this sense, 
our estimate for the critical flavor number might be a lower
bound to the true value of $\Nfc$. A larger critical flavor number
could, for instance, arise from a nontrivial interplay of dynamical
vector and scalar channels driving the system to criticality also
for larger flavor number.
Thus, we consider our findings for $\Nfc$ to be
still compatible with the lattice results \cite{Christofi:2007ye},
pointing to $\Nfc\simeq6.6$. In any case however, it would be
very interesting to see in how far the inclusion of momentum-dependent
terms in the vector sector would modify the IR flow of the large-$\Nf$
theory. This is in particular true for the term $\propto \bar\zeta_k
V_\mu V_\mu \partial_\nu V_\nu$, which has no analogue in scalar field
theories and thus could lead to a qualitatively different IR behavior
as compared to the latter. Furthermore, at this point we also cannot
exclude with certainty that this term does not change the UV structure
of our theory in the large-$\Nf$ limit. This deserves further
investigation.

Whereas most~\cite{Gomes:1990ed, Itoh:1994cr, Sugiura:1996xk, Kondo:1995jn,
Kim:1996xza, DelDebbio:1997dv, DelDebbio:1999xg, Hands:1999id, Christofi:2007ye}
(but not all~\cite{Hong:1993qk}) of the previous studies agree at least on the
very existence of a critical flavor number at order $\Nfc \sim \mathcal O(2\dots
7)$, the nature of the phase transition has so far been a substantially delicate
issue. Based on our detailed predictions, in particular for the transition at
fixed $\Nf < \Nfc$ as a function of the coupling, we believe that it may now be
possible to resolve the discrepancies in the literature. For that purpose, we
propose an independent investigation of the critical behavior of the 3d Thirring
model, for instance, by means of a Monte Carlo simulation with a lattice action
exhibiting manifest $\mathrm U(2\Nf)$ symmetry.

\begin{acknowledgments}
Helpful discussions with J.~Braun, S.~Hands, I.~Herbut, M.~Huber, D.~Roscher,
D.~Scherer, L.~von Smekal, B.~Wellegehausen, and A. Wipf are gratefully
acknowledged. 
This work has been supported by the DFG under GRK~1523, FOR~723, and
Gi~328/5-2.
\end{acknowledgments}

\appendix*


\section{Loop integrals}
\label{app:threshold-functions}
For self-containedness, we give here the explicit results for the loop
integrals for linear and sharp cutoff.  The regulator functions $R_k$
present in the Wetterich equation~\eqref{eq:wetterich-equation} may be
written in terms of dimensionless shape functions $r_k$ via
\begin{align}
R_{\phi,k}(q) & = Z_{\phi,k} q^2 r_{\phi,k}(q^2), \\
R_{\psi,k}(q) & = - Z_{\psi,k} \slashed{q} r_{\psi,k}(q^2),
\end{align}
with collective bosonic and fermionic fields $\phi= (\phi^{ab},V_\mu,\dots)$ and $\psi = (\psi^a,\dots)$, respectively. The linear cutoff, which satisfies an optimization criterion \cite{Litim:2001up}, is defined as
\begin{align}
r_{\phi,k}^\text{opt}(q^2) & = 
\left(\frac{k^2}{q^2}-1\right) \Theta(k^2 - q^2), \\
r_{\psi,k}^\text{opt}(q^2) & = 
\left(\sqrt{\frac{k^2}{q^2}} - 1\right) \Theta(k^2 - q^2).
\end{align}
The sharp cutoff is defined as the $a\rightarrow \infty$ limit of the class of regulators given by
\begin{align}
r_{\phi,k}^\text{sc}(q^2) & = 
a \left(\frac{k^2}{q^2}-1\right) \Theta(k^2 - q^2), \\
r_{\psi,k}^\text{sc}(q^2) & =
\left(\sqrt{a\left(\frac{k^2}{q^2} - \frac{a-1}{a}\right)} - 1 \right) \Theta(k^2 - q^2),
\end{align}
where we demand for definiteness that the sharp-cutoff limit $a \rightarrow \infty$ is to be taken \emph{after} the integration over the internal momentum $q$, i.e., after the substitution into the threshold functions~\cite{Reuter:2001ag}.

The one-loop structure of the Wetterich equation guarantees that the flow
equations can always be written in terms of single integrals---the threshold
functions, which encode the details of the regularization scheme. Their
definitions are
\begin{widetext}
\begin{align}
\ell_0^{\mathrm{(B/F)}d}(\omega;\eta_{\phi/\psi}) & = 
\frac{1}{2} k^{-d} \TildeDt \IntX x^{\frac{d}{2}-1}
\log\left[P_{\phi/\psi}(x)+\omega k^2\right], \displaybreak[0] \\
\ell_{n}^{\mathrm{(B/F)}d}(\omega;\eta_{\phi/\psi}) & =
\frac{(-1)^n}{(n-1)!} \partial_\omega^n
\ell_0^{\mathrm{(B/F)}d}(\omega;\eta_{\phi/\psi}) 
= -\frac{1}{2} k^{2n-d} \TildeDt \IntX x^{\frac{d}{2}-1} \left[
P_{\phi/\psi}(x)+\omega k^2\right]^{-n}, \displaybreak[0] \\
\ell_{n_1,n_2}^{\mathrm{(BB)}d}(\omega_1,\omega_2;\eta_\phi,\eta_V) & =
-\frac{1}{2} k^{2(n_1+n_2)-d} \TildeDt \IntX x^{\frac{d}{2}-1}
\left[P_\phi(x) + \omega_1 k^2\right]^{-n_1}
\left[P_V(x) + \omega_2 k^2\right]^{-n_2}, \displaybreak[0] \\
\ell_{n_1,n_2}^{\mathrm{(FB)}d}(\omega_1,\omega_2;\eta_\psi,\eta_{\phi}) & =
-\frac{1}{2} k^{2(n_1+n_2)-d} \TildeDt \IntX x^{\frac{d}{2}-1}
\left[P_\psi(x) + \omega_1 k^2\right]^{-n_1}
\left[P_{\phi}(x) + \omega_2 k^2\right]^{-n_2}, \displaybreak[0] \\
\ell_{n_1,n_2,n_3}^{\mathrm{(FBB)}d}(\omega_1,\omega_2,\omega_3;\eta_\psi,
\eta_\phi) & =
-\frac{1}{2} k^{2(n_1+n_2+n_3)-d} \TildeDt \IntX x^{\frac{d}{2}-1}
\left[P_\psi(x) + \omega_1 k^2\right]^{-n_1}
\left[P_{\phi}(x) + \omega_2 k^2\right]^{-n_2}
\nonumber \\ &\hspace{25em} \times
\left[P_{\phi}(x) + \omega_3 k^2\right]^{-n_3},
\displaybreak[0] \\
m_{2,2}^{\mathrm{(B)}d}(\omega_1,\omega_2;\eta_\phi) & =
-\frac{1}{2} k^{6-d} \TildeDt \IntX x^{\frac{d}{2}} 
\left[\partial_{x} \frac{1}{P_\phi(x) + \omega_1 k^2}\right]
\left[\partial_{x} \frac{1}{P_\phi(x) + \omega_2 k^2}\right], \displaybreak[0] \\
m_2^{\mathrm{(F)}d}(\omega;\eta_{\psi}) & =
-\frac{1}{2} k^{6-d} \TildeDt \IntX x^{\frac{d}{2}} 
\left[\partial_{x} \frac{1} {P_{\psi}(x) + \omega k^2}\right]^2, \displaybreak[0] \\
m_4^{\mathrm{(F)}d}(\omega;\eta_\psi) & = 
- \frac{1}{2} k^{4-d} \TildeDt \IntX x^{\frac{d}{2}+1} \left[
\partial_{x} \frac{1+r_\psi(x)}{P_\psi(x) + \omega k^2}
\right]^2, \displaybreak[0] \\
m_{1,2}^{\mathrm{(FB)}d}(\omega_1,\omega_2;\eta_\psi,\eta_{\phi/V}) & =
\frac{1}{2} k^{4-d} \TildeDt \IntX x^{\frac{d}{2}}
\frac{1+r_\psi(x)}{P_\psi(x) + \omega_1 k^2}
\partial_x \frac{1}{P_{\phi/V}(x) + \omega_2 k^2},
\end{align}
with $n_i \in \mathbbm N$ and where we have suppressed the scale index $k$ for the sake of simplicity. Moreover, we have abbreviated the (inverse) regularized propagator parts by
\begin{align}
P_{\phi}(x) & \coloneqq x \left[1 + r_{\phi}(x)\right], &
P_\psi(x) & \coloneqq x \left[1 + r_\psi(x)\right]^2.
\end{align}
For the integrations, we have substituted $q^2 \mapsto x$, viz.,
\begin{align}
\int \frac{\mathrm{d}^dq}{(2\pi)^d} = 4 v_d \int \mathrm dq
\,q^{d-1} = 2 v_d \int \mathrm{d}x\, x^{\frac{d}{2}-1}
\end{align}
with $v_d \coloneqq \frac{1}{4} \text{Vol}(S^{d-1})/ (2\pi)^d = [2^{d+1} \pi^{d/2} \Gamma(d/2)]^{-1}$. $\TildeDt$~is defined to act only on the regulator's $t$-dependence,
\begin{align}
\TildeDt & \coloneqq 
\sum_{\Phi=\phi,V,\psi} \int \mathrm d x'
\frac{\partial_t\left[Z_\Phi r_\Phi(x')\right]}{Z_\Phi}
\frac{\delta}{\delta r_\Phi(x')} 
\\
& \phantom{:}= \int \mathrm d x' x' \left\{
\frac{\partial_t\left[Z_\phi r_\phi(x')\right]}{Z_\phi}
\frac{\delta}{\delta P_\phi(x')} +
\frac{\partial_t\left[Z_V r_V(x')\right]}{Z_V}
\frac{\delta}{\delta P_V(x')}
%
%
+ 2 \left[1 + r_\psi(x')\right] 
\frac{\partial_t\left[Z_\psi r_\psi(x')\right]}{Z_\psi}
\frac{\delta}{\delta P_\psi(x')}
\right\}.
\end{align}
Both the linear and sharp regulators have the very convenient feature that all loop integrals can be performed explicitly. For the linear cutoff the results are
\begin{align}
\ell_{0}^{\mathrm{(B)}d}(\omega;\eta_\phi) & =
\frac{2}{d}\left( 1 - \frac{\eta_\phi}{d+2}\right) \frac{1}{1+\omega}, \displaybreak[0] \\
\ell_{n}^{\mathrm{(B)}d}(\omega;\eta_\phi) & =
\frac{2}{d}\left( 1 - \frac{\eta_\phi}{d+2}\right) \frac{n}{(1+\omega)^{n+1}},
\displaybreak[0] \\
\ell_{0}^{\mathrm{(F)}d}(\omega;\eta_\psi) & =
\frac{2}{d}\left( 1 - \frac{\eta_\psi}{d+1}\right) \frac{1}{1+\omega},\displaybreak[0] \\
\ell_{n}^{\mathrm{(F)}d}(\omega;\eta_\psi) & = 
\frac{2}{d}\left( 1 - \frac{\eta_\psi}{d+1}\right) \frac{n}{(1+\omega)^{n+1}}, \displaybreak[0] \\
\ell_{n_1,n_2}^{\mathrm{(BB)}d} (\omega_1,\omega_2;\eta_\phi,\eta_V) & =
\frac{2}{d}
\left[
\left(1-\frac{\eta_\phi}{d+2}\right) \frac{n_1}{1+\omega_1} +
\left(1-\frac{\eta_V}{d+2}\right) \frac{n_2}{1+\omega_2} 
\right]
\frac{1}{(1+\omega_1)^{n_1} (1+\omega_2)^{n_2}}, \displaybreak[0] \\
\ell_{n_1,n_2}^{\mathrm{(FB)}d} (\omega_1,\omega_2;\eta_\psi,\eta_{\phi}) & =
\frac{2}{d}
\left[
\left(1-\frac{\eta_\psi}{d+1}\right) \frac{n_1}{1+\omega_1} +
\left(1-\frac{\eta_{\phi}}{d+2}\right) \frac{n_2}{1+\omega_2} 
\right]
\frac{1}{(1+\omega_1)^{n_1} (1+\omega_2)^{n_2}}, \displaybreak[0] \\
\ell_{n_1,n_2,n_3}^{\mathrm{(FBB)}d}
(\omega_1,\omega_2,\omega_3;\eta_\psi,\eta_\phi) & =
\frac{2}{d}
\left[
\left(1-\frac{\eta_\psi}{d+1}\right) \frac{n_1}{1+\omega_1} +
\left(1-\frac{\eta_{\phi}}{d+2}\right) 
\left(\frac{n_2}{1+\omega_2} + \frac{n_3}{1+\omega_3} \right)
\right]
\nonumber \\ & \hspace*{20em}
\times\frac{1}{(1+\omega_1)^{n_1} (1+\omega_2)^{n_2} (1+\omega_3)^{n_3}}, \displaybreak[0] \\
m_{2,2}^{\mathrm{(B)}d}(\omega_1,\omega_2;\eta_\phi) & = 
\frac{1}{(1+\omega_1)^2 (1+\omega_2)^2}, \displaybreak[0] \\
m_2^{\mathrm{(F)}d}(\omega;\eta_{\psi}) & = 
\frac{1}{(1+\omega)^4}, \displaybreak[0] \\
m_4^{\mathrm{(F)}d}(\omega;\eta_\psi) & = 
\frac{1}{(1+\omega)^4} + \frac{1-\eta_\psi}{d-2} \frac{1}{(1+\omega)^3}
- \left( \frac{1-\eta_\psi}{2d-4} + \frac{1}{4}\right) \frac{1}{(1+\omega)^2}, \\
m_{1,2}^{\mathrm{(FB)}d}(\omega_1,\omega_2;\eta_\psi,\eta_{\phi/V}) & =
\left(1-\frac{\eta_{\phi/V}}{d+1}\right)
\frac{1}{(1+\omega_1)(1+\omega_2)^2}.
\end{align}
For the sharp cutoff we find
\begin{align}
\ell_{0}^{\mathrm{(B/F)}d}(\omega;\eta_{\phi/\psi}) & = 
- \log(1+\omega) + \ell_0^{\mathrm{(B/F)}d}(0;\eta_{\phi/\psi}), \\
\ell_{n}^{\mathrm{(B/F)}d}(\omega;\eta_{\phi/\psi}) & = 
\frac{1}{(1+\omega)^n}, \\
\ell_{n_1,n_2}^{\mathrm{(BB)}d} (\omega_1,\omega_2;\eta_\phi,\eta_V) & =
\frac{1}{(1+\omega_1)^{n_1} (1+\omega_2)^{n_2}}, \\
\ell_{n_1,n_2}^{\mathrm{(FB)}d} (\omega_1,\omega_2;\eta_\psi,\eta_{\phi/V}) & =
\frac{1}{(1+\omega_1)^{n_1} (1+\omega_2)^{n_2}}, \\
\ell_{n_1,n_2,n_3}^{\mathrm{(FBB)}d}
(\omega_1,\omega_2,\omega_3;\eta_\psi,\eta_\phi) & =
\frac{1}{(1+\omega_1)^{n_1} (1+\omega_2)^{n_2} (1+\omega_3)^{n_3}}, \\
m_{2,2}^{\mathrm{(B)}d}(\omega_1,\omega_2;\eta_\phi) & = 
\frac{1}{(1+\omega_1)^2 (1+\omega_2)^2}, \\
m_2^{\mathrm{(F)}d}(\omega;\eta_{\psi}) & = 
\frac{1}{(1+\omega)^4}, \\
m_4^{\mathrm{(F)}d}(\omega;\eta_\psi) & = 
\frac{1}{(1+\omega)^4}, \\
m_{1,2}^{\mathrm{(FB)}d}(\omega_1,\omega_2;\eta_\psi,\eta_{\phi}) & =
\frac{1}{(1+\omega_1)(1+\omega_2)^2}.
\end{align}
\end{widetext}
%


\bibliographystyle{hep-revtex}
\bibliography{bibliography}{}

\end{document}